\documentclass[12pt]{article}
\usepackage{axodraw}
\usepackage{psfig}
\newcommand{\be}{\begin{equation}}
\newcommand{\ee}{\end{equation}}
\newcommand{\sss}[1]{\mbox{\scriptsize #1}}

\newcommand{\Li}{\mbox{${\mbox L}{\mbox i}_{2}$}}
\newcommand{\real}{{\cal\mbox{Re\,}}}
\newcommand{\imag}{{\cal\mbox{Im\,}}}
\newcommand{\la}{\langle}
\newcommand{\ra}{\rangle}
\newcommand{\M}{{\cal M}}
\newcommand{\J}{{\cal J}}
\newcommand{\I}{{\cal I}}
\newcommand{\OO}{{\cal O}}
\newcommand{\PP}{{\cal P}}
\newcommand{\C}{{\cal C}}
\newcommand{\E}{{\cal E}}
\newcommand{\e}{\varepsilon}
\newcommand{\D}{{\cal D}}
\newcommand{\lsim}
{\;\raisebox{-.3em}{$\stackrel{\displaystyle <}{\sim}$}\;}

\newcommand{\es}{\varepsilon \hspace{-0.48em}/}
\newcommand{\ks}{k\hspace{-0.52em}/\hspace{0.04em}}
\newcommand{\ps}{p\hspace{-0.42em}/}
\newcommand{\qs}{q\hspace{-0.5em}/}
\newcommand{\Vs}{V\hspace{-0.77em}/\hspace{0.11em}}

\newcommand{\GeV}{\unskip\,\mathrm{GeV}}
\newcommand{\MeV}{\unskip\,\mathrm{MeV}}
\newcommand{\TeV}{\unskip\,\mathrm{TeV}}

\begin{document}
\pagestyle{empty}
\begin{flushright}
{DTP/98/90}\\
{INLO-PUB 11/98}
\end{flushright}
\vspace*{5mm}
\begin{center}
  {\bf 
   RADIATIVE CORRECTIONS TO PAIR PRODUCTION \\ 
   OF UNSTABLE PARTICLES: RESULTS FOR \boldmath $E^{+}E^{-}\to 4$ FERMIONS
  } \\
\vspace*{1cm} 
{\bf W.~Beenakker}$^{*)}$\\ 
\vspace{0.3cm}
Physics Department, University of Durham, Durham DH1 3LE, England\\
\vspace{0.5cm}
{\bf F.A.~Berends} \ \  
{\bf and}  \ \ 
{\bf A.P.~Chapovsky}$^{\dagger)}$\\
\vspace{0.3cm}
Instituut--Lorentz, University of Leiden, The Netherlands\\
\vspace*{2cm}  
                                {\bf ABSTRACT} \\ \end{center}
\vspace*{5mm}
\noindent
Radiative corrections to processes that involve the production and
subsequent decay of unstable particles are complex due to various theoretical
and practical problems. The so-called double-pole approximation 
offers a way out of these problems. This method is applied to the 
reaction $e^{+}e^{-} \to W^{+}W^{-} \to 4\ $fermions, which allows us to
address all the key issues of dealing with unstable particles, 
like gauge invariance, interactions between different stages of the
reaction, and overlapping resonances. Within the double-pole approximation
the complete $\OO(\alpha)$ electroweak corrections are evaluated for this
off-shell $W$-pair production process. Examples of the effect of these
corrections on a number of distributions are presented. These comprise mass 
and angular distributions as well as the photon-energy spectrum.
\vspace*{3cm}\\ 
\begin{flushleft}
November 1998
\end{flushleft}
\noindent 
\rule[.1in]{16.5cm}{.002in}

\noindent
$^{*)}$Research supported by a PPARC Research Fellowship.\\
$^{\dagger)}$Research supported by the Stichting FOM.
\vspace*{0.3cm}

\vfill\eject

\setcounter{page}{1}
\pagestyle{plain}



\section{Introduction}
\label{sec:intro}

In order to test the Standard Model (SM) of electroweak interactions, 
millions of $Z$ bosons have been produced and studied at LEP1~\cite{lep1}. 
High-precision measurements of the $Z$-boson parameters have been performed 
by measuring
\be
        e^{+}e^{-} \to Z \to \bar{f}f
\ee
and comparing the results with the theoretical predictions. This requires 
theoretical precision calculations for the reaction
\be
        e^{+}e^{-} \to \bar{f}f,
\ee
involving {\it all} lowest-order diagrams and the radiative corrections (RC) 
to them~\cite{lep1-calc}.

Measuring the parameters of the $W$ boson provides further tests of the SM. 
In particular, the accuracy of the $W$-boson mass ($M_{W}$) has to be 
improved, since it can give indirect information on the Higgs sector and 
on physics beyond the SM. At LEP2, $W$ bosons can be studied in the 
$W$-pair production reaction
\be 
\label{intro/ee->ww->4f}
        e^{+}e^{-} \to W^{+}W^{-} \to 4 \ \mbox{fermions}.
\ee
Besides the mass $M_{W}$ also the Yang--Mills form of the triple gauge-boson
couplings (TGC) $ZW^{+}W^{-}$ and $\gamma W^{+}W^{-}$ can be tested in
reaction (\ref{intro/ee->ww->4f}), where they directly manifest themselves 
in the lowest-order cross-section. Deviations from the SM Yang--Mills 
couplings can be searched for most effectively by a detailed investigation
of angular distributions~\cite{lep2-TGC}.

Just like in the case of the $Z$ boson, the precise determination of the 
$W$-boson parameters requires both accurate experiments and accurate 
theoretical predictions. Since the statistics at LEP2 is much smaller than
at LEP1, the theoretical calculations do not have to be as precise as those 
for LEP1. However, just the Born prediction for process 
(\ref{intro/ee->ww->4f}), involving only three diagrams, is not 
sufficient~\cite{lep2-ww}. Therefore, one would like to have an idea 
how the cross-sections for the four-fermion process
\be
\label{intro/ee->4f}
        e^{+}e^{-} \to 4 \ \mbox{fermions}
\ee
are affected by the inclusion of the remaining lowest-order diagrams 
and the RC to the {\it complete} set of lowest-order diagrams.

The question of the complete lowest-order calculation of process
(\ref{intro/ee->4f}) has been studied in the 
literature~\cite{lep2-ww,background}. 
Roughly speaking, those diagrams that contain a single $W$ boson are a
factor of ${\OO}(\Gamma_{W}/M_{W})$ smaller than those for the $W$-pair 
production process (\ref{intro/ee->ww->4f}). Diagrams that do not contain 
a $W$ boson at all are down by $\OO(\Gamma_{W}^{2}/M_{W}^{2})$.
Note, however, that some diagrams that do not contain two $W$ bosons can 
nevertheless be large, e.g.~as a result of the exchange of almost real 
photons. Besides such special cases, the lowest-order
``background'' diagrams in (\ref{intro/ee->4f}), i.e.~the non-$W$-pair
diagrams, will give at most a correction of $\OO(\Gamma_{W}/M_{W})$
to the Born cross-section of process (\ref{intro/ee->ww->4f}).

The RC to process (\ref{intro/ee->ww->4f}) are a priori of $\OO(\alpha)$,
i.e.~they are genericly of the same order as the lowest-order background
diagrams. In analogy to the lowest-order case, the RC to the background
diagrams are at most of $\OO(\alpha\Gamma_{W}/M_{W})$.

In view of the above estimates, the most relevant corrections to the 
Born cross-section of process (\ref{intro/ee->ww->4f}) are divided into
two classes:
\begin{itemize}
  \item[1.]     $\OO(\alpha)$ RC to the $W$-pair process 
                (\ref{intro/ee->ww->4f}).
  \item[2.]     $\OO(\Gamma_{W}/M_{W})$ lowest-order background 
                contributions from the full four-fermion process
                (\ref{intro/ee->4f}).
\end{itemize}
At present the first class of corrections has not been fully studied in the 
literature. What has been discussed quantitatively and qualitatively is 
merely a subset of $\OO(\alpha)$ effects:
\begin{itemize}
  \item[1.1.]   Initial-state radiation (ISR). Since ISR corrections are 
                enhanced by collinear-photon effects, they are large and
                even higher-order contributions should be taken into 
                account \cite{lep2-ww}--\cite{ww-review}.
  \item[1.2.]   Final-state radiation (FSR). Similarly FSR can be sizeable,
                in particular for distributions in the 
                fermion-pair invariant masses. This QED effect has recently 
                been pointed out in the literature~\cite{fsr}. 
                Again higher-order corrections are non-negligible for these
                $W$ line-shape distributions.
  \item[1.3.]   Whereas the effects 1.1 and 1.2 are enhanced by logarithms
                originating from collinear photons, also non-enhanced
                $\OO(\alpha)$ effects have been studied. One of those 
                belongs to the class of the so-called non-factorizable 
                corrections~\cite{Me96}--\cite{nf-corr/ddr}, 
                i.e.~corrections that at first sight do not seem to have two 
                overall $W$ propagators as factors. As such one would expect
                these corrections to be suppressed by additional powers of
                $\Gamma_{W}/M_{W}$. In the special case of semi-soft photonic 
                corrections, however, the suppression does not take place 
                and an $\OO(\alpha)$ correction survives. This $\OO(\alpha)$ 
                correction has been calculated and turns out to be relatively 
                small. In the vicinity of the $W$-pair threshold also the
                Coulomb interaction between the unstable $W$ bosons has been
                studied in great detail~\cite{coulomb}. The corrections are 
                relatively large, but higher orders are not required.  
  \item[1.4.]   The effects 1.1--1.3 are all of QED origin.
                Complete electroweak RC to process (\ref{intro/ee->ww->4f})
                have not been applied yet, but some attempts have been made 
                to take into account the dominant effects. 
                An overall effect of electroweak corrections has been
                considered by using $G_{\mu}$ as coupling constant instead of 
                $\alpha$. From other calculations it is known that a $G_{\mu}$
                parametrization of the lowest-order term reduces the size of 
                the one-loop non-photonic RC considerably. In addition to this
                overall effect, the full electroweak RC to stable $W$-pair 
                production~\cite{prod-corr}--\cite{wwgamma} and on-shell $W$ 
                decay~\cite{decay-corr} are already known for quite some time.
\end{itemize}

The purpose of this paper is twofold. In the first place we will present
a complete quantitative evaluation of $\OO(\alpha)$ electroweak RC to 
$W$-pair-mediated four-fermion production. 
To this end we calculate all $\OO(\alpha)$ factorizable corrections and add  
them to our previous results on non-factorizable corrections.
In this way the gap in the above list of corrections 1.1--1.4 will be
filled, and the exact size of all $\OO(\alpha)$ corrections to reaction 
(\ref{intro/ee->ww->4f}) will be known. In the light of the physics 
motivation given above, this practical result is clearly wanted.

Secondly, this paper discusses in detail the so-called pole scheme.
This scheme offers a way to avoid theoretical problems associated with the
gauge-invariant treatment of reactions that involve the production and
subsequent decay of unstable particles. The study of the $W$-pair case
serves as a relevant example of this method and can be used as a
guideline for other cases. Since there are more unstable-particle production 
processes to be studied at future accelerators, the relevance of the 
presented study goes beyond the $W$-pair case.

The outline of the paper is as follows.
In Sect.~\ref{sec:gauge} various gauge-invariance issues will be discussed. 
In Sect.~\ref{sec:pole-scheme} the pole-scheme treatment is described,
with special emphasis on the so-called double-pole approximation (DPA),
since it is the basic ingredient for our calculation of the $\OO(\alpha)$ RC.
The comparison with the exact lowest-order evaluation in 
Sect.~\ref{sec:born_num} gives an estimate of the accuracy of this 
approximation. The discussion of the RC in the DPA for 
the $W$-pair production process (\ref{intro/ee->ww->4f}) is presented in 
Sect.~\ref{sec:corr}, whereas the corresponding numerical results can be 
found in Sect.~\ref{sec:plots}.
In Sect.~\ref{sec:concl} we give the conclusions of our study.


\section{Gauge-invariant treatment of unstable gauge bosons}
\label{sec:gauge}

\subsection{Lowest order}

The above-described processes, with or without RC, all involve 
fermions in the initial and final state and unstable gauge bosons as 
intermediate particles. Sometimes a photon is also present in the final 
state. If complete sets of graphs contributing to a given process are taken 
into account, the associated matrix elements are in principle gauge-invariant, 
i.e.~they are independent of gauge fixing and respect Ward identities.
This is, however, not guaranteed for incomplete sets of graphs like the ones 
corresponding to the off-shell $W$-pair production process 
(\ref{intro/ee->ww->4f}). Indeed this process was found to violate the
$SU(2)$ Ward identities~\cite{ww-review}. 
 
In addition, the unstable gauge bosons that appear as intermediate 
particles can give 
rise to poles $1/(p^2-M^2)$ in physical observables if they are treated as 
stable particles. This can be cured by introducing the finite decay widths for 
these gauge bosons. In field theory, such widths arise naturally from the 
imaginary parts of higher-order diagrams describing the gauge-boson 
self-energies, resummed to all orders. However, in doing a Dyson summation 
of self-energy graphs, we are singling out only a very limited subset of all 
possible higher-order diagrams. It is therefore not surprising that one often 
ends up with a result that violates Ward identities and/or retains some gauge
dependence resulting from incomplete higher-order contributions. 

Since the latter gauge breaking is caused by the finite decay width and is, 
as such, in principle suppressed by powers of $\Gamma/M$, 
one might think that it is of academic nature. For LEP1 observables we 
indeed know that gauge breaking can be negligible for all practical purposes. 
However, the presence of small scales can amplify the gauge-breaking terms. 
This is for instance the case for almost real space-like 
photons~\cite{BW70,BHF1} or longitudinal gauge bosons ($V_L$) at high 
energies~\cite{BHF2}, involving scales of $\OO(p_{_B}^2/E_{_B}^2)$ 
for $B=\gamma,V_L$. The former plays an important role in TGC studies in 
the reaction $e^+e^- \to e^-\bar{\nu}_{e}u\bar{d}$, where the electron may 
emit a virtual photon with an invariant mass $p_{\gamma}^2$ as small as 
$m_e^2$. The latter determines the high-energy behaviour of the generic 
reaction $e^+e^- \to 4\ $fermions.
In these situations the external current coupled to the photon or to the 
longitudinal gauge boson becomes approximately proportional to $p_{_B}$. 
Sensible theoretical predictions, with a proper dependence on $p_{\gamma}^2$ 
and a proper high-energy behaviour, are only possible if the amplitudes with 
external currents replaced by the corresponding gauge-boson momenta fulfil 
appropriate Ward identities. 

So, how should one go about including the finite decay widths?
The simplest approach is the so-called ``fixed-width scheme'', involving the 
systematic replacement $1/(p_V^2-M_V^2) \to 1/(p_V^2-M_V^2+i M_V\Gamma_V)$, 
where $\Gamma_V$ denotes the physical width of the gauge boson $V$ with mass 
$M_V$ and momentum $p_V$. Since in perturbation theory the propagator for 
space-like momenta does not develop an imaginary part, the introduction of a 
finite width also for $p_V^2<0$ has no physical motivation and in fact violates
unitarity, i.e. the cutting equations. This can be cured by using a
running width $i M_V\Gamma_V(p_V^2)$ instead of the constant one
$i M_V\Gamma_V$ (``running-width scheme''). However, as in general the 
resonant diagrams are not gauge-invariant by themselves, the introduction of 
a constant or running width destroys gauge invariance.  

A truly gauge-invariant scheme evidently has to be a bit more sophisticated 
than this. It should be stressed, however, that any such scheme is arbitrary 
to a greater or lesser extent: since the Dyson summation must necessarily be 
taken to all orders of perturbation theory, and we are not able to compute the 
complete set of {\it all} Feynman diagrams to {\it all} orders, the various 
schemes differ even if they lead to formally gauge-invariant results. Bearing 
this in mind, we need besides gauge invariance some physical motivation for 
choosing a particular scheme. In this context two options can be mentioned.
The first option is the so-called ``pole scheme''~\cite{pole-scheme}.
In this scheme one decomposes the complete amplitude by expanding around the 
poles. As the physically observable residues of the poles are gauge-invariant, 
gauge invariance is not broken if the finite width is taken into account in 
the pole terms $\propto 1/(p_V^2-M_V^2)$. Note that the leading terms in such 
an expansion play a special role in view of their close relation to on-shell 
production and decay of the unstable particles. This point will be
explained in more detail in the following sections.
 
The second option is based on the philosophy of trying to determine and 
include the minimal set of Feynman diagrams that is necessary for compensating 
the gauge violation caused by the self-energy graphs. This is obviously the 
theoretically most satisfying solution, but it may cause an increase in the 
complexity of the matrix elements and consequently a slowing down of the 
numerical calculations. For the gauge bosons we are guided by the observation 
that the lowest-order decay widths are exclusively given by the imaginary 
parts of the fermion loops in the one-loop self-energies. It is therefore 
natural to perform a Dyson summation of these fermionic one-loop self-energies 
and to include the other possible one-particle-irreducible fermionic one-loop 
corrections (``fermion-loop scheme'')~\cite{BHF1,BHF2}. For the LEP2 process 
$e^+e^- \to 4\ $fermions this amounts to adding the fermionic corrections to
the triple gauge-boson vertex. The complete set of fermionic
contributions forms a gauge-independent subset and obeys all Ward
identities exactly, even with resummed propagators~\cite{BHF2}. 

The above arguments, although general, apply in particular to tree diagrams.
Therefore an additional discussion for RC is required.

\subsection{Radiative corrections}

The next question that should be addressed involves the interplay between
RC and gauge invariance. After all, the RC are indispensable for coming up 
with reliable theoretical predictions for the LEP2 process 
$e^+e^- \to 4\ $fermions.

As far as real-photon corrections are concerned, not much changes as compared
to the lowest-order case. Still both the pole scheme and fermion-loop scheme
yield gauge-invariant results. However, in the fermion-loop-scheme treatment 
of the process $e^+e^- \to 4f\gamma$ the full set of fermionic corrections to 
the quartic gauge-boson vertex emerges. This evidently is too much complexity 
for a tree-level calculation. The pole-scheme, with its close relation to
on-shell subprocesses, remains relatively simple. As we shall see later
on, some subtleties arise when photons are radiated from a virtual $W$ boson,
because this $W$ boson may give rise to two poles.

The implementation of the one-loop RC adds an additional 
level of complexity by the sheer number of contributions ($10^3$\,--\,$10^4$) 
that have to be evaluated. By employing a gauge-invariant lowest-order 
finite-width scheme it is possible to cover the most important electroweak 
effects, like running couplings and leading QED corrections (see previous 
section), which are controlled by factorization theorems. However, there 
is still the question about the remaining corrections, which can be large, 
especially at high energies~\cite{lep2-ww,ww-review,ww-app}.

In order to include these corrections one might attempt to extend the 
fermion-loop scheme. At present this solution is not yet workable in view of
the fact that a gauge-invariant inclusion of the one-loop corrections to the 
decay width in turn requires the inclusion of (the imaginary parts of) some 
two-loop corrections. Moreover, the number of one-loop contributions that have
to be evaluated remains large.

As a more appealing and economic strategy we discuss in the next section how 
the RC can be calculated in an approximated pole-scheme expansion.


\section{The pole scheme in double-pole approximation}
\label{sec:pole-scheme}

As mentioned above, the pole scheme consists in decomposing the complete 
amplitude by expanding around the poles of the unstable particles. 
The residues in this expansion are physically observable and therefore 
gauge-invariant. The pole-scheme expansion can be viewed as a gauge-invariant 
prescription for performing an expansion in powers of $\Gamma_V/M_V$. It should
be noted that there is no unique definition of the residues. Their calculation
involves a mapping of off-shell matrix elements with off-shell kinematics on 
on-resonance matrix elements with restricted kinematics. This mapping,
however, is not unambiguously fixed. After all, it involves more 
than just the invariant masses of the unstable particles and one thus has to 
specify the variables that have to be kept fixed in the mapping.
The resulting implementation dependence manifests itself in differences of 
subleading nature, e.g.~$\OO(\Gamma_V/M_V)$ suppressed deviations in the 
leading pole-scheme residue. In special regions of phase space, where the
matrix elements vary rapidly, the implementation dependence can take 
noticeable proportions. This happens in particular near phase-space
boundaries, like thresholds.

In order to make these statements a bit more transparent, we sketch the 
pole-scheme method for a single unstable particle.
In this case the Dyson resummed lowest-order matrix element is given by
\begin{eqnarray}
  \label{pole-scheme}
  \M^\infty 
     &=& \frac{W(p_V^2,\omega)}{p_V^2-\tilde{M}_V^2}\,\sum_{n=0}^{\infty}
         \,\Biggl( \frac{-\tilde{\Sigma}_V(p_V^2)}{p_V^2-\tilde{M}_V^2} 
           \Biggr)^n 
      =\ \frac{W(p_V^2,\omega)}{p_V^2-\tilde{M}_V^2+\tilde{\Sigma}_V(p_V^2)}
         \nonumber \\[1mm]
     &=& \frac{W(M^2,\omega)}{p_V^2-M^2}\,\frac{1}{Z(M^2)} + \Biggl[  
         \frac{W(p_V^2,\omega)}{p_V^2-\tilde{M}_V^2+\tilde{\Sigma}_V(p_V^2)}
         - \frac{W(M^2,\omega)}{p_V^2-M^2}\,\frac{1}{Z(M^2)} \Biggr],
\end{eqnarray}
where $\tilde{\Sigma}_V(p_V^2)$ is the unrenormalized self-energy of the 
unstable particle $V$ with momentum $p_V$ and unrenormalized mass 
$\tilde{M}_V$. The renormalized quantity $M^2$ is the pole in the complex 
$p_V^2$ plane, whereas $Z(M^2)$ denotes the wave-function factor:
\be
  \label{pole-scheme/def}
  M^2-\tilde{M}_V^2+\tilde{\Sigma}_V(M^2) = 0, \quad\quad
  Z(M^2) = 1+\tilde{\Sigma}'_V(M^2). 
\ee
The first term in the last expression of Eq.~(\ref{pole-scheme}) represents the
single-pole residue, which is closely related to on-shell production and 
decay of the unstable particle. The second term between the square brackets
has no pole and can be
expanded in powers of $\,p_V^2-M^2$. The argument $\omega$ denotes the
dependence on the other variables, i.e.~the implementation dependence.
After all, the unstable particle is always accompanied by other
particles in the production and decay stages. For instance, consider the
LEP1 reaction $e^+e^- \to \bar{f}f$. In the mapping $p_Z^2 \to M^2$ one can 
either keep $\,t=(p_{e^-}-p_{f})^2=-p_Z^2(1-\cos\theta)/2\,$ fixed or
$\cos\theta$. In the former mapping $\cos\theta_{\sss{pole}}$ is obtained 
from the on-shell relation $\,\cos\theta_{\sss{pole}} = 1+2t/M^2$, whereas in 
the latter mapping $\,t_{\sss{pole}} = -M^2(1-\cos\theta)/2$. It may
be that a particular mapping leads to an unphysical point in the
`on-shell' phase space. In the present example $t_{\sss{pole}}$ will
always be physical when $\cos\theta$ is kept fixed in the mapping.
However, since $|\cos\theta_{\sss{pole}}| > 1$ for $t<-\real M^2$, it
is clear that mappings with fixed Mandelstam variables harbour the potential
risk of producing such unphysical phase-space points.%
\footnote{In the resonance region, $|p_Z^2-M^2| \ll |M^2|$, the unphysical
          `on-shell' phase-space points occur near the edge of the off-shell
          phase space, since $t<-\real M^2$ requires $\cos\theta \approx -1$.}
This can have repercussions on the convergence of the pole-scheme expansion.
Therefore we choose in our calculations only implementations that are
free of unphysical on-shell phase-space points. 

It should be noted that the mass and width of the $W$ and $Z$ bosons
are usually defined in the so-called on-shell scheme:
\be
  M_V^2-\tilde{M}_V^2+\real\tilde{\Sigma}_V(M_V^2) = 0, \quad\quad
  Z_{\sss{OS}}(M_V^2) = 1+\real\tilde{\Sigma}'_V(M_V^2), \quad\quad
  M_V\Gamma_V = \frac{\imag\tilde{\Sigma}_V(M_V^2)}{Z_{\sss{OS}}(M_V^2)}.
\ee 
Both schemes can be related according to (see e.g.~Ref.~\cite{BHF2}):
\begin{eqnarray}
  \label{CP-OS}
  M^2 &=& (M_V^2-iM_V\Gamma_V)
          \Biggl[ 1 - \frac{\Gamma_V^2}{M_V^2}
                  + \OO\biggl( \frac{\Gamma_V^3}{M_V^3} \biggr)
          \Biggr],
          \nonumber \\[1mm]
  (p_V^2-M^2)\,Z(M^2) 
      &=& \biggl( p_V^2-M_V^2+ip_V^2\,\frac{\Gamma_V}{M_V} \biggr)\,
          \biggl[ Z_{\sss{OS}}(M_V^2)+\OO(\alpha^2) \biggr].
\end{eqnarray}  
As we are aiming for $\OO(\alpha)$ precision in our study, the differences
between both schemes can be ignored. For the same reason $ip_V^2\Gamma_V/M_V$
can be replaced by $iM_V\Gamma_V$, since the difference only induces
$\OO(\alpha^2)$ corrections to the cross-sections.
 
The at present only workable approach for evaluating the RC 
to resonance-pair-production processes, like $W$-pair 
production, involves the so-called double-pole approximation (DPA). This 
approximation restricts the complete pole-scheme expansion to the term with the
highest degree of resonance. In the case of $W$-pair production only the 
double-pole residues are hence considered. The intrinsic error associated with
this procedure is $\alpha\Gamma_W/(\pi M_W) \lsim 0.1\%$, except far off 
resonance, where the pole-scheme expansion cannot be viewed as an 
effective expansion in powers of $\Gamma_V/M_V$, and close to phase-space 
boundaries, where the DPA cannot be trusted to produce the dominant 
contributions. In the latter situations also the implementation
dependence of the double-pole residues can lead to enhanced errors. 
Close to the nominal (on-shell) $W$-pair threshold, for instance, the 
intrinsic error is effectively enhanced by a factor $M_W/(\sqrt{s}-2M_W)$. 
In view of this it is wise to apply the DPA only if the energy is several 
$\Gamma_W$ above the threshold. 

\begin{figure}
  \begin{center}
  \begin{picture}(200,130)(0,-10)
    \ArrowLine(43,58)(25,40)        \Text(7,40)[lc]{$e^+$}
    \ArrowLine(25,90)(43,72)        \Text(7,92)[lc]{$e^-$}
    \Photon(50,65)(150,95){1}{12}   \Text(100,105)[]{$W$}  
    \Photon(50,65)(150,35){1}{12}   \Text(100,25)[]{$W$}
    \ArrowLine(155,98)(175,110)     \Text(190,110)[rc]{$f_1'$}
    \ArrowLine(175,80)(155,92)      \Text(190,80)[rc]{$\bar{f}_1$}
    \ArrowLine(175,50)(155,38)      \Text(190,50)[rc]{$\bar{f}_2'$}
    \ArrowLine(155,32)(175,20)      \Text(190,20)[rc]{$f_2$}
    \DashLine(75,115)(75,15){5}     \Text(40,5)[]{production}
    \DashLine(125,115)(125,15){5}   \Text(150,5)[]{decays}
    \GCirc(50,65){10}{1}
    \GCirc(100,80){10}{0.8}
    \GCirc(100,50){10}{0.8}
    \GCirc(150,95){10}{1}
    \GCirc(150,35){10}{1}
  \end{picture}
  \end{center}
  \caption[]{The generic structure of the factorizable $W$-pair 
             contributions. The shaded circles indicate the Breit--Wigner 
             resonances, whereas the open circles denote the Green functions 
             for the production and decay subprocesses up to $\OO(\alpha)$ 
             precision.}
  \label{fig:WWfact}
\end{figure}
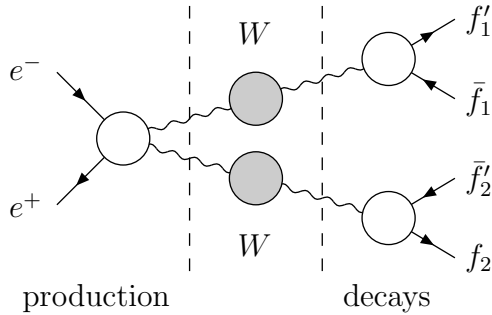

\begin{figure}
  \begin{center}
  \unitlength .7pt\small\SetScale{0.7}
  \begin{picture}(120,100)(0,0)
  \ArrowLine(30,50)( 5, 95)
  \ArrowLine( 5, 5)(30, 50)
  \Photon(30,50)(90,80){2}{6}
  \Photon(30,50)(90,20){2}{6}
  \GCirc(30,50){10}{0}
  \Vertex(90,80){1.2}
  \Vertex(90,20){1.2}
  \ArrowLine(90,80)(120, 95)
  \ArrowLine(120,65)(105,72.5)
  \ArrowLine(105,72.5)(90,80)
  \Vertex(105,72.5){1.2}
  \ArrowLine(120, 5)( 90,20)
  \ArrowLine( 90,20)(105,27.5)
  \ArrowLine(105,27.5)(120,35)
  \Vertex(105,27.5){1.2}
  \Photon(105,27.5)(105,72.5){2}{4.5}
  \put(92,47){$\gamma$}
  \put(55,73){$W$}
  \put(55,16){$W$}
  \end{picture}
  \quad\quad
  \begin{picture}(120,100)(0,0)
  \ArrowLine(30,50)( 5, 95)
  \ArrowLine( 5, 5)(30, 50)
  \Photon(30,50)(90,80){2}{6}
  \Photon(30,50)(90,20){2}{6}
  \Vertex(60,35){1.2}
  \GCirc(30,50){10}{0}
  \Vertex(90,80){1.2}
  \Vertex(90,20){1.2}
  \ArrowLine(90,80)(120, 95)
  \ArrowLine(120,65)(105,72.5)
  \ArrowLine(105,72.5)(90,80)
  \Vertex(105,72.5){1.2}
  \ArrowLine(120, 5)(90,20)
  \ArrowLine(90,20)(120,35)
  \Photon(60,35)(105,72.5){2}{5}
  \put(87,46){$\gamma$}
  \put(63,11){$W$}
  \put(38,22){$W$}
  \put(55,73){$W$}
  \end{picture}
  \quad\quad 
  \begin{picture}(160,100)(0,0)
  \ArrowLine(30,50)( 5, 95)
  \ArrowLine( 5, 5)(30, 50)
  \Photon(30,50)(90,80){-2}{6}
  \Photon(30,50)(90,20){2}{6}
  \Vertex(60,65){1.2}
  \GCirc(30,50){10}{0}
  \Vertex(90,80){1.2}
  \Vertex(90,20){1.2}
  \ArrowLine(90,80)(120, 95)
  \ArrowLine(120,65)(105,72.5)
  \ArrowLine(105,72.5)(90,80)
  \Vertex(105,27.5){1.2}
  \ArrowLine(120, 5)(90,20)
  \ArrowLine(105,27.5)(120,35)
  \ArrowLine(90,20)(105,27.5)
  \Photon(60,65)(105,27.5){-2}{5}
  \put(84,55){$\gamma$}
  \put(63,78){$W$}
  \put(40,68){$W$}
  \put(55,16){$W$}
  \end{picture}
%
  \\[2ex]
  \unitlength .7pt\small\SetScale{0.7}
  \begin{picture}(240,100)(0,0)
  \ArrowLine(30,50)( 5, 95)
  \ArrowLine( 5, 5)(30, 50)
  \Photon(30,50)(90,80){2}{6}
  \Photon(30,50)(90,20){2}{6}
  \GCirc(30,50){10}{0}
  \Vertex(90,80){1.2}
  \Vertex(90,20){1.2}
  \ArrowLine(90,80)(120, 95)
  \ArrowLine(120,65)(105,72.5)
  \ArrowLine(105,72.5)(90,80)
  \ArrowLine(120, 5)( 90,20)
  \ArrowLine( 90,20)(120,35)
  \Vertex(105,72.5){1.2}
  \PhotonArc(120,65)(15,150,270){2}{3}
  \put(55,73){$W$}
  \put(55,16){$W$}
  \put(99,47){$\gamma$}
  \DashLine(120,0)(120,100){6}
  \PhotonArc(120,35)(15,-30,90){2}{3}
  \Vertex(135,27.5){1.2}
  \ArrowLine(150,80)(120,95)
  \ArrowLine(120,65)(150,80)
  \ArrowLine(120, 5)(150,20)
  \ArrowLine(150,20)(135,27.5)
  \ArrowLine(135,27.5)(120,35)
  \Vertex(150,80){1.2}
  \Vertex(150,20){1.2}
  \Photon(210,50)(150,80){2}{6}
  \Photon(210,50)(150,20){2}{6}
  \ArrowLine(210,50)(235,95)
  \ArrowLine(235, 5)(210,50)
  \GCirc(210,50){10}{0}
  \put(177,73){$W$}
  \put(177,16){$W$}
  \end{picture}
  \end{center}
  \caption[]{Examples for virtual (top) and real (bottom) non-factorizable 
             corrections to $W$-pair production. The black circles denote 
             the lowest-order Green functions for the production of the
             virtual $W$-boson pair.}
  \label{fig:WWnf}
\end{figure}
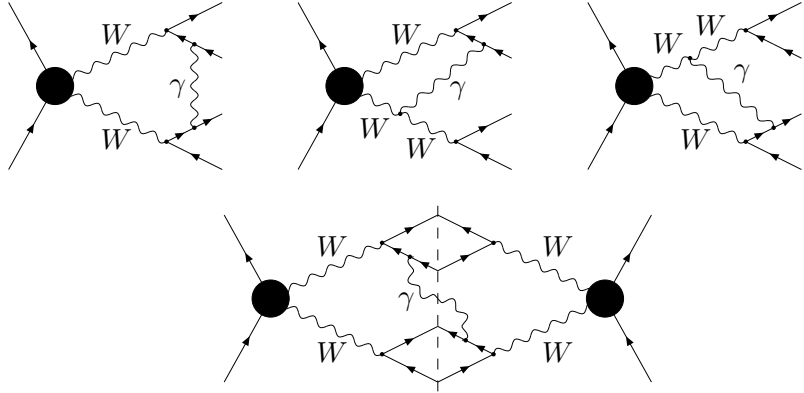

In the DPA one can identify two types of contributions. One type comprises all 
diagrams that are strictly reducible at both unstable $W$-boson lines
(see Fig.~\ref{fig:WWfact}).
These corrections are therefore called factorizable and can be attributed 
unambiguously either to the production of the $W$-boson pair or to one of 
the subsequent decays. The second type consists of all diagrams in which the 
production and/or decay subprocesses are not independent and which therefore
do not seem to have two overall $W$ propagators as factors 
(see Fig.~\ref{fig:WWnf}). 
We refer to these effects as non-factorizable corrections.%
\footnote{It should be noted that the exact split-up between factorizable and
          non-factorizable radiative corrections requires a precise 
          (gauge-invariant) definition. We will come back to this point 
          in Sect.~\ref{sec:corr}.}
In the DPA the non-factorizable corrections arise 
exclusively from the exchange or emission of photons with 
$E_\gamma \lsim \OO(\Gamma_W)$. Hard photons as well as massive-particle 
exchanges do not lead to double-resonant contributions. The physical picture 
behind all of this is that in the DPA the $W$-pair process can be viewed 
as consisting of several subprocesses: the production of the $W$-boson
pair, the propagation of the $W$ bosons, and the subsequent decay of the 
unstable $W$ bosons. The production and decay are ``hard'' subprocesses, which 
occur on a relatively short time interval, $\OO(1/M_{W})$. They are in general
distinguishable as they are well separated by a relatively big propagation 
interval, $\OO(1/\Gamma_{W})$. Consequently, the corresponding amplitudes have 
certain factorization properties. The same holds for the RC
to the subprocesses. The only way the various stages can be interconnected
is via the radiation of soft photons with energy of $\OO(\Gamma_{W})$. These
photons induce relatively long range interactions and thereby allow the 
various subprocesses to communicate with each other.

Within the DPA the generic form of the virtual and soft-photonic  
$\OO(\alpha)$ RC to off-shell $W$-pair production can now be cast 
in the following gauge-invariant form:
\be
        \label{relcorr}
        d\sigma_{\sss{DPA}}
        =
        d\sigma^{0}_{\sss{DPA}} \bigl(1+\delta_{\sss{DPA}}\bigr),
\ee
where $d\sigma^{0}_{\sss{DPA}}$ is the lowest-order differential 
cross-section in DPA. The hard-photon effects in DPA are added separately, 
in view of their dependence on the phase-space of the hard photons. Their
contribution strongly depends on the distribution that is studied,
i.e.~on the integrations that have to be performed. More details will be 
given in Sect.~\ref{sec:corr}. Since one also knows the exact Born 
cross-sections for off-shell $W$-pair production [$d\sigma^{0}_{\sss{WW}}$]
and for the background diagrams contained in the four-fermion process
(\ref{intro/ee->4f}) [$d\sigma^{0}_{\sss{bkg}}$], one can also add those
to the above expression. The final gauge-invariant result up to 
$\OO(\alpha)$ or, equivalently, $\OO(\Gamma_{W}/M_{W})$ precision reads
\be
        \label{pole+bkg}
        d\sigma
        =
        d\sigma^{0}_{\sss{DPA}} \bigl(1+\delta_{\sss{DPA}}\bigr)
        +
        \bigl( d\sigma^{0}_{\sss{WW}} - d\sigma^{0}_{\sss{DPA}} \bigr)     
        +
        d\sigma^{0}_{\sss{bkg}}.
\ee
The purpose of this paper is to give a detailed discussion of 
$\delta_{\sss{DPA}}$, i.e.~the $\OO(\alpha)$ corrections to 
$d\sigma^{0}_{\sss{DPA}}$. In order to make contact with experimental 
cross-sections the other terms, 
$d\sigma^{0}_{\sss{WW}}-d\sigma^{0}_{\sss{DPA}}$ and $d\sigma^{0}_{\sss{bkg}}$,
are also relevant. The full gauge-invariant Born term, including all diagrams,
has been discussed in the literature~\cite{background}. It has also been 
compared with the non-gauge-invariant cross-section $d\sigma^{0}_{\sss{WW}}$,
calculated in the unitary gauge. In many cases $d\sigma^{0}_{\sss{WW}}$
gives numerically a good approximation to the complete Born cross-section.
Moreover, in practice it is often extracted from the data in the experimental 
analyses. Therefore it is useful to present a numerical comparison between 
$d\sigma_{\sss{WW}}^{0}$ and $d\sigma^{0}_{\sss{DPA}}$. This will be done in 
Sect.~\ref{sec:born_num}.

\subsection{The double-pole approximation: conventions and an example}

As is clear from the above-given discussion of the DPA, a specific prescription
has to be given for the calculation of the DPA residues. Or, in other words, we
have to fix the implementation of the mapping of the full off-shell phase space
on the kinematically restricted (on-resonance) one. We have opted to always
extract {\it pure double-pole residues}. This means in particular that after 
the integration over decay kinematics and invariant masses has been
performed the on-shell cross-section should be recovered. 

In the rest of this subsection we will explain our method in more detail
by applying it to the lowest-order reaction
\be
\label{ee->ww->4f}
        e^{+}(q_1)\,e^{-}(q_2)\to W^{+}(p_1)\,W^{-}(p_2)
        \to
        \bar{f}_1(k_1)f_1'(k_1')\,f_2(k_2)\bar{f}_2'(k_2'),
\ee
involving only those diagrams that contain as factors the Breit--Wigner 
propagators for the $W^{+}$ and $W^{-}$ bosons. Here $\bar{f}_1$ and $f_1'$ 
are the decay products of the $W^+$ boson, and $f_2$ and $\bar{f}_2'$ those 
of the $W^-$ boson. It should be noted that a large part 
of the RC in DPA to this reaction can be treated in a way 
similar to the lowest-order case, which is therefore a good starting point.
The amplitude for process (\ref{ee->ww->4f}) takes the form
\be
\label{pole/A}
        \M =
        \sum_{\lambda_{1},\lambda_{2}}
        \Pi_{\lambda_{1}  \lambda_{2}}(M_{1}, M_{2}) \
        \frac{\Delta^{(+)}_{\lambda_{1}}(M_{1})}{D_{1}} \
        \frac{\Delta^{(-)}_{\lambda_{2}}(M_{2})}{D_{2}} \ ,
\ee
where any dependence on the helicities of the initial- and final-state
fermions has been suppressed, and 
\be
        D_{i} = M_{i}^{2}-M_{W}^{2}+i M_{W}\Gamma_{W},\quad\quad
        M_{i}^{2} = (k_{i}+k_{i}')^{2}.
\ee
The quantities $\Delta^{(+)}_{\lambda_{1}}(M_{1})$ and  
$\Delta^{(-)}_{\lambda_{2}}(M_{2})$ are the off-shell $W$-decay amplitudes 
for specific spin-polarization states $\lambda_{1}$ (for the $W^+$) and
$\lambda_{2}$ (for the $W^-$), with $\lambda_i=(-1,0,+1)$. 
The off-shell $W$-pair production amplitude
$\Pi_{\lambda_{1}\lambda_{2}}(M_{1}, M_{2})$ depends on the invariant
fermion-pair masses $M_{i}$ and on the polarizations $\lambda_{i}$
of the virtual $W$ bosons. In the limit $M_{i} \to M_{W}$ the amplitudes 
$\Pi$ and $\Delta^{(\pm)}$ go over into the on-shell production and decay 
amplitudes.

The choice of the polarization states labelled by $\lambda_{i}$ is in 
principle free. The amplitude $\M$ is obtained by summing over the 
polarizations and is therefore independent of such a specific choice.
As it turns out, it will be convenient to use different choices in
different parts of the RC calculation (see Apps.~\ref{app:virt} and 
\ref{app:hard_rad}).

In the cross-section the above factorization leads to
\be
\label{pole/|A|^2}
        \sum\limits_{\sss{fermion helicities}} |\M|^{2}\ \ 
        =
        \sum\limits_{\lambda_1,\lambda_2,\lambda_1',\lambda_2'}
        \PP_{[\lambda_{1}\lambda_{2}][\lambda_{1}'\lambda_{2}']}(M_{1},M_{2}) \
        \frac{\D_{\lambda_{1}\lambda_{1}'}(M_{1})}{|D_{1}|^{2}} \
        \frac{\D_{\lambda_{2}\lambda_{2}'}(M_{2})}{|D_{2}|^{2}}.
\ee
In Eq.~(\ref{pole/|A|^2}) the production part is given by a $9\times9$~density 
matrix
\be
\label{pole/prod}
\PP_{[\lambda_{1}\lambda_{2}][\lambda_{1}'\lambda_{2}']}(M_{1},M_{2})\ \ 
         =
         \sum\limits_{\sss{$e^{\pm}$ helicities}}
         \Pi_{\lambda_{1}\lambda_{2}}(M_{1},M_{2}) \
         \Pi_{\lambda_{1}'\lambda_{2}'}^{*}(M_{1},M_{2}).
\ee
Similarly the decay part is governed by $3\times3$~density matrices
\be
\label{pole/decay}
         \D_{\lambda_{i}\lambda_{i}'}(M_{i})\ \ 
         =
         \sum\limits_{\sss{fermion helicities}}
         \Delta_{\lambda_{i}}(M_{i}) \
         \Delta_{\lambda_{i}'}^{\!*}(M_{i}),
\ee
where the summation is performed over the helicities of the final-state
fermions.
 
It is clear that Eq.~(\ref{pole/prod}) is closely related to the absolute 
square of the matrix element for stable unpolarized $W$-pair production. 
In that case the cross-section contains the trace of the above density matrix
\be
\label{pole/prod_on}
        \mbox{\bf Tr} \ \PP(M_{W},M_{W}) 
        =
        \sum\limits_{\lambda_1,\lambda_2}
        \PP_{[\lambda_{1}\lambda_{2}][\lambda_{1}\lambda_{2}]}(M_{W},M_{W})\ \ 
        = 
        \sum\limits_{\sss{all polarizations}}
        |\Pi_{\lambda_{1} \lambda_{2}}(M_{W}, M_{W})|^2.
\ee
The decay of an unpolarized on-shell $W$ boson is determined by
\be
\label{pole/dec_on}
        \mbox{\bf Tr} \ \D(M_{W}) 
        =
        \sum\limits_{\lambda_{i}} \D_{\lambda_{i}\lambda_{i}}(M_{W})\ \ 
        =
        \sum\limits_{\sss{all polarizations}}
        |\Delta_{\lambda_{i}}(M_{W})|^{2}.
\ee
Note, however, that also the off-diagonal elements of $\PP(M_{W},M_{W})$ and 
$\D(M_{W})$ are required for determining Eq.~(\ref{pole/|A|^2}) in the limit
$M_{i} \to M_{W}$. 

As a next step it is useful to describe the kinematics of process
(\ref{ee->ww->4f}) in a factorized way, i.e.~using the invariant masses 
$M_{1}$ and $M_{2}$ of the fermion pairs. The differential cross-section 
takes the form
\be
        d\sigma
        =
        \frac{1}{2s} \sum |\M|^{2}\, d\Gamma_{4f}
        =
        \frac{1}{2s} \sum |\M|^{2}\, d\Gamma_{\sss{pr}}\cdot 
        d\Gamma_{\sss{dec}}^{+} \cdot d\Gamma_{\sss{dec}}^{-}   
        \cdot \frac{dM_{1}^{2}}{2\pi} \cdot \frac{dM_{2}^{2}}{2\pi}, 
\ee
where $d\Gamma_{4f}$ indicates the complete four-fermion phase-space factor
and $s=(q_1+q_2)^2$ the centre-of-mass energy squared. The phase-space factors
for the production and decay subprocesses, $d\Gamma_{\sss{pr}}$ and 
$d\Gamma_{\sss{dec}}^{\pm}$, read 
\begin{eqnarray}
     d\Gamma_{\sss{pr}}\
     &=& \frac{1}{(2\pi)^{2}}\, \delta(q_{1}+q_{2}-p_{1}-p_{2})\,
         \frac{d \vec{p}_{1}}{2p_{10}}\,
         \frac{d \vec{p}_{2}}{2p_{20}}, 
         \nonumber \\
     d\Gamma_{\sss{dec}}^{+}
     &=& \frac{1}{(2\pi)^{2}}\, \delta(p_{1}-k_{1}-k_{1}')\,
         \frac{d \vec{k}_{1}}{2k_{10}}\,
         \frac{d \vec{k}_{1}'}{2k_{10}'}, 
         \nonumber \\      
     d\Gamma_{\sss{dec}}^{-}
     &=& \frac{1}{(2\pi)^{2}}\, \delta(p_{2}-k_{2}-k_{2}')\,
         \frac{d \vec{k}_{2}}{2k_{20}}\,
         \frac{d \vec{k}_{2}'}{2k_{20}'}.
\end{eqnarray}
When the factorized form for $\sum|\M|^{2}$ is inserted one obtains
\begin{eqnarray}
\label{pole/15}
  d\sigma
  &=& \frac{1}{2s}\sum\limits_{\lambda_1,\lambda_2,\lambda_1',\lambda_2'}
      \PP_{[\lambda_{1}\lambda_{2}][\lambda_{1}'\lambda_{2}']}(M_{1},M_{2})\,
      d\Gamma_{\sss{pr}}
      \times
      \D_{\lambda_{1}\lambda_{1}'}(M_{1})\, d\Gamma_{\sss{dec}}^{+}
      \times
      \D_{\lambda_{2}\lambda_{2}'}(M_{2})\, d\Gamma_{\sss{dec}}^{-}
      \times
      \nonumber \\
  & & \times\, 
      \frac{1}{2\pi}\,\frac{d M_{1}^{2}}{|D_{1}|^{2}}
      \times
      \frac{1}{2\pi}\,\frac{d M_{2}^{2}}{|D_{2}|^{2}}.
\end{eqnarray}

As mentioned before, the definition of the DPA residues is not unique.
To define them we first organize the four-fermion kinematics in a special way. 
In the laboratory (LAB) frame we write the four-fermion phase space in terms 
of a solid production angle for the $W^{+}$ boson and solid decay angles for 
two of the final-state fermions, one originating from the $W^{+}$ boson and 
one from the $W^{-}$ boson. These angles will be kept fixed at any time
during the process of defining the DPA residues. For later use we explicitly
write down our conventions for the momenta, invariants, and phase-space 
factors. The momenta read
\begin{eqnarray}
\label{pole/kinematics}
        q_{1} = E(1,\sin\theta,0,\cos\theta), 
        &\quad &
        q_{2} = E(1,-\sin\theta,0,-\cos\theta),
        \\
        p_{1} = E_{1} (1,0,0,\frac{p}{E_{1}}),
        &\quad &
        p_{2} = E_{2} (1,0,0,-\frac{p}{E_{2}}),
        \nonumber \\
        k_{1} = E_{3} (1,\sin\theta_{3}\cos\phi_{3},
                       \sin\theta_{3}\sin\phi_{3},\cos\theta_{3}),
        &\quad &
        k_{2} = E_{4} (1,\sin\theta_{4}\cos\phi_{4},
                       \sin\theta_{4}\sin\phi_{4},-\cos\theta_{4}),
        \nonumber
\end{eqnarray}
with
\begin{eqnarray}
\label{pole/energies}
     && E = \frac{1}{2}\,\sqrt{s}, \quad\quad
        E_{1,2} = \frac{1}{2\sqrt{s}}\bigl(s+M_{1,2}^{2}-M_{2,1}^{2}\bigr),
        \quad\quad
        E_{3,4} = \frac{1}{2}\,\frac{M_{1,2}^2}{E_{1,2}-p\cos\theta_{3,4}},
        \nonumber \\
     && p = \frac{1}{2\sqrt{s}}\,\lambda^{1/2} (\sqrt{s},M_{1},M_{2}),
        \quad\quad 
        \lambda (\sqrt{s},M_{1},M_{2})
        = 
        \bigl[s-(M_{1}+M_{2})^{2}\bigr]\bigl[s-(M_{1}-M_{2})^{2}\bigr].
        \quad\quad
\end{eqnarray}
The momenta of the other final-state particles follow from 
$k_{i}' = p_{i} - k_{i}$.
The masses of the fermions are neglected whenever possible. This, hence, 
excludes situations in which the fermion masses are needed to regularize
singularities from the radiation of collinear photons.
The Mandelstam invariants are defined as
\be
        s = (q_{1}+q_{2})^{2},\quad\quad
        t = (p_{1}-q_{1})^{2},\quad\quad
        u = (p_{2}-q_{1})^{2} = M_{1}^{2} + M_{2}^{2} - s - t.  
\ee 
From all this it should be clear that the invariant masses $M_i$ only appear
explicitly in the energies and velocities of the $W$ bosons and their
decay products.

In this notation the production phase-space factor reads 
\be
        d\Gamma_{\sss{pr}} 
        =
        \frac{1}{8\pi}\,
        \frac{p}{E}\,
        \frac{d\Omega}{4\pi},
\ee
with $d\Omega$ denoting the solid angle between the $W^+$ boson and the 
positron. The decay phase-space factors are given by
\be
        d\Gamma_{\sss{dec}}^{+}
        =
        \frac{1}{8\pi}\,
        \frac{M_{1}^{2}}{(E_{1}-p \cos\theta_{3})^2}\,
        \frac{d\Omega_{3}}{4\pi}
\ee
and a similar expression for $d\Gamma_{\sss{dec}}^{-}$. Here $d\Omega_{3}$
denotes the solid decay angle between the $W^+$ boson and $\bar{f}_1$.

Our choice of the DPA residues amounts to a two-step procedure for fixing the
invariant masses $M_{1}$ and $M_{2}$, appearing in the
four-fermion kinematics and the amplitudes
$\Pi(M_{1},M_{2})$, $\Delta^{(+)}(M_{1})$, and $\Delta^{(-)}(M_{2})$.
The first step is the replacement
\be
 M_{i}^{2} \to M_{W}^{2} - i M_{W} \Gamma_{W},
\ee
i.e.~the residue is taken at the Breit--Wigner poles [see discussion
below Eq.~(\ref{CP-OS})]. Note that this replacement, of course, does not
apply to the Breit--Wigner resonances themselves. The phase-space 
conventions displayed above fix our choice for the implementation. 
The solid angles are kept fixed, whereas the energies and velocities become
complex [as can be seen from Eq.~(\ref{pole/energies})]. Note that the 
so-obtained set of momenta preserves momentum conservation. This protects
the DPA residues against effectively crossing the four-fermion phase-space 
boundaries, which might lead to a reduced quality of the DPA. 

For practical purposes, however, it is messy to evaluate the $\OO(\alpha)$
RC to the amplitudes $\Pi(M_{1},M_{2})$, $\Delta^{(+)}(M_{1})$, and 
$\Delta^{(-)}(M_{2})$ at the complex Breit--Wigner poles. This would require 
the analytic continuation of the one-loop expressions to the second Riemann 
sheet. As such we approximate the DPA residues by using  
\be
\label{pole/17}
        M_{i} = M_{W}.
\ee
The error introduced by the {\it on-shell} approximation (\ref{pole/17}) 
is of order $\OO(\Gamma_{W}/M_{W})$. When this error comes on top of the 
$\OO(\alpha)$ RC it can be neglected, since terms of 
$\OO(\alpha\Gamma_{W}/M_{W})$ are neglected anyhow in the DPA approach. 
By having fixed the solid angles in the mapping, the on-shell phase-space 
points defined with Eq.~(\ref{pole/17}) remain physical. From this point of
view it is a sound implementation procedure. A procedure that fixes
Mandelstam variables in the mapping involves phase-space regions where
it may be regarded as being unsound, as has been indicated in the example 
below Eq.~(\ref{pole-scheme/def}). In such cases it is preferable to
set the cross-section to zero rather than evaluating it for unphysical values.

The thus-obtained amplitudes become (gauge-invariant) on-shell ones, whereas 
the four-fermion phase space is reduced to the phase space of two fermion 
pairs with invariant masses $M_{W}$. Since the DPA forces us to only consider 
collider energies that are several $\Gamma_{W}$ above the on-shell $W$-pair
threshold, the $W$-boson velocities stay well defined in our approximation.
The only place where the invariant masses $M_i$ still show up is in the 
Breit--Wigner resonances $D_i$, which can be pulled out as overall factors.
The integration over the Breit--Wigner resonances need not be restricted 
to the physical region $M_{i}>0$, $M_{1}+M_{2}<\sqrt{s}$. Since the DPA is
anyhow not valid far off resonance, we can just as well integrate over the 
full range of the distributions, $(-\infty, +\infty)$, which guarantees that
the on-shell results are recovered when the decay kinematics and invariant 
masses are integrated out. This means that the following integral will
be used:
\be
  \label{mass_integration}
  \frac{1}{2\pi} \int\limits_{-\infty}^{\infty} d M_{i}^{2} 
  \frac{1}{|D_{i}|^{2}}
  =
  \frac{1}{2M_{W}\Gamma_{W}}.
\ee

So far we have explained how to calculate the canonical multidifferential
cross-section $d\sigma/(d\Omega d\Omega_3 d\Omega_4 dM_1^2 dM_2^2)$ in the DPA.
In case one needs a multidifferential cross-section in other variables one
should relate that cross-section to the canonical one by means of a Jacobian.
In order to obtain this Jacobian the off-shell kinematical relations 
(\ref{pole/kinematics}) and (\ref{pole/energies}) should be used.


\section{A numerical comparison of different Born cross-sections}
\label{sec:born_num}

In order to have an idea of the differences between the exact Born 
cross-section $d\sigma^{0}_{\sss{WW}}$, corresponding to process 
(\ref{intro/ee->ww->4f}), and its DPA limit $d\sigma^{0}_{\sss{DPA}}$,
we now present a brief numerical comparison. 

First we discuss the set of parameters used to produce the plots throughout 
this paper. To facilitate cross-checks with the results presented in the
literature, we adopt the LEP2 input-parameter scheme advocated 
in Ref.~\cite{lep2-ww}. In this scheme the Fermi constant $G_{\mu}$, the 
fine-structure constant $\alpha$, and the masses of the light fermions%
\footnote{The masses of the light quarks are adjusted in such a way that the
          experimentally measured hadronic vacuum polarization is reproduced.}
and $W,Z$ bosons are the independent input parameters. The mass of the top
quark, $m_t$, follows from the Standard Model prediction for muon decay
\be
\label{plots/Gf}
  G_{\mu} 
  = 
  \frac{\alpha \pi}{\sqrt{2}M_W^2 (1-M_W^2/M_Z^2)}\frac{1}{1-\Delta r}.
\ee
The quantity $\Delta r$ denotes the loop corrections to muon decay. 
It is zero at tree level, but when loop corrections are included it depends
on the input parameters as well as on $m_t$, the Higgs-boson mass
$M_H$, and the strong coupling $\alpha_S$.%
\footnote{The so-obtained top-quark mass will become $\alpha_S$-
          and $M_H$-dependent. It can be confronted with the direct
          measurements at Fermilab and the indirect ones from LEP in 
          order to obtain indirect information on $M_H$.}

In analogy to Ref.~\cite{lep2-ww} we use in our numerical evaluations the 
following set of (slightly outdated) input parameters:
\begin{eqnarray*}
  \label{num/input}
      \alpha  = 1/137.0359895,  
  &\quad\quad &
      G_{\mu} = 1.16639\times 10^{-5}\GeV^{-2},
      \\
      M_Z     = 91.1884\GeV,
  &\quad\quad &
      M_W     = 80.26\GeV,
\end{eqnarray*}
\begin{displaymath}
      m_e      = 0.51099906\MeV,
  \quad\quad  
      m_{\mu}  = 105.658389\MeV,
  \quad\quad  
      m_{\tau} = 1.7771\GeV,
\end{displaymath}
\begin{eqnarray}
      m_u  =  47\MeV,
  &\quad\quad &
      m_d  =  47\MeV,
      \nonumber \\
      m_s  =  150\MeV,
  &\quad\quad &
      m_c  = 1.55\GeV,
      \nonumber \\
      m_b  = 4.7\GeV,
\end{eqnarray}
and choose
\be
      M_H = 300 \GeV,
  \quad\quad  
      \alpha_S(M_Z^2) = 0.123.
\ee
The solution of Eq.~(\ref{plots/Gf}), using a calculation of $\Delta r$ that
contains all the known higher-order effects, gives the value for the top-quark
mass~\cite{lep2-ww}
\be
      m_t = 165.26\GeV.
\ee

As was already mentioned in Sect.~\ref{sec:intro}, the use of $G_{\mu}$ 
instead of $\alpha$ in the lowest-order cross-sections very often reduces the
size of the one-loop non-photonic RC considerably. This so-called 
$G_{\mu}$-parametrization consists in the replacement
\be
  d\sigma = d\sigma^0 (1+\delta^{\sss{1-loop}}) 
  \quad \to \quad   
  \frac{d\sigma^0}{(1-\Delta r)^n}\,(1+\delta^{\sss{1-loop}}
                                     -n\Delta r^{\sss{1-loop}})
  \equiv 
  d\bar{\sigma}^0 (1+\bar{\delta}^{\sss{1-loop}}), 
\ee
where $d\sigma^0 \propto \alpha^n$ and according to Eq.~(\ref{plots/Gf})
$d\bar{\sigma}^0 \propto G_{\mu}^n$. The results presented in this paper
are all calculated in this parametrization.

Another important parameter featuring in our calculations is the width of 
the $W$ boson. As explained below Eq.~(\ref{CP-OS}), we will use the
calculated on-shell width. Since we want the Breit--Wigner resonances to be
as close to reality as possible, we will always use the $\OO(\alpha)$- and 
$\OO(\alpha_S)$-corrected width $\Gamma_W$, regardless of the fact 
that we sometimes consider lowest-order distributions.  
Using the above set of input parameters we find in the 
$G_{\mu}$-parametrization
\be
\label{num/width}
      \Gamma_W = 2.08174\GeV.
\ee
For future use we note that the lowest-order $W$-boson width in this 
parametrization reads $\Gamma_W^0 = 2.03540 \GeV$. It is also relevant to
stress that the $\OO(\alpha)$ corrections to the leptonic partial
widths $\Gamma_{W\to\ell\nu_{\ell}}$ are small and negative ($\sim -0.3\%$). 
The $\OO(\alpha_S)$ corrections to the hadronic partial widths are positive, 
leading to the positive overall correction to the total $W$-boson width.

\begin{figure}
  \unitlength 1cm
  \begin{center}
  \begin{picture}(13.4,8)
  \put(-0.2,6.5){\makebox[0pt][c]{\boldmath $\sigma_{\sss{\bf tot}}$}}
  \put(-0.2,5.4){\makebox[0pt][c]{\bf [pb]}}
  \put(13.5,-0.2){\makebox[0pt][c]{\bf\boldmath $2E$ [$\GeV$]}}
  \put(-1,-6){\includegraphics{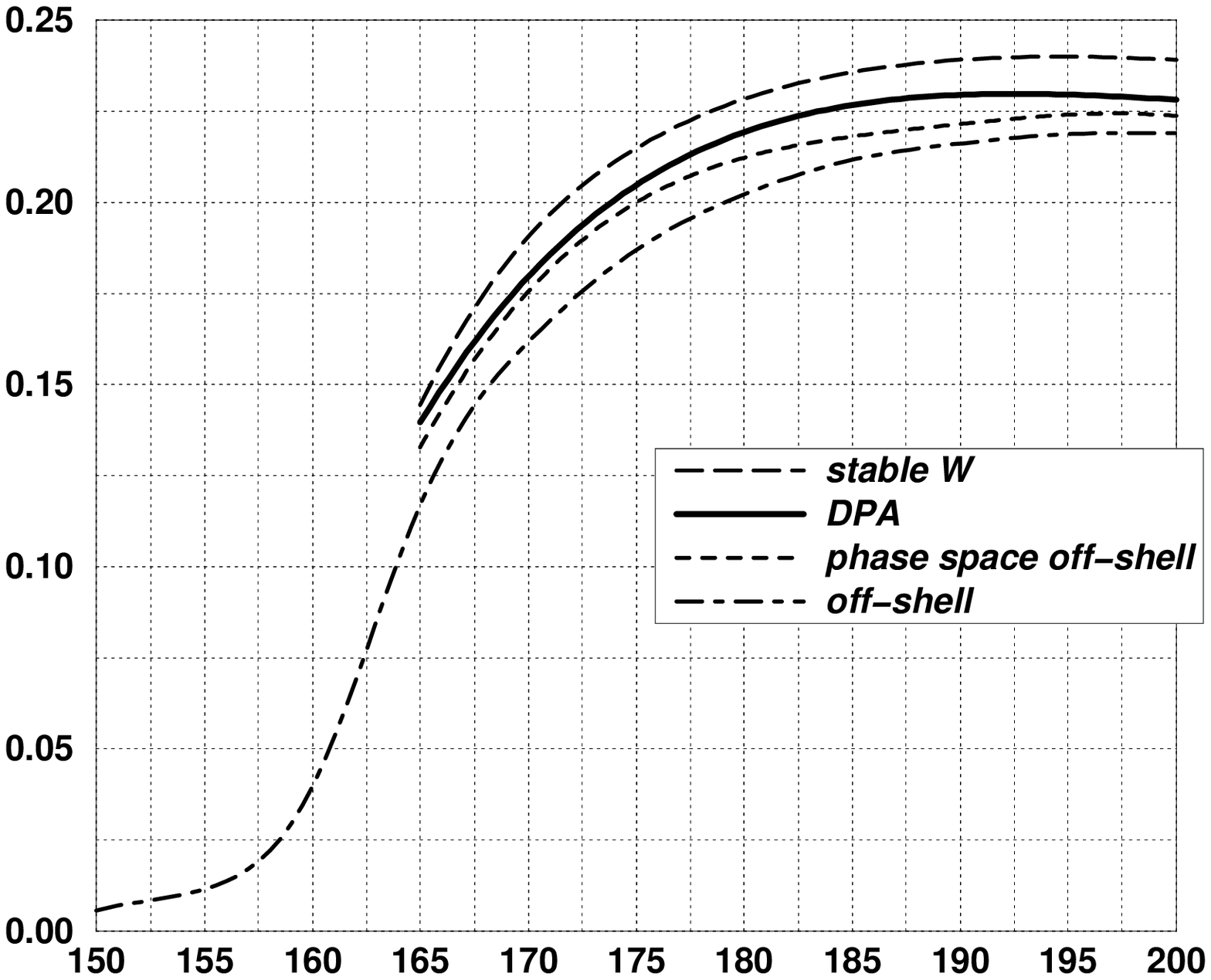}}
  \end{picture}
  \end{center}
  \caption[]{Comparison of different Born approximations for the total 
             cross-section $\sigma_{\sss{tot}}$ as a function of the 
             accelerator energy. The four curves correspond to the cases 
             i)\,--\,iv) introduced in the text.}
\label{fig:born}
\end{figure}

\begin{figure}
  \unitlength 1cm
  \begin{center}
  \begin{picture}(13.4,8)
  \put(-0.3,6.4){\makebox[0pt][c]{\boldmath 
                 $\frac{\displaystyle d\sigma}{\displaystyle d\cos\theta}$}}
  \put(-0.3,5.2){\makebox[0pt][c]{\bf [pb]}}
  \put(13.5,-0.2){\makebox[0pt][c]{\boldmath $\cos\theta$}}
  \put(-1,-6){\includegraphics{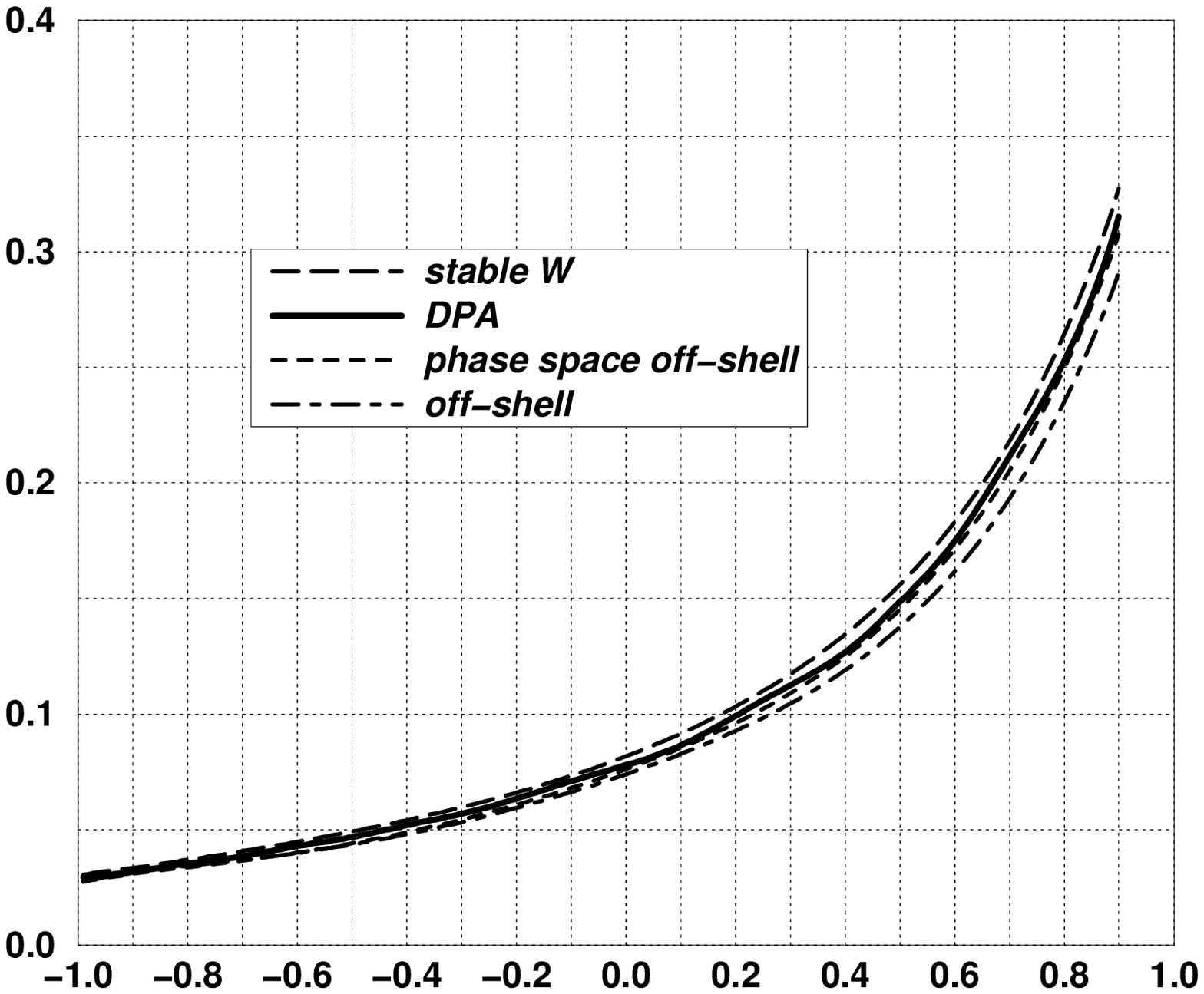}}
  \end{picture}
  \end{center}
  \caption[]{The same curves as in the previous plot, this time for the 
             lowest-order production-angle distribution 
             $\,d\sigma/d\cos\theta\,$ at $\,2E=184\GeV$.}
\label{fig:born/prod}
\end{figure}

Having fixed the input, we now compare $d\sigma^{0}_{\sss{WW}}$ and
$d\sigma^{0}_{\sss{DPA}}$ for the total cross-section $\sigma_{\sss{tot}}$
(in Fig.~\ref{fig:born}) and the differential cross-section 
$d\sigma/d\cos\theta$ (in Fig.~\ref{fig:born/prod}), where $\theta$ is the
polar angle between the $W^+$ boson and the positron in the LAB frame
[see Eq.~(\ref{pole/kinematics})]. The latter distribution is given 
for $2E=184\GeV$, whereas $\sigma_{\sss{tot}}$ is presented for a
range of LEP2 energies. We select one particular purely leptonic final state, 
$\mu^+\nu_{\mu} \tau^-\bar{\nu}_{\tau}$. In view of the massless treatment of 
the final-state fermions and the universal lowest-order interaction between the
fermions and the $W$ bosons, the results for the various final states
can be obtained by multiplying the purely leptonic result by a 
factor $N_C^{f_1}\,|V_{f_1' f_1}|^2\,N_C^{f_2}\,|V_{f_2' f_2}|^2$. 
Here $V_{f_i' f_i}$ is the mixing matrix and $N_C^{f_i}$ the colour factor.
For leptons only $V_{\nu_{\ell} \ell} = 1$ is non-vanishing and 
$N_C^{\ell}=1$.

We consider the following four cases:
\begin{itemize} 
  \item[i)]   The calculation for stable $W$ bosons, multiplied by the
              branching ratio $(\Gamma^0_{W\to\ell\nu_{\ell}}/\Gamma^0_W)^2$.
  \item[ii)]  The DPA calculation, where in Eq.~(\ref{pole/15}) the on-shell 
              approximation is applied to both the matrix elements and the 
              four-fermion phase space. The $M_i^2$ integrations over the 
              Breit--Wigner resonances are extended to the full range
              $(-\infty,+\infty)$, i.e.~Eq.~(\ref{mass_integration}) is used.
  \item[iii)] The calculation where the matrix element (\ref{pole/A})
              is on-shell, but the four-fermion phase space in 
              Eq.~(\ref{pole/15}) is not. The $M_i^2$ integrations are
              performed in the physical region.
  \item[iv)]  The off-shell calculation according to Eq.~(\ref{pole/15}), with
              the $M_i^2$ integrations performed in the physical region.
              This corresponds to $d\sigma^{0}_{\sss{WW}}$.
\end{itemize}
In cases ii), iii), and iv) the $W$-boson propagators contain the width
$\Gamma_W$, as given in Eq.~(\ref{num/width}). The matrix element in case
iv) is not gauge-invariant, it is calculated in the unitary gauge.

For the total cross-section (see Fig.~\ref{fig:born}) cases i) and ii) differ 
by a fixed overall factor, as was to be expected from the Breit--Wigner
integrals. The overall factor is determined by 
$(\Gamma_W/\Gamma^0_W)^2 = 1.04605$ and is found to be 1.04609. 
The agreement with this overall factor is a check on the numerical integration 
over the decay angles. For the production-angle distribution
(see Fig.~\ref{fig:born/prod}) the same overall factor is observed. 

One of the ingredients of Eq.~(\ref{pole+bkg}) is the difference 
$d\sigma_{WW}^0-d\sigma_{\sss{DPA}}^0$. By comparing cases ii) and iv)
in Fig.~\ref{fig:born} one 
observes that this difference varies between $18\%$ and $5\%$, when
going from $165\GeV$ to $200\GeV$. This is in good agreement with the expected 
$\OO(\Gamma_W/\Delta E)$ precision of the DPA limit, where $\Delta E$ is 
defined as the distance in energy to the $W$-pair threshold,
$\Delta E=\sqrt{s}-2M_W$. Larger differences arise when the comparison
between $\sigma_{WW}^0$ and $\sigma_{\sss{DPA}}^0$ is carried out at much 
higher energies. The reason is that $\sigma_{WW}^0$ is not $SU(2)$ 
gauge-invariant and does not fall off sufficiently fast. By properly combining
$\sigma_{WW}^0$ and the background contributions $\sigma_{\sss{bkg}}^0$ 
the $SU(2)$ gauge invariance can be restored and $\sigma_{\sss{DPA}}^{0}$ again
turns out to be a good approximation to the total (four-fermion) cross-section.


\section{Radiative corrections in the double-pole approximation}
\label{sec:corr}

In this section we discuss the complete $\OO(\alpha)$ RC to process
(\ref{ee->ww->4f}) in the context of the DPA. 
As mentioned in Sect.~\ref{sec:pole-scheme}, the most economic way of 
calculating the RC up to $\OO(\alpha)$ precision involves 
the DPA. In this approximation all virtual corrections can be classified 
into two groups: factorizable and non-factorizable corrections. 
The factorizable corrections are directly linked to the electroweak one-loop 
RC to the on-shell production and on-shell decay of the $W$ bosons. 
The remaining non-factorizable virtual corrections can be 
viewed as describing interactions between different stages of the off-shell 
process. They will only originate from certain photonic loop diagrams,
as stated in Sect.~\ref{sec:pole-scheme}. The real-photon corrections 
can also be classified in factorizable and non-factorizable corrections, 
although the various regimes for the photon energy require some special 
attention.  

In the following we will describe the calculation of all these corrections in
more detail and comment on the accuracy and applicability of the results. 
At this point we remind the reader that throughout the calculations of the RC
in the DPA two additional approximations are used. First, whenever possible 
we consider the initial- and final-state fermions to be massless, 
i.e.~excluding the cases in which the fermion masses are needed to regularize 
singularities from the radiation of collinear photons. The error of this 
approximation is at most $\OO(\alpha\,m_{\tau}/M_W)$ or 
$\OO(\alpha\,|V_{cb}|\,m_{b}/M_W)$, which is beyond the accuracy of our
calculation. Second, we assume that the accelerator energy 
is sufficiently far (read: several $\Gamma_W$) above the threshold for  
on-shell $W$-pair production. Close to threshold the DPA cannot be trusted to 
produce the dominant contributions and therefore our approach breaks down. 
The accuracy of this ``far-from-threshold'' approximation is 
$\OO(\alpha\,\Gamma_{W}/\Delta E)$, where $\Delta E$ is the distance in 
energy to the $W$-pair threshold, $\Delta E=\sqrt{s}-2M_{W}$.

\subsection{Virtual corrections}
\label{sec:corr/virt}

As a first step we discuss how to separate the virtual corrections into a sum 
of factorizable and non-factorizable virtual corrections. The diagrammatic 
split-up according to reducible and irreducible $W$-boson lines is an 
illustrative way of understanding the different nature of the two classes of
corrections, but since the double-resonant diagrams are not gauge-invariant 
by themselves the precise split-up needs to be defined properly. 

We can make use of the fact that there are effectively two scales in the 
problem: $M_{W}$ and $\Gamma_{W}$. Let us now consider virtual corrections 
coming from photons with different energies:
\begin{itemize}
  \item  soft photons, $E_{\gamma} \ll \Gamma_{W}$, 
  \item  semi-soft photons, $E_{\gamma} = \OO(\Gamma_{W})$,
  \item  hard photons, $\Gamma_{W} \ll E_{\gamma} = \OO(M_W)$.
\end{itemize}
Only soft and semi-soft photons contribute to both factorizable and 
non-factorizable corrections. The latter being defined to describe interactions
between different stages of the off-shell process. The reason for this is that
only these photons can induce relatively long-range interactions and thereby 
allow the various subprocesses, which are separated by a propagation interval 
of $\OO(1/\Gamma_W)$, to communicate with each other. Virtual corrections 
involving the exchange of hard photons or massive particles contribute 
exclusively to the factorizable corrections. In view of the short range of the 
interactions induced by these particles, their contribution to the 
non-factorizable corrections are suppressed by at least
$\OO(\Gamma_{W}/M_{W})$.

As hard photons contribute to the factorizable corrections only, we only need
to define a split-up for soft and semi-soft photons.
It is impossible to do this in a consistent gauge-invariant way on the basis
of diagrams. As we will see, it might happen that only part of some particular
diagram should be attributed to the non-factorizable corrections, the rest 
being of factorizable nature. 

The matrix element for soft and semi-soft photons can be written as a product 
of the lowest-order matrix element in DPA ($\M^0_{\sss{DPA}}$) and conserved 
(semi)soft-photon currents. These currents can be decomposed into production 
and decay currents with the help of a partial-fraction decomposition for 
virtual-photon emission from a $W$-boson line: 
\begin{eqnarray}
\label{decomp}
  \frac{1}{[p^2-M^2][p^2+2pk-M^2+io]} 
  &=& \frac{1}{2pk+io}\,
      \Biggl(\frac{1}{p^2-M^2}- \frac{1}{p^2+2pk-M^2+io} \Biggr)
      \nonumber \\[1mm]
  &=& \frac{1}{2pk+io}\,\Biggl(\frac{1}{D}- \frac{1}{D+2pk+io} \Biggr).
\end{eqnarray}
Here $M^2 = M_W^2 - i M_W\Gamma_W$, $k$ is the loop momentum of the exchanged
photon, $p+k$ is the momentum of the $W$ boson
inside the integral, and $p$ is the momentum of the $W$ boson outside 
the integral. The infinitesimal imaginary part $+io$ is needed to ensure
a proper incorporation of causality.
In this way one obtains a sum of two resonant $W$-boson propagators multiplied 
by an ordinary on-shell eikonal factor. This decomposition allows a 
gauge-invariant split-up of the matrix element in terms of one contribution 
where the photon is effectively emitted from the production part, and two 
others where the photon is effectively emitted from one of the two decay 
parts. The squares of the three contributions can be identified as 
factorizable corrections, whereas the interference terms are of 
non-factorizable nature.

\begin{figure} 
  \begin{center} {\unitlength  .7pt\small\SetScale{0.7}
  \begin{picture}(170,100)(0,0)
  \ArrowLine(30,50)( 5, 95)
  \ArrowLine( 5, 5)(30, 50)
  \Photon(30,50)(90,80){2}{6}
  \Photon(30,50)(90,20){2}{6}
  \Photon(70,30)(70,70){2}{4}
  \Vertex(70,30){1.2}
  \Vertex(70,70){1.2}
  \GCirc(30,50){10}{0}
  \Vertex(90,80){1.2}
  \Vertex(90,20){1.2}
  \ArrowLine(90,80)(120, 95)
  \ArrowLine(120,65)(90,80)
  \ArrowLine(120, 5)( 90,20)
  \ArrowLine( 90,20)(120,35)
  \put(76,47){$\gamma$}
  \put(45,68){$W$}
  \put(45,20){$W$}
  \put(72,83){$W$}
  \put(70,6){$W$}
  \Text(65,-15)[]{(a)}
  \end{picture} 
  \quad\quad
  \begin{picture}(170,100)(0,0)
  \ArrowLine(30,50)( 5, 95)
  \ArrowLine( 5, 5)(30, 50)
  \Photon(30,50)(90,80){2}{6}
  \Photon(30,50)(90,20){2}{6}
  \PhotonArc(90,80)(16.8,210,330){2}{3}
  \Vertex(75,72.5){1.2}
  \GCirc(30,50){10}{0}
  \Vertex(90,80){1.2}
  \Vertex(90,20){1.2}
  \Vertex(105,72.5){1.2}
  \ArrowLine(90,80)(120,95)
  \ArrowLine(120,65)(105,72.5)
  \ArrowLine(105,72.5)(90,80)
  \ArrowLine(120, 5)( 90,20)
  \ArrowLine( 90,20)(120,35)
  \put(76,52){$\gamma$}
  \put(43,67){$W$}
  \put(70,83){$W$}
  \put(53,14){$W$}
  \put(126,94){$\nu_{\mu}$}
  \put(126,62){$\mu^+$}
  \Text(65,-15)[]{(b)}
  \end{picture} }
  \end{center}
  \caption[]{Examples of one-loop diagrams that contribute to both 
             factorizable and non-factorizable corrections.}
  \label{fig:mixed}
\end{figure}
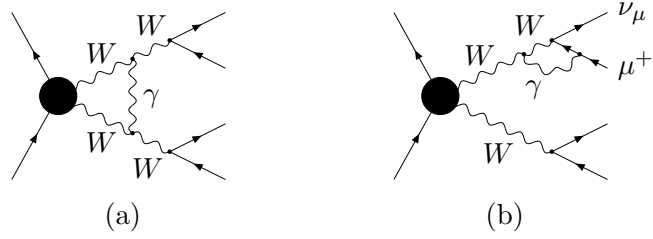

In order to illustrate our method, we explicitly apply it to two special
one-loop contributions (see Fig.~\ref{fig:mixed}). 
The first one [diagram (a)] is the so-called Coulomb graph, involving 
photon exchange between the two $W$ bosons. The corresponding semi-soft 
matrix element reads
\begin{eqnarray}
\label{Ma}
  \M_a &=& i e^2 \M^0_{\sss{DPA}} \int \frac{d^4 k}{(2\pi)^4 [k^2+io]} 
           \frac{4p_1p_2}{[D_1+2kp_1+io][D_2-2kp_2+io]}
           \nonumber \\
       &=& i e^2 \M^0_{\sss{DPA}} \int \frac{d^4 k}{(2\pi)^4 [k^2+io]}\,
           \frac{4p_1p_2}{[2kp_1+io][-2kp_2+io]}\,
           \Biggl\{ 1 - \frac{D_1}{D_1+2kp_1+io} 
           \nonumber \\
       & &          -\,\frac{D_2}{D_2-2kp_2+io}
                    + \frac{D_1 D_2}{[D_1+2kp_1+io][D_2-2kp_2+io]}
           \Biggr\}.
\end{eqnarray}
The first term in the last expression gives rise to a factorizable (on-shell)
contribution to the production stage, whereas the other three terms are 
counted as non-factorizable contributions. These three terms are classified as
$(\mbox{prod}\times\mbox{dec}^+)$, $(\mbox{prod}\times\mbox{dec}^-)$, and 
$(\mbox{dec}^+\times\mbox{dec}^-)$, respectively. In other words, the 
Coulomb graph contributes both to the usual Coulomb effect in on-shell
$W$-pair production and to a non-factorizable part.
For the photonic interaction between the $W^+$ boson and its $\mu^+$ 
decay product [diagram (b)] we obtain
\begin{eqnarray}
\label{Mb}
  \M_b &=& -\,i e^2 \M^0_{\sss{DPA}} \int \frac{d^4 k}{(2\pi)^4 [k^2+io]} 
           \frac{4p_1k_1}{[2kk_1+io][D_1+2kp_1+io]}
           \nonumber \\
       &=& -\,i e^2 \M^0_{\sss{DPA}} \int \frac{d^4 k}{(2\pi)^4 [k^2+io]}\,
           \frac{4p_1k_1}{[2kp_1+io][2kk_1+io]}\,
           \Biggl\{ 1 - \frac{D_1}{D_1+2kp_1+io} \Biggr\}.
\end{eqnarray}
Again the first term is a factorizable contribution, belonging to the $W^+$
decay stage, whereas the second term is a non-factorizable contribution
of the type $(\mbox{prod}\times\mbox{dec}^+)$. As is
clear from these examples, the non-factorizable contributions always
involve the Breit--Wigner ratios $D_i/(D_i\pm 2kp_i)$, which effectively 
remove the overall $W$ propagator $1/D_i$. The more energetic the 
exchanged photon is, the more suppressed such a ratio will be in the 
vicinity of the $M_i^2$ resonance. In fact, for $k_0>\Lambda$ the 
non-factorizable contributions are suppressed by
$\OO(M_W\Gamma_W/[E\Lambda])$ (see App.~\ref{app:semi-soft_functions}).

Every one-loop diagram with a semi-soft photon can be treated in this way.
By collecting all terms that contain the ratios $D_{i}/[D_{i}\pm 2kp_{i}]$ 
the formula for non-factorizable corrections is obtained. As one can 
see explicitly below, this expression is gauge-invariant. Since the 
expression contains those diagrams where irreducible $W$-boson lines are 
present, it can be viewed as a gauge-invariant extension of the set of
$W$-irreducible diagrams.

The so-defined non-factorizable corrections read 
\be
\label{corr/virt/nf_mtrx}
 \M_{\sss{nf}}^{\sss{virt}}
 =
 i\M^{0}_{\sss{DPA}}
 \int\frac{d^{4}k}{(2\pi)^{4}[k^{2}+io]}
 \biggl[
 \bigl(\J^{\mu}_{0}+\J^{\mu}_{\oplus}\bigr)\J_{+,\,\mu}+
 \bigl(\J^{\mu}_{0}+\J^{\mu}_{\ominus}\bigr)\J_{-,\,\mu}+
 \J^{\mu}_{+}\J_{-,\,\mu}
 \biggr].
\ee
The currents are given by
$$
 \J_{0}^{\mu}
 =
 e\Biggl[
  \frac{p_{1}^{\mu}}{kp_{1}+io}
 +\frac{p_{2}^{\mu}}{-kp_{2}+io}
 \Biggr],
$$
\be
 \J_{\oplus}^{\mu}
 = -\,
 e\Biggl[
  \frac{q_{1}^{\mu}}{kq_{1}+io}
 -\frac{q_{2}^{\mu}}{kq_{2}+io}
 \Biggr],
 \ \ \
 \J_{\ominus}^{\mu}
 = +\,
 e\Biggl[
  \frac{q_{1}^{\mu}}{-kq_{1}+io}
 -\frac{q_{2}^{\mu}}{-kq_{2}+io}
 \Biggr]
\ee
for photon emission from the production stage of the process, and 
\begin{eqnarray}
 \J_{+}^{\mu}
 &=& -\,e\Biggl[ \frac{p_{1}^{\mu}}{kp_{1}+io}
                + Q_{f_1}\frac{k_{1}^{\mu}}{kk_{1}+io}
                - Q_{f_1'}\frac{{k_{1}'}^{\mu}}{kk_{1}'+io}
        \Biggr]\frac{D_{1}}{D_{1}+2kp_{1}},
        \nonumber \\
 \J_{-}^{\mu}
 &=& -\,e\Biggl[ \frac{p_{2}^{\mu}}{-kp_{2}+io}
                + Q_{f_2}\frac{k_{2}^{\mu}}{-kk_{2}+io}
                - Q_{f_2'}\frac{{k_{2}'}^{\mu}}{-kk_{2}'+io}
        \Biggr]\frac{D_{2}}{D_{2}-2kp_{2}}
\end{eqnarray}
for photon emission from the decay stages of the process.
Here $Q_{f}$ stands for the charge of fermion $f$ in units of $e$.
After having defined the gauge-invariant currents, the $+io$ terms can be
dropped from $D_1+2kp_1$ and $D_2-2kp_2$, since $\imag D_i>0$.
Note the difference in the sign of the $io$ parts appearing 
in the currents $\J_{\ominus}$ and $\J_{\oplus}$. These signs actually
determine which interference terms give rise to a non-vanishing 
non-factorizable contribution after virtual and real-photon corrections
have been added. As can be seen from Eq.~(\ref{Mb}), in the upper hemisphere
of the complex $k_0$ plane there is only one pole: the so-called 
`photon pole', originating from the photon propagator $1/[k^2+io]$.
When virtual and real-photon corrections are combined, such a non-factorizable
contribution will vanish~\cite{nf-cancellations}. 
For the Coulomb graph, Eq.~(\ref{Ma}), this is
not the case, as also poles from the other propagators are present in both
hemispheres. As a result of such considerations only a very limited subset
of `final-state' interferences survives~\cite{nf-corr/bbc,nf-corr/ddr}:
the virtual corrections corresponding to Figs.~\ref{fig:WWnf} and 
\ref{fig:mixed}(a) as well as the associated real-photon corrections. 

The virtual factorizable corrections consist of all hard contributions and the
above-indicated part of the semi-soft ones.
The so-defined factorizable corrections have the nice feature that they can be 
expressed in terms of corrections to on-shell subprocesses, i.e.~the production
of two on-shell W bosons and their subsequent on-shell decays. 
The corresponding matrix element can be expressed in the same way as described
in Sect.~\ref{sec:pole-scheme}:
\begin{eqnarray}
  \label{mvirt}
  \M_{\sss{fact}}^{\sss{virt}} = \sum_{\lambda_{1},\lambda_{2}}
        \Pi_{\lambda_{1}  \lambda_{2}}(M_{1}, M_{2}) \
        \frac{\Delta^{(+)}_{\lambda_{1}}(M_{1})}{D_{1}} \
        \frac{\Delta^{(-)}_{\lambda_{2}}(M_{2})}{D_{2}}.
\end{eqnarray}
Here two of the amplitudes are taken at lowest order, whereas the remaining
one contains all possible one-loop contributions, including the $W$
wave-function factors that appear in Eq.~(\ref{pole-scheme}). 
In this way the well-known on-shell RC to the production and decay of pairs 
of W bosons~\cite{prod-corr}--\cite{decay-corr} appear 
as basic building blocks of the factorizable corrections.%
\footnote{Note that the complete density matrix is 
          required in this case, in contrast to the pure on-shell calculation
          which involves the trace of the density matrix only.} 
In the semi-soft limit the virtual factorizable corrections to the 
production stage, contained in $\Pi$, will cancel against the corresponding 
real-photon corrections. Non-vanishing contributions from $\Pi$ occur as 
soon as the $k^2$ terms in the propagators
cannot be neglected anymore. An example of this is the factorizable
correction from the Coulomb graph, given in Eq.~(\ref{Ma}).   
For the on-shell (factorizable) part of the Coulomb effect photons with
momenta $k_0=\OO(\Delta E)$ and $\,|\vec{k}|=\OO(\sqrt{M_{W}\Delta E}\,)$
are important~\cite{coulomb}, i.e.~$k^{2}$ cannot be neglected
in the propagators of the unstable particles. Since we stay well away from
the $W$-pair threshold ($\Delta E \gg \Gamma_W$), this situation occurs 
outside the realm of the
semi-soft photons. This fits nicely into the picture of the production stage
being a ``hard'' subprocess, governed by relatively short time scales
as compared with the much longer time scales required for the 
non-factorizable corrections, which interconnect the different subprocesses.

\subsection{Real-photon radiation}

In this subsection we discuss the aspects of real-photon radiation in the DPA.
To this end we consider the process
\be
\label{ee->ww->4fa}
      e^{+}(q_1)\,e^{-}(q_2) \to W^{+}(p_1)\,W^{-}(p_2)\,\bigl[\gamma(k)\bigr]
      \to
      \bar{f}_1(k_1)f_1'(k_1')\,f_2(k_2)\bar{f}_2'(k_2')\,\gamma(k),
\ee  
where in the intermediate state there may or may not be a photon. We will 
show how to extract the gauge-invariant double-pole residues in 
different situations. The exact cross-section for process (\ref{ee->ww->4fa}) 
can be written in the following form
\be
        \label{sig_rad}
        d\sigma
        =
        \frac{1}{2s}
        |\M_{\gamma}|^{2} 
        d\Gamma_{4f\gamma} 
        =
        \frac{1}{2s} 
        \Biggl[
        2\real\biggl( \M_{0}\M_{+}^{*} + \M_{0}\M_{-}^{*} 
                    + \M_{+}\M_{-}^{*} \biggr)
        + |\M_{0}|^2 + |\M_+|^2 + |\M_{-}|^2
        \Biggr] 
        d\Gamma_{4f\gamma},
\ee
where $d\Gamma_{4f\gamma}$ indicates the complete five-particle phase-space
factor, and the matrix elements $\M_{0}$ and $\M_{\pm}$ correspond
to the diagrams where the photon is attached to the 
production or decay stage of the three $W$-pair diagrams, respectively.
This split-up can be achieved with the help of the real-photon version of
the partial-fraction decomposition (\ref{decomp}).
Each contribution to the cross-section can be written in terms of polarization 
density matrices, which originate from the amplitudes
\be
\label{prodrad}
        \M_{0} =
        \Pi_{\gamma}(M_{1}, M_{2}) \
        \frac{\Delta^{(+)}(M_{1})}{D_{1}} \
        \frac{\Delta^{(-)}(M_{2})}{D_{2}} \ ,
\ee
\be
\label{dec+rad}
        \M_{+} =
        \Pi(M_{1\gamma}, M_{2}) \
        \frac{\Delta^{(+)}_{\gamma}(M_{1\gamma})}{D_{1\gamma}} \
        \frac{\Delta^{(-)}(M_{2})}{D_{2}} \ ,
\ee
\be
\label{dec-rad}
        \M_{-} =
        \Pi(M_{1}, M_{2\gamma}) \
        \frac{\Delta^{(+)}(M_{1})}{D_{1}} \
        \frac{\Delta^{(-)}_{\gamma}(M_{2\gamma})}{D_{2\gamma}} \ ,
\ee
where all polarization indices for the W bosons and the photon have been 
suppressed, and 
\be
        D_{i\gamma}=D_{i}+2kk_{i}+2kk_{i}', 
        \ \ \ 
        M_{i\gamma}^{2}=M_{i}^{2}+2kk_{i}+2kk_{i}',
        \ \ \ 
        M_{i}^{2} = (k_{i}+k_{i}')^{2}.
\ee
The matrix elements $\Pi_{\gamma}$ and $\Delta^{(\pm)}_{\gamma}$ describe the
production and decay of the $W$ bosons accompanied by the radiation of a 
photon. The matrix elements without subscript $\gamma$ have been 
introduced in Eq.~(\ref{pole/A}).

In the calculation of the Born matrix element and virtual corrections only two 
poles could be identified in the amplitudes, originating from the 
Breit--Wigner propagators $1/D_{i}$. The pole-scheme expansion was performed 
around these two poles. In contrast, the bremsstrahlung matrix element has 
four in general different poles, originating from the four Breit--Wigner 
propagators $1/D_{i}$ and $1/D_{i\gamma}$. As mentioned above, the matrix 
element can be rewritten as a sum of three matrix elements ($\M_0,\M_+,\M_-$),
each of which only contain two Breit--Wigner propagators. For these three
individual matrix elements the pole-scheme expansion is fixed, as before, to 
an expansion around the corresponding two poles. However, when calculating 
cross-sections [see Eq.~(\ref{sig_rad})] the mapping of the five-particle 
phase space introduces a new type of ambiguity. The interference terms in 
Eq.~(\ref{sig_rad}) involve two different double-pole expansions 
simultaneously. As such there is no natural choice for the 
phase-space mapping in those cases. The resulting ambiguity (implementation 
dependence) might have important repercussions on the quality of the DPA
calculation and therefore deserves some special attention. 
In this context the three earlier-defined regimes for the photon energy play 
a role:
\begin{itemize}
  \item for hard photons [$E_{\gamma}\gg \Gamma_W$] the Breit--Wigner poles of 
        the W-boson resonances before and after photon radiation are well 
        separated in phase space (see $M_{i\gamma}^{2}$ and $M_{i}^{2}$ 
        defined above). As a result, the interference terms in
        Eq.~(\ref{sig_rad}) can be neglected. 
        This leads to three {\it distinct}
        regions of on-shell contributions, where the photon can be assigned 
        unambiguously to the W-pair-production subprocess or to one of the 
        two decays. This assignment is determined by the pair of invariant 
        masses (out of $M_{i}^{2}$ and $M_{i\gamma}^{2}$) that is in the 
        $M_{W}^{2}$ region. Therefore, the double-pole residue
        can be expressed as the sum of the three on-shell contributions 
        without increasing the intrinsic error of the DPA. Note that in the 
        same way it is also possible to experimentally assign the photon to 
        one of the subprocesses, since misassignment errors are suppressed,
        assuming for convenience that all final-state momenta can
        ideally be measured.
  \item for semi-soft photons [$E_{\gamma} = {\cal O}(\Gamma_W)$] the 
        Breit--Wigner poles are relatively close together in phase space, 
        resulting in a substantial overlap of the line shapes. The assignment 
        of the photon is now subject to larger errors. Moreover, since the 
        interference terms in Eq.~(\ref{sig_rad}) cannot be neglected,
        the issue of the phase-space mapping has to be addressed. In the 
        following we give a proper prescription for calculating the DPA 
        residues and discuss their quality. 
  \item for soft photons [$E_{\gamma}\ll \Gamma_W$] the Breit--Wigner poles 
        are on top of each other, resulting in a pole-scheme expansion that is 
        identical to the one without the photon.
\end{itemize}

\subsubsection{Hard photons}

Let us first consider the hard-photon regime in more detail. Due to the fact 
that the poles are well separated in the hard-photon regime, it is clear that
the interference terms are suppressed and can be neglected:
\be
        d\sigma
        =
        \frac{1}{2s}
        \biggl[ |\M_{0}|^{2}+|\M_{+}|^{2}+|\M_{-}|^{2}\biggr]
        d\Gamma_{4f\gamma}.
\ee
Note that each of the three terms has two poles, originating from two 
resonant propagators. However, these poles are different for different 
terms. The phase-space factor can be rewritten in three equivalent ways.
The first is
\be
        d\Gamma_{4f\gamma} = d\Gamma_{\sss{0}}^{\gamma} = 
        d\Gamma_{\sss{pr}}^{\gamma}\cdot 
        d\Gamma_{\sss{dec}}^{+} \cdot d\Gamma_{\sss{dec}}^{-}   
        \cdot   
        \frac{dM_{1}^{2}}{2\pi} \cdot \frac{dM_{2}^{2}}{2\pi},
\ee
with
\be
        d\Gamma_{\sss{pr}}^{\gamma}
        =
        \frac{1}{(2\pi)^{2}}\,
        \delta(q_{1}+q_{2}-p_{1}-p_{2}-k)\, \frac{d\vec{p}_{1}}{2p_{10}}\, 
        \frac{d\vec{p}_{2}}{2p_{20}}\,    \frac{d\vec{k}}{(2\pi)^{3}2k_{0}}.
\ee
The two others are 
\be
        d\Gamma_{4f\gamma} = d\Gamma_{+}^{\gamma} = 
        d\Gamma_{\sss{pr}}\cdot 
        d\Gamma_{\sss{dec}}^{+\gamma} \cdot d\Gamma_{\sss{dec}}^{-}     
        \cdot   
        \frac{dM_{1\gamma}^{2}}{2\pi} \cdot \frac{dM_{2}^{2}}{2\pi},
\ee
with
\be
        d\Gamma_{\sss{dec}}^{+\gamma}
        =
        \frac{1}{(2\pi)^{2}}\,
        \delta(p_{1}-k_{1}-k_{1}'-k)\, \frac{d\vec{k}_{1}}{2k_{10}}\, 
        \frac{d\vec{k}_{1}'}{2k_{10}'}\,  \frac{d\vec{k}}{(2\pi)^{3}2k_{0}},
\ee
and a similar expression for $d\Gamma_{-}^{\gamma}$.
The phase-space factors $d\Gamma_{\sss{pr}}$ and $d\Gamma_{\sss{dec}}^{\pm}$
are just the lowest-order ones. The cross-section can then be written in the 
following equivalent form
\be
\label{corr/hard}
        d\sigma
        =
        \frac{1}{2s}
        \biggl[ 
         |\M_{0}|^{2}\,d\Gamma_{0}^{\gamma}
        +|\M_{+}|^{2}\,d\Gamma_{+}^{\gamma}
        +|\M_{-}|^{2}\,d\Gamma_{-}^{\gamma}
        \biggr].
\ee
In order to extract gauge-invariant quantities, the DPA limit should be 
taken. This amounts to taking the limit $p_{1,2}^{2}\to M_{W}^{2}$, using a
prescription that resembles the one presented in Sect.~\ref{sec:pole-scheme}. 
Note however that $p_{1,2}$ can be different according to the 
$\delta$-functions in the decay parts of the different phase-space factors. 
To be specific, the production term $|\M_{0}|^{2}$ has poles at 
$p_{i}^{2}=M_{i}^{2}=M_{W}^{2}$, $|\M_{+}|^{2}$ has poles at 
$p_{1}^{2}=M_{1\gamma}^{2}=M_{W}^{2}$ and $p_{2}^{2}=M_{2}^{2}=M_{W}^{2}$, and 
$|\M_{-}|^{2}$ has poles at $p_{1}^{2}=M_{1}^{2}=M_{W}^{2}$ and
$p_{2}^{2}=M_{2\gamma}^{2}=M_{W}^{2}$.
Again we fix solid angles in the mapping, including the solid angle of
the photon. Since the energy range of the photon in the on-shell kinematics
of the $W$ bosons is different from the off-shell case, it may happen that the
energy of the photon in an off-shell four-fermion\,--\,one-photon event with 
certain solid angles lies outside the on-shell phase space with the same solid 
angles. A possible procedure is to assign a zero cross-section to those 
events. Since the events are anyhow rare, being close to the edge of the 
off-shell phase space, this procedure constitutes an acceptable and simple 
solution. An alternative way to avoid unphysical on-shell phase-space points
would be to write the photon energy as a fraction of the maximum attainable 
photon energy for given invariant masses $\sqrt{p_i^2}$ of the resonant 
$W$ bosons and given solid angles:
\be
\label{ph_energy}
  E_{\gamma} = x_{\gamma} E_{\gamma}^{\sss{max}}(\sqrt{p_1^2},\sqrt{p_2^2}, 
  \mbox{angles}\,).
\ee
In this way the photon energy is projected on the interval $[0,1]$.
The maximum photon energy $E_{\gamma}^{\sss{max}}$ depends on the specific 
term in Eq.~(\ref{corr/hard}). 
Subsequently the fraction $x_{\gamma}$ and the afore-mentioned solid
angles are kept fixed during the mapping from off- to on-shell events.
Then $E_{\gamma}$~for the on-shell case is found from Eq.~(\ref{ph_energy}),
where $p_{i}^{2}$ are replaced by $M_{W}^{2}$.

It should be stressed that in the on-shell phase space there is no ambiguity
concerning the treatment of the photon. One obtains in the DPA limit three 
gauge-invariant on-shell contributions to Eq.~(\ref{corr/hard}).
The calculation of the corresponding matrix elements is presented in 
App.~\ref{app:hard_rad}.

\subsubsection{Semi-soft and soft photons}

We complete our survey of the different photon-energy regimes by
considering semi-soft and soft photons.
The split-up of factorizable and non-factorizable real-photon corrections
proceeds in the same way as described in the previous subsection for
virtual corrections. The result reads in semi-soft approximation
\be
\label{corr/real/mtrx}
        d\sigma
        =
        \frac{1}{2s}|\M_{\gamma}|^{2} d\Gamma_{\sss{4f}\gamma}
        \approx 
        -\, d\sigma_{\sss{DPA}}^{0}
        \frac{d\vec{k}}{(2\pi)^{3}2k_{0}}
        \Biggl[
        2\real\biggl(\I_{0}^{\mu}\I_{+,\,\mu}^{*}+\I_{0}^{\mu}\I_{-,\,\mu}^{*}
              +\I_{+}^{\mu}\I_{-,\,\mu}^{*}\biggr)
         +|\I_{0}^2|+|\I_{+}^2|+|\I_{-}^2|
         \Biggr].
\ee
The gauge-invariant currents $\I_{0}$ and $\I_{\pm}$ are given by
\begin{eqnarray}
\label{corr/real/currents}
 \I_{0}^{\mu}
 &=& e\Biggl[ \frac{p_{1}^{\mu}}{kp_{1}}-\frac{p_{2}^{\mu}}{kp_{2}}
              -\frac{q_{1}^{\mu}}{kq_{1}} + \frac{q_{2}^{\mu}}{kq_{2}}
      \Biggr],
      \nonumber \\
 \I_{+}^{\mu} 
 &=& -\,e\Biggl[ \frac{p_{1}^{\mu}}{kp_{1}}
               +Q_{f_1}\frac{k_{1}^{\mu}}{kk_{1}}
               -Q_{f_1'}\frac{{k_{1}'}^{\mu}}{kk_{1}'}
       \Biggr]\frac{D_{1}}{D_{1}+2kp_{1}},
      \nonumber \\ 
 \I_{-}^{\mu} 
 &=& +\,e\Biggl[ \frac{p_{2}^{\mu}}{kp_{2}}
               + Q_{f_2}\frac{k_{2}^{\mu}}{kk_{2}}
               - Q_{f_2'}\frac{{k_{2}'}^{\mu}}{kk_{2}'}
       \Biggr]\frac{D_{2}}{D_{2}+2kp_{2}}.
\end{eqnarray}
The first three interference terms in Eq.~(\ref{corr/real/mtrx})
correspond to the real non-factorizable corrections. 
The last three squared terms in Eq.~(\ref{corr/real/mtrx}) belong to the 
factorizable real-photon corrections. They constitute the semi-soft limit of 
Eq.~(\ref{corr/hard}). Note also that the currents are the same for both 
possible expressions for $p_i$, i.e.~it does not matter whether 
$p_i=k_i+k_i'\,$ or $\,p_i=k_i+k_i'+k$.

As mentioned before, the DPA residues have to be defined properly in the 
presence of overlapping Breit--Wigner resonances in the semi-soft regime.
The above equation (\ref{corr/real/mtrx}) specifies such a procedure: 
the Born cross-section for a four-particle phase space is 
factored out and does not depend on the photon momentum. The factor between 
the square brackets is the usual soft-photon factor, except that the rapid 
variation of the Breit--Wigner resonances has been kept, leading to the 
ratios $D_{i}/D_{i\gamma}$ which take into account the shift in the 
Breit--Wigner distributions due to the photon. In the vicinity
of the $M_{i}^{2}$ resonance these ratios are negligible for 
hard photons, unity for soft photons, and of $\OO(1)$ for semi-soft photons.

This prescription is by no means unique. In principle one could have
chosen any of the two Breit--Wigner resonances, $1/D_i\,$ or $\,1/D_{i\gamma}$,
to define the phase-space mapping for the interference terms. Or in other 
words, both on-shell phase-space limits, $M_i^2=M_W^2\,$ or 
$\,M_{i\gamma}^2=M_W^2$, constitute equally plausible DPA mappings. 
The differences between both phase-space mappings are of $\OO(k)$.
Since the interference terms are only non-negligible in the semi-soft regime,
it is conceivable that the implementation dependence associated with these
different on-shell limits remains of $\OO(\Gamma_W/M_W)$. Let us,
for instance, consider the $M_i^2$ distribution in the vicinity of the
resonance. Any $\OO(k)$ shift in the factor multiplying $1/D_i$ would result 
in additional terms of $\OO(k/M_W)$, i.e.~at worst $\OO(\Gamma_W/M_W)$ in the 
semi-soft limit. A similar shift in the factor belonging to $1/D_{i\gamma}$ 
results in additional contributions to the DPA residues that are suppressed by 
$\OO(\Gamma_W/E)$ for {\it all} values of the photon energy, as is explained 
in App.~\ref{app:semi-soft_functions}. Therefore one can conclude that
overlapping Breit--Wigner resonances do not necessarily imply a reduced
quality of the DPA. 

Based on this observation, we have the freedom to choose a suitable procedure
for semi-soft photons. The fact whether hard-photon contributions are 
suppressed or not will serve as our guideline for fixing the choice. 
Whenever the hard-photon effects yield vanishing contributions to certain 
observables, we will use the semi-soft currents of Eqs.~(\ref{corr/real/mtrx})
and (\ref{corr/real/currents}). As a consequence, the non-factorizable
corrections are always calculated with the help of the semi-soft interference 
terms. In order to specify our approach, we indicate in the following
how the above-defined matrix elements and currents are used in
the various distributions of experimental interest.

\subsubsection{Real-photon contributions to various distributions}

As in all RC, the role of the bremsstrahlung process is twofold. In the first 
place the off-shell $W$-pair process (\ref{intro/ee->ww->4f}) has at 
lowest-order level all kinds of distributions to which one would like to
calculate $\OO(\alpha)$ RC. These RC consist of virtual and real-photon 
contributions. An integration over all allowed photon energies should be 
carried out, i.e.~the radiated photon should be treated inclusively.

The second application of the bremsstrahlung process involves the evaluation
of exclusive photon distributions. Since the photon has to be detected it 
should be sufficiently hard. How hard depends on the experimental resolution.
An example of such an exclusive photon distribution is the photon-energy
spectrum $d\sigma/dE_{\gamma}$. This distribution receives contributions from
the three hard-photon terms in Eq.~(\ref{corr/hard}) as well as from the 
semi-soft interference terms of Eq.~(\ref{corr/real/mtrx}). The latter terms
of course only contribute for photon energies that are not too hard.

In the next section results will be given for various distributions. 
For exclusive photons we present the photon-energy spectrum 
$d\sigma/dE_{\gamma}$. For the inclusive photon distributions
we discuss in the following how the bremsstrahlung part is treated. Depending 
on the distribution, different approximations can be made.

As indicated above, the calculation of RC to distributions of the off-shell 
$W$-pair process (\ref{intro/ee->ww->4f}) involves an integration over the 
photon phase space that is left available when fixing other kinematic 
variables. This means that the photon phase
space for hard, semi-soft, and soft photons will in general be integrated 
over. The soft-photon part is standard and should be combined with the
virtual corrections. How the non-soft part should be treated depends
on the distribution one likes to study. In general one considers 
distributions of the form $d\sigma_{\sss{DPA}}/dX$, where $dX$ stands for 
$dM_{1}^{2}dM_{2}^{2}$ and a product of solid angles for $W$-boson
production and decay. For the following discussion it does not matter 
whether these solid angles are kept or integrated out. However, it does 
matter whether one wants a double distribution in $M_{1}^{2}$ and $M_{2}^{2}$,
a single distribution in $M_{1}^{2}$, or no distribution in $M_{i}^{2}$ at 
all. In other words, the treatment differs depending on whether one 
integrates over $M_{i}^{2}$ or not.

Let us first discuss the procedure for the real-photon corrections to the 
double Breit--Wigner distribution $\,d\sigma/dM_{1}^{2}dM_{2}^{2}\,$ in the 
vicinity of the peak, i.e.~one is not interested in the tails of this peaked 
distribution. In Table~\ref{tab:1} we specify which expressions are
used in the different regimes for the photon energy. For this specification 
we use the positions of the Breit--Wigner poles as listed below 
Eq.~(\ref{corr/hard}).
\begin{table}
\[
  \begin{array}{||c||c|c|c|c||}    \hline \hline
    \raisebox{-3mm}{photon energy} 
    & \multicolumn{4}{|c||}{\raisebox{-1mm}{contributions to 
                                            $d\sigma/dM_{1}^{2}dM_{2}^{2}$}}\\ 
    \cline{2-5}
         & \mbox{prod.}+\gamma    
         & W^{+}\mbox{ decay}+\gamma  
         & W^{-}\mbox{ decay}+\gamma 
         & \mbox{non-factorizable} \\
    \hline  \hline
    \mbox{hard}       & |\M_{0}|^{2}\ (\ref{corr/hard})       
                      & \to 0 
                      & \to 0 
                      & \to 0
                      \\   
    \mbox{semi-soft}  & |\M_{0}|^{2}\ (\ref{corr/hard})       
                      & |\I_{+}^{2}|\ \ \ \ (\ref{corr/real/mtrx})           
                      & |\I_{-}^{2}|\ \ \ \ (\ref{corr/real/mtrx})
                      & \mbox{interference\ \ \ \ \ \ \ (\ref{corr/real/mtrx})}
                      \\
    \mbox{soft}       & |\I_{0}^{2}|\ \ \ \ (\ref{corr/real/mtrx})  
                      & |\I_{+}^{2}|_{\sss{soft}}\ (\ref{corr/real/mtrx}) 
                      & |\I_{-}^{2}|_{\sss{soft}}\ (\ref{corr/real/mtrx})
                      & \mbox{soft interference\ (\ref{corr/real/mtrx})}
                      \\
    \hline \hline
  \end{array}
\]
\caption[]{Formulae for the bremsstrahlung matrix elements for the 
           distribution $d\sigma/dM_{1}^{2}dM_{2}^{2}$.}
\label{tab:1}
\end{table}%

The entries of Table~\ref{tab:1} can be explained as follows. The real-photon 
corrections to the distribution $\,d\sigma/dM_{1}^{2}dM_{2}^{2}\,$ come in the 
first place from the production part, where the hard-photon matrix element
squared $|\M_{0}|^{2}$ contains the overall Breit--Wigner line shapes in 
$M_{1}^{2}$ and $M_{2}^{2}$. This term should be taken from 
Eq.~(\ref{corr/hard}) and should be integrated over the photon phase space.
In the soft-photon limit the usual soft-photon
factorization in terms of $|\I_{0}^{2}|$ is obtained. This explains the 
second column in Table.~\ref{tab:1}, where the usual RC to stable $W$-pair 
production appear, except that one now calculates the full density matrices 
rather than merely the trace. 

The second type of contribution involves the $W$-boson decays with additional
photon. When the photon is hard these contributions tend to zero as
$\Gamma_W^2/k_0^2$ (see App.~\ref{app:semi-soft_functions}), since they do 
not contain a double Breit--Wigner distribution in both $M_{1}^{2}$ and 
$M_{2}^{2}$ according to the list below Eq.~(\ref{corr/hard}). For (semi)soft 
photons, however, one gets overlapping Breit--Wigner resonances and the 
$|\I_{\pm}^{2}|$ terms of Eq.~(\ref{corr/real/mtrx}) can be used.
These (semi)soft-photon factors have to be integrated while keeping 
$D_{i}$ fixed, i.e.~we keep the variables $M_{i}^{2}$ fixed. The corresponding
integrals can be found in App.~\ref{app:semi-soft_functions}. The integration 
over the semi-soft photon momenta leads to contributions from 
$M_{i\gamma}^{2}$ values that are higher then $M_{i}^{2}$, resulting in a 
distortion of the original $M_{i}^{2}$ Breit--Wigner distribution. 
This final-state-radiation effect has been discussed recently in the 
literature and turns out to be quite sizeable~\cite{fsr}. The reason why the 
distortion is so large lies in the fact that the final-state collinear
singularities [$\propto\frac{\alpha}{\pi}\,Q_{f}^{2}\log(m_f^2/M_W^2)$] do 
not vanish, even not for fully inclusive photons. After all, a fixed value 
of $M_{i}^{2}$ makes it impossible to sum over all degenerate final states
by a mere integration over the photon momentum.%
\footnote{The usual cancellation of final-state collinear singularities 
          is achieved only if also $M_{i}^{2}$ is integrated over.}
On top of that the IR poles $\propto 1/k$ induce a further enhancement of
the collinear logarithms by a factor $\log(D_i/M_W^2)$. 

So far we have included all factorizable corrections. To this one should
add the non-factorizable corrections. Again effectively only the (semi)soft 
region is relevant. For more information on these non-factorizable corrections
we refer to the detailed discussion in the 
literature~\cite{nf-corr/bbc,nf-corr/ddr,rheinsberg98}.

\begin{table}
\[
  \begin{array}{||c||c|c|c|c||}    \hline \hline
    \raisebox{-3mm}{photon energy} 
    & \multicolumn{4}{|c||}{\raisebox{-1mm}{contributions to 
                                            $d\sigma/dM_{1}^{2}$}}\\ 
    \cline{2-5}
         & \mbox{prod.}+\gamma    
         & W^{+}\mbox{ decay}+\gamma  
         & W^{-}\mbox{ decay}+\gamma 
         & \mbox{non-factorizable} \\
    \hline  \hline
    \mbox{hard}       & |\M_{0}|^{2}\ (\ref{corr/hard})       
                      & \to 0 
                      & |\M_{-}|^{2}\ (\ref{corr/hard})
                      & \to 0
                      \\   
    \mbox{semi-soft}  & |\M_{0}|^{2}\ (\ref{corr/hard})        
                      & |\I_{+}^{2}|\ \ \ \ (\ref{corr/real/mtrx})           
                      & |\M_{-}|^{2}\ (\ref{corr/hard})
                      & \mbox{interference\ \ \ \ \ \ \ (\ref{corr/real/mtrx})}
                      \\
    \mbox{soft}       & |\I_{0}^{2}|\ \ \ \ (\ref{corr/real/mtrx}) 
                      & |\I_{+}^{2}|_{\sss{soft}}\ (\ref{corr/real/mtrx}) 
                      & |\I_{-}^{2}|_{\sss{soft}}\ (\ref{corr/real/mtrx})
                      & \mbox{soft interference\ (\ref{corr/real/mtrx})}
                      \\
    \hline \hline
  \end{array}
\]
\caption[]{Formulae for the bremsstrahlung matrix elements for the 
           distribution $d\sigma/dM_{1}^{2}$.}
\label{tab:2}
\end{table}%
As the next relevant distribution we discuss the real-photon corrections to 
the single Breit--Wigner distribution $d\sigma/dM_{1}^{2}$, obtained from
the previous case by integrating over $M_{2}^{2}$. To this end we make a 
similar table (Table~\ref{tab:2}) and discuss the necessary changes. 
Based on the discussion below Eq.~(\ref{corr/hard}), it is clear that a 
Breit--Wigner distribution in $M_{1}^{2}$ is present explicitly in 
$|\M_{0}|^{2}$ and $|\M_{-}|^{2}$, both for hard and semi-soft photons.  
These two terms should be taken from Eq.~(\ref{corr/hard}) and should
be integrated over the photon phase space and $M_{2}^{2}$. 
The $|\M_{+}|^{2}$ term does not have a Breit--Wigner distribution in 
$M_{1}^{2}$ as long as the photon is hard. Therefore it effectively
only contributes in the (semi)soft regime, like in the previous case.
Therefore the $|\I_{+}^{2}|$ term in Eq.~(\ref{corr/real/mtrx}) is used and 
the photon integration is performed while keeping $M_{1}^{2}$ fixed.
The non-factorizable corrections contribute in the same way as described for 
the previous case. 

We conclude by considering pure angular distributions. In this case the 
picture is simple. All contributions should be taken from 
Eq.~(\ref{corr/hard}) and should be integrated over the photon phase 
space and the invariant masses $M_{i}^{2}$. The non-factorizable corrections
vanish in this situation, which is typical for these interference 
effects~\cite{nf-cancellations}.


\section{Numerical results}
\label{sec:plots}

All the corrections discussed in the previous sections were combined
into a Fortran program. In this section we present some numerical results.
In particular we study the implications of the RC on various distributions.
For the numerical evaluations we use the $G_{\mu}$-parametrization
(see Sect.~\ref{sec:born_num}) and the input parameters listed in 
Eqs.~(\ref{num/input})--(\ref{num/width}).

\subsection{One-dimensional distributions}

We start off by considering various one-dimensional distributions. Where
applicable both the lowest-order distribution ($d\sigma_{\sss{DPA}}^0$) and 
the corrected one ($d\sigma_{\sss{DPA}}$) will be presented, followed by a
separate plot for the relative correction factor $\delta_{\sss{DPA}}$
[see Eq.~(\ref{relcorr})].

\subsubsection{The photon-energy spectrum}

\begin{figure}
  \unitlength 1cm
  \begin{center}
  \begin{picture}(13.4,8)
  \put(-0.2,6.9){\makebox[0pt][c]{\boldmath $\frac{\displaystyle d\sigma}
                                 {\displaystyle dE_{\gamma}}$}}
  \put(-0.2,5.8){\makebox[0pt][c]{\bf [fb]}}
  \put(13.5,-0.2){\makebox[0pt][c]{\bf\boldmath $E_{\gamma}$ [GeV]}}
  \put(-1,-6){\includegraphics{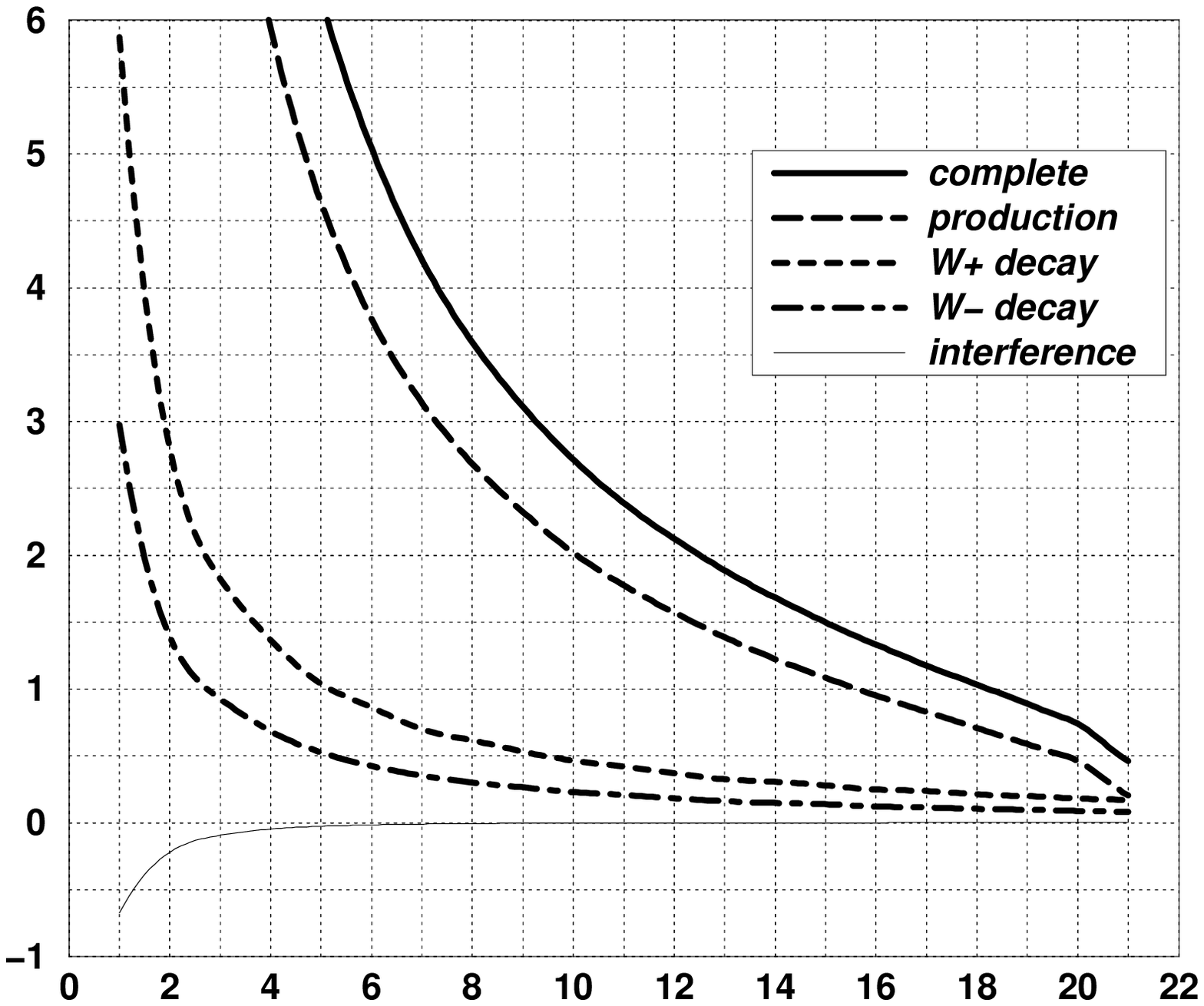}}
  \end{picture}
  \end{center}
  \caption[]{The photon-energy distribution $d\sigma/dE_{\gamma}$ for the
             $\mu^+\nu_{\mu} \tau^-\bar{\nu}_{\tau} \gamma$ final state at 
             $2E=184\GeV$. In addition the separate production and decay 
             contributions are given.}
\label{fig:rad1}
\end{figure}%
\begin{figure}
  \unitlength 1cm
  \begin{center}
  \begin{picture}(13.4,8)
  \put(-0.2,6.8){\makebox[0pt][c]{\boldmath $\frac{\displaystyle d\sigma}
                                 {\displaystyle dE_{\gamma}}$}}
  \put(-0.2,5.7){\makebox[0pt][c]{\bf [fb]}}
  \put(13.5,-0.2){\makebox[0pt][c]{\bf\boldmath $E_{\gamma}$ [GeV]}}
  \put(-1,-6){\includegraphics{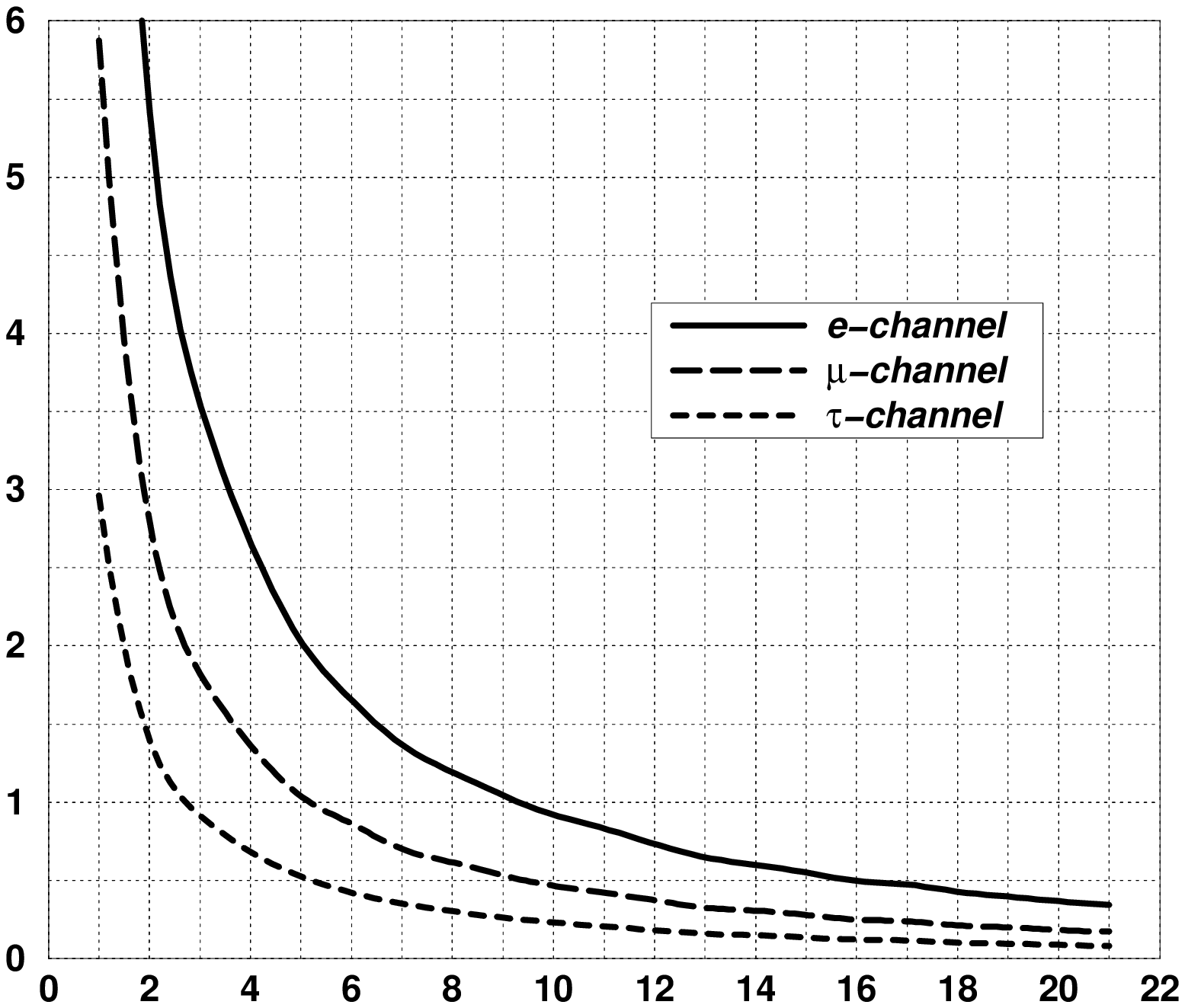}}
  \end{picture}
  \end{center}
  \caption[]{Decay contributions to the photon spectrum from different
             leptonic $W$ decays.}
\label{fig:rad2}
\end{figure}%
Since real-photon bremsstrahlung contributes to the RC to various 
distributions, it is useful to evaluate the photon-energy spectrum 
$d\sigma/dE_{\gamma}$ separately as well. In DPA it gets contributions from 
the three terms in Eq.~(\ref{corr/hard}) and the semi-soft interference terms 
of Eq.~(\ref{corr/real/mtrx}). The photon spectrum originating from the 
production stage is the same for all final states, but the spectra from the
decay stages depend on the specific final state. In Fig.~\ref{fig:rad1} the
DPA photon-energy distribution $d\sigma/dE_{\gamma}$ is shown for the 
$\mu^+\nu_{\mu} \tau^-\bar{\nu}_{\tau} \gamma$ final state at $2E=184\GeV$, 
together with the production and decay parts of the spectrum. 
In Fig.~\ref{fig:rad2} we separately list the three possible leptonic 
radiative-decay contributions, originating from $W \to \ell\nu_{\ell}\gamma$ 
($\ell=e,\mu,\tau$). The substantial differences are caused by the
explicit fermion-mass dependence for collinear photon radiation 
(see App.~\ref{app:collrad}).

\subsubsection{The total cross-section as a function of energy}

\begin{figure}
  \unitlength 1cm
  \begin{center}
  \begin{picture}(13.4,8)
  \put(-0.2,6.7){\makebox[0pt][c]{\boldmath $\sigma_{\sss{\bf tot}}$}}
  \put(-0.2,5.6){\makebox[0pt][c]{\bf [pb]}}
  \put(13.5,-0.2){\makebox[0pt][c]{\bf\boldmath $2E$ [GeV]}}
  \put(-1,-6){\includegraphics{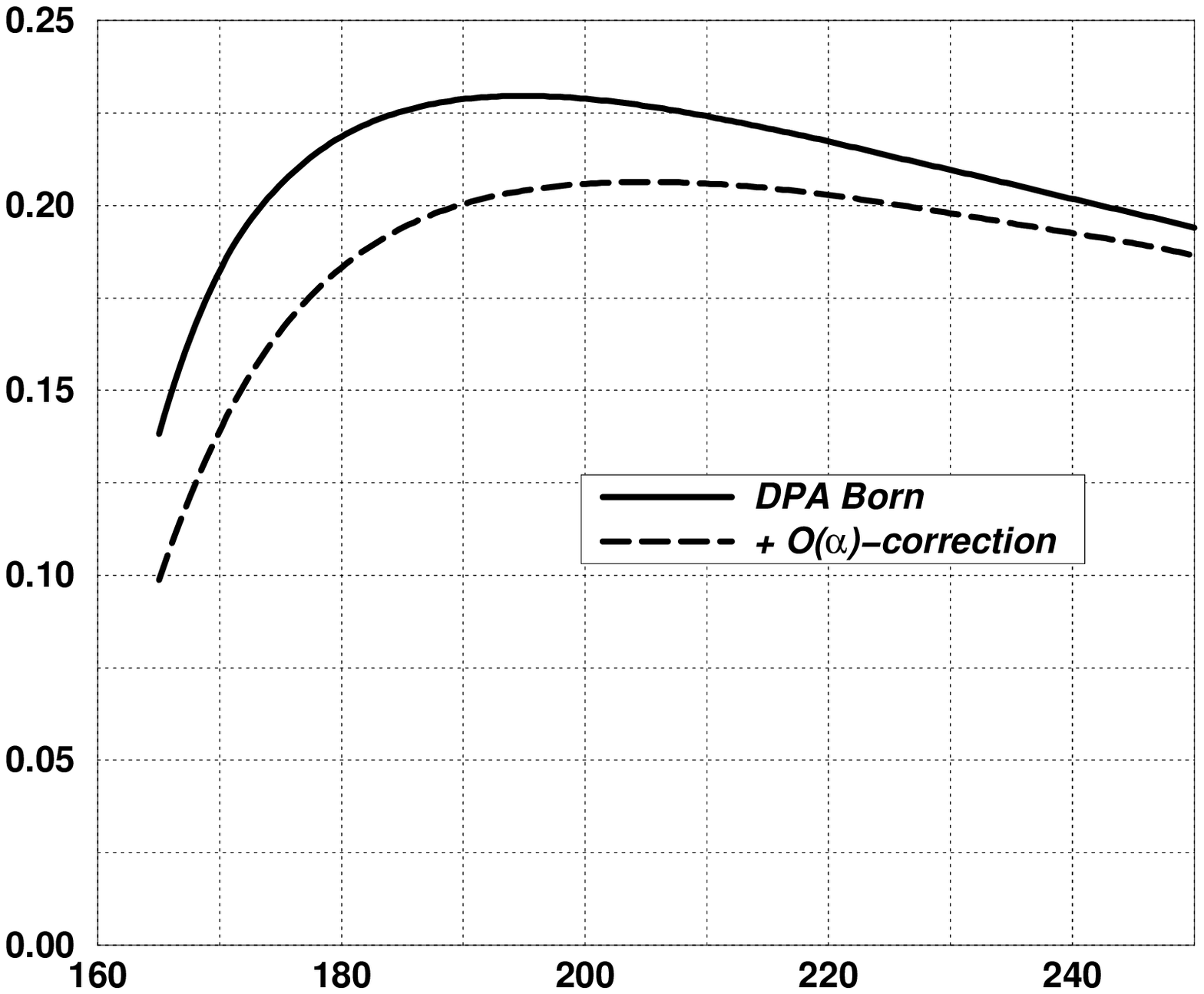}}
  \end{picture}
  \end{center}
  \caption[]{The energy dependence of the total cross-section 
             $\sigma_{\sss{tot}}$ for the
             $\mu^+\nu_{\mu} \tau^-\bar{\nu}_{\tau}$ final state.}
\label{fig:incl}
\end{figure}%
\begin{figure}
  \unitlength 1cm
  \begin{center}
  \begin{picture}(13.4,8)
  \put(-0.2,6.7){\makebox[0pt][c]{\boldmath $\delta$}}
  \put(-0.2,5.7){\makebox[0pt][c]{\bf [\%]}}
  \put(13.5,-0.2){\makebox[0pt][c]{\bf\boldmath $2E$ [GeV]}}
  \put(-1,-6){\includegraphics{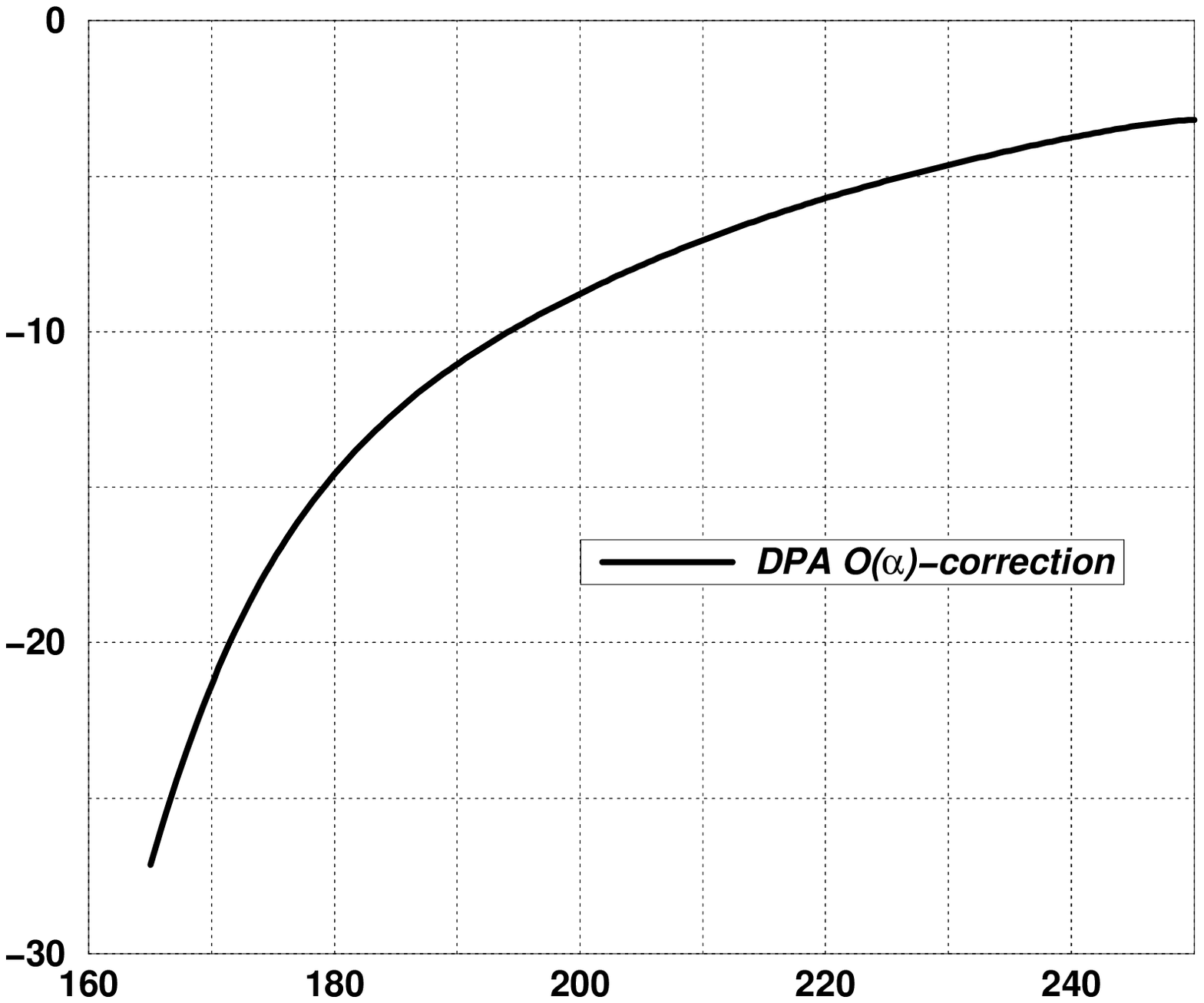}}
  \end{picture}
  \end{center}
  \caption[]{The relative correction factor corresponding to
             Fig.~\protect\ref{fig:incl}.}
\label{fig:incl/rel}
\end{figure}%
In Fig.~\ref{fig:incl} we compare the total cross-section with and without RC
for the $\mu^+\nu_{\mu} \tau^-\bar{\nu}_{\tau}$ final state. The corresponding
relative correction factor $\delta$ is given in Fig.~\ref{fig:incl/rel}. 
As a check of our calculation we have verified that the corrected
cross-section coincides within the integration errors with the corrected
cross-section for stable $W$ bosons multiplied by the corrected branching ratio
$(\Gamma_{W\to\ell\nu_{\ell}}/\Gamma_W)^2$. 

The observed RC are large and negative, especially close to the $W$-pair 
threshold. This is mainly caused by real-photon ISR, which 
effectively lowers the available $W$-pair energy, combined with the fact that
near the $W$-pair threshold the cross-section is rapidly decreasing with 
decreasing energy.

\subsubsection{Production-angle distribution}

\begin{figure}
  \unitlength 1cm
  \begin{center}
  \begin{picture}(13.4,8)
  \put(-0.3,6.8){\makebox[0pt][c]{\boldmath $\frac{\displaystyle d\sigma}
                                 {\displaystyle d\cos\theta}$}}
  \put(-0.3,5.6){\makebox[0pt][c]{\bf [pb]}}
  \put(13.5,-0.2){\makebox[0pt][c]{\boldmath $\cos\theta$}}
  \put(-1,-6){\includegraphics{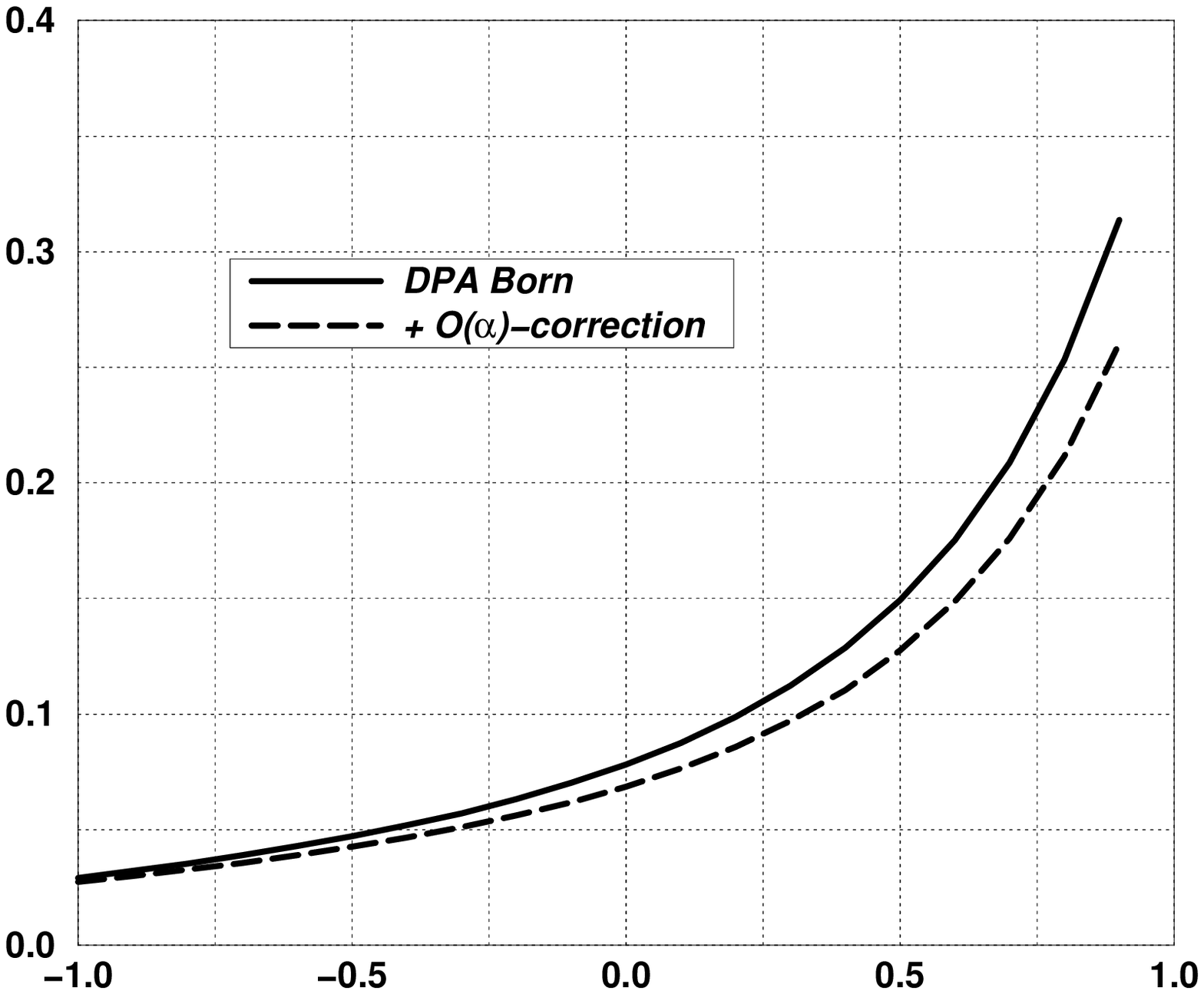}}
  \end{picture}
  \end{center}
  \caption[]{The production-angle distribution $\,d\sigma/d\cos\theta\,$  
             for the $\mu^+\nu_{\mu} \tau^-\bar{\nu}_{\tau}$ final state
             at $2E=184\GeV$.}
\label{fig:prod}
\end{figure}%
\begin{figure}
  \unitlength 1cm
  \begin{center}
  \begin{picture}(13.4,8)
  \put(-0.2,6.1){\makebox[0pt][c]{\boldmath $\delta$}}
  \put(-0.2,5.1){\makebox[0pt][c]{\bf [\%]}}
  \put(13.5,-0.2){\makebox[0pt][c]{\boldmath $\cos\theta$}}
  \put(-1,-6){\includegraphics{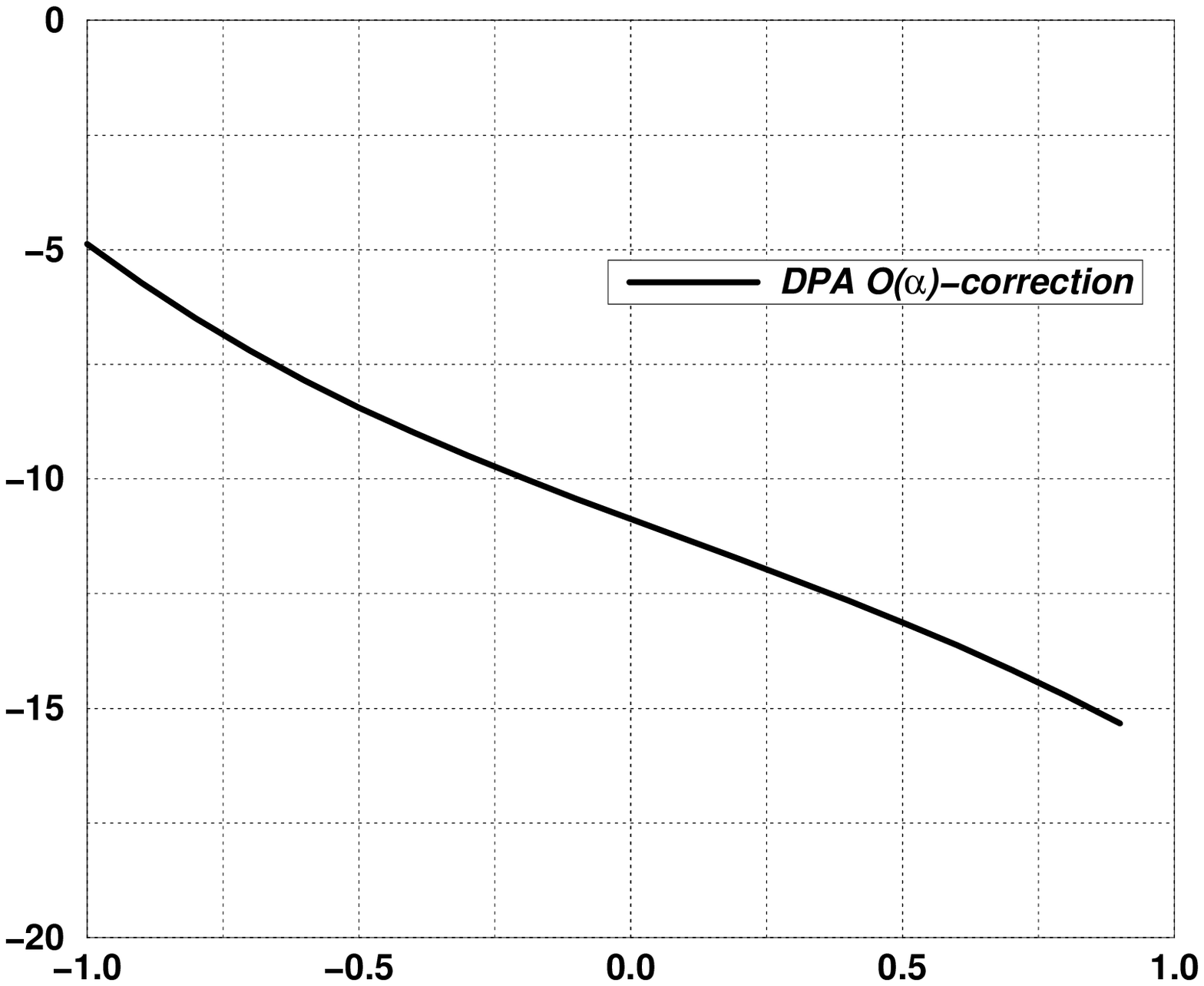}}
  \end{picture}
  \end{center}
  \caption[]{The relative correction factor corresponding to
             Fig.~\protect\ref{fig:prod}.}
\label{fig:prod/rel}
\end{figure}%
In Fig.~\ref{fig:prod} we plot the production-angle distribution 
$\,d\sigma/d\cos\theta\,$ for the $\mu^+\nu_{\mu} \tau^-\bar{\nu}_{\tau}$ 
final state at $2E=184\GeV$. The relative correction factor displayed in 
Fig.~\ref{fig:prod/rel} differs substantially from the $-12.8\%$ normalization
effect that was observed for the 
total cross-section. In the forward direction it is slightly more negative,
whereas in the backward direction it is substantially less negative. A proper
understanding of this distortion is quite important, since non-standard triple 
gauge-boson couplings might lead to exactly this type of signature.
The origin of the distortion can be traced back to hard-photon ISR.
Such hard-photon emissions boost the centre-of-mass (CM) system
of the $W$ bosons, causing a migration of events to take place from regions
with large cross-sections in the CM system (forward direction) to
regions with small cross-sections in the LAB system (backward direction).
The more peaked the distribution is, the stronger the boost effects will be.

\subsubsection{Invariant-mass distribution}

\begin{figure}
  \unitlength 1cm
  \begin{center}
  \begin{picture}(13.4,8)
  \put(-0.3,6.4){\makebox[0pt][c]{\boldmath $\frac{\displaystyle d\sigma}
                                 {\displaystyle dM_1^2}$}}
  \put(-0.3,5.2){\makebox[0pt][c]{\bf [fb]}}
  \put(13.7,-0.2){\makebox[0pt][c]{\bf\boldmath $M_1$ [GeV]}}
  \put(-1,-6){\includegraphics{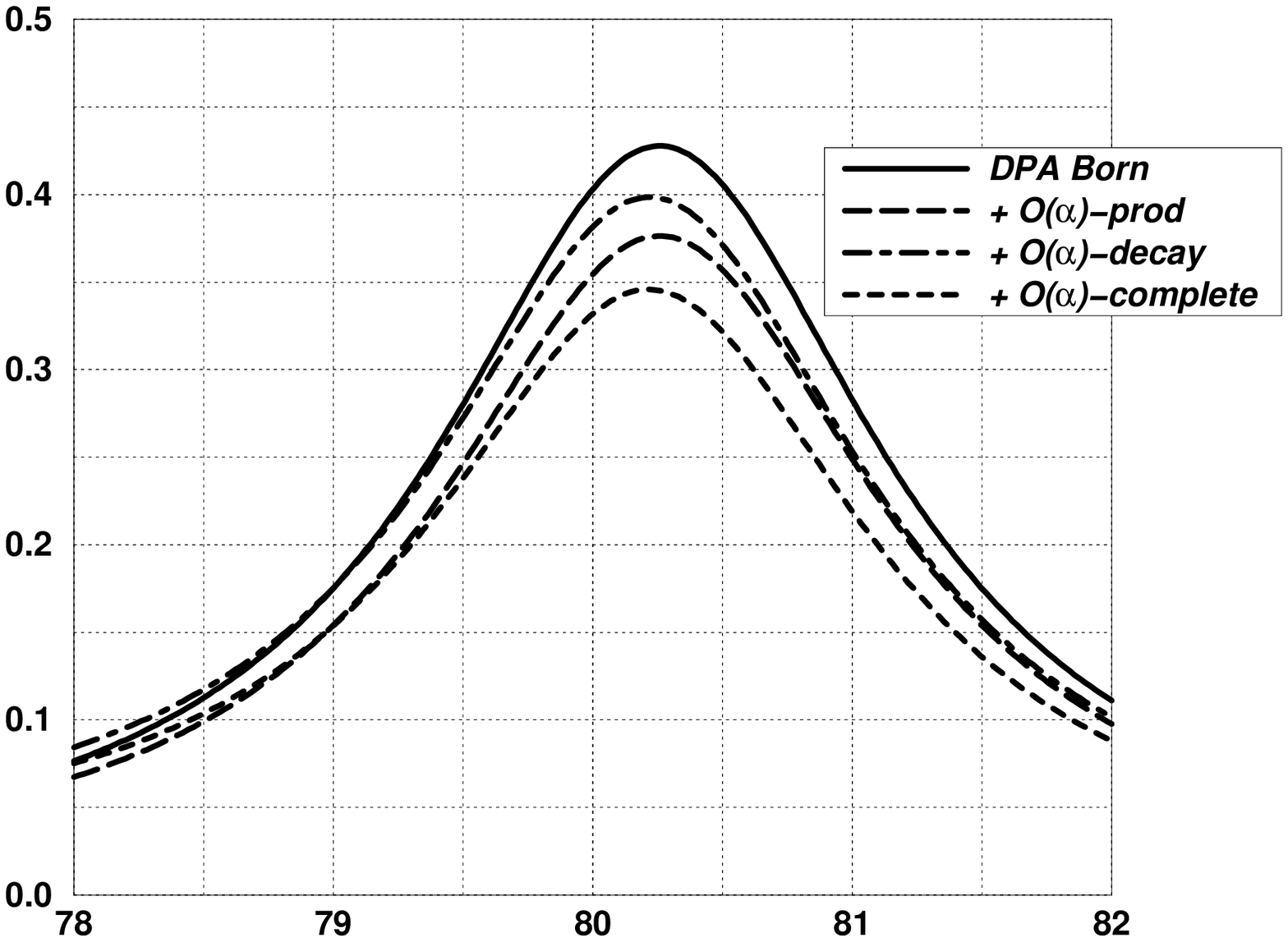}}
  \end{picture}
  \end{center}
  \caption[]{The invariant-mass distribution $d\sigma/dM_1^2$ for the
             $\mu^+\nu_{\mu} \tau^-\bar{\nu}_{\tau}$ final state
             at $2E=184\GeV$.} 
\label{fig:mass}
\end{figure}%
\begin{figure}
  \unitlength 1cm
  \begin{center}
  \begin{picture}(13.4,8)
  \put(-0.2,6.7){\makebox[0pt][c]{\boldmath $\delta$}}
  \put(-0.2,5.7){\makebox[0pt][c]{\bf [\%]}}
  \put(13.7,-0.2){\makebox[0pt][c]{\bf\boldmath $M_1$ [GeV]}}
  \put(-1,-6){\includegraphics{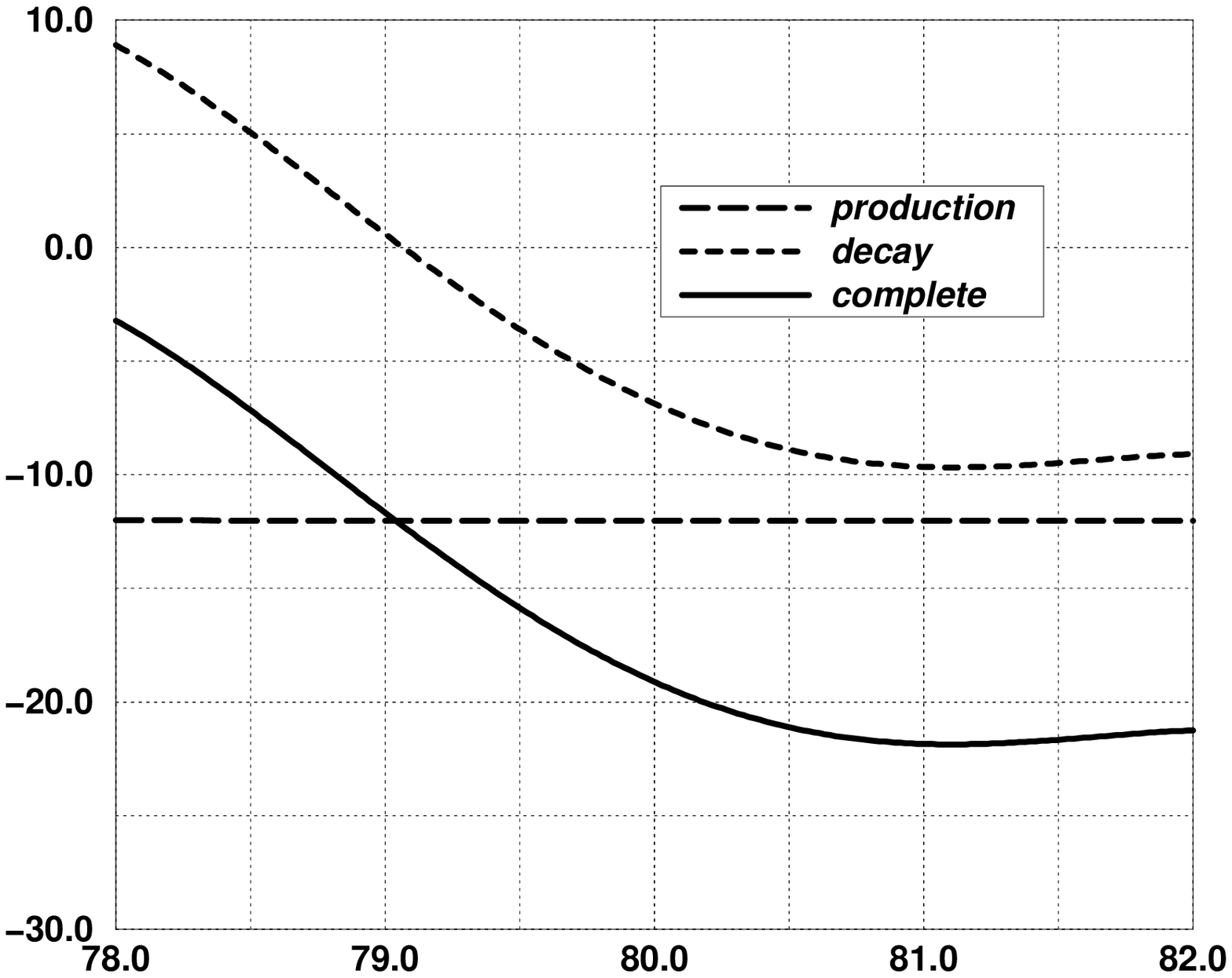}}
  \end{picture}
  \end{center}
  \caption[]{The relative correction factors $\delta$ corresponding to
             Fig.~\protect\ref{fig:mass}.}
\label{fig:mass/rel}
\end{figure}%
In Fig.~\ref{fig:mass} we compare the Breit--Wigner line shape 
$d\sigma/dM_1^2$ with and without RC for the 
$\mu^+\nu_{\mu} \tau^-\bar{\nu}_{\tau}$ final state at $2E=184\GeV$. 
Since the corrected line shape receives completely different contributions 
from the production and decay stages, these parts are displayed separately. 
The corresponding relative correction factors are presented in 
Fig.~\ref{fig:mass/rel}. 

\begin{figure}
  \unitlength 1cm
  \begin{center}
  \begin{picture}(13.4,8)
  \put(-0.3,6.3){\makebox[0pt][c]{\boldmath $\frac{\displaystyle d\sigma}
                                 {\displaystyle dM_1^2}$}}
  \put(-0.3,5.1){\makebox[0pt][c]{\bf [fb]}}
  \put(13.7,-0.2){\makebox[0pt][c]{\bf\boldmath $M_1$ [GeV]}}
  \put(-1,-6){\includegraphics{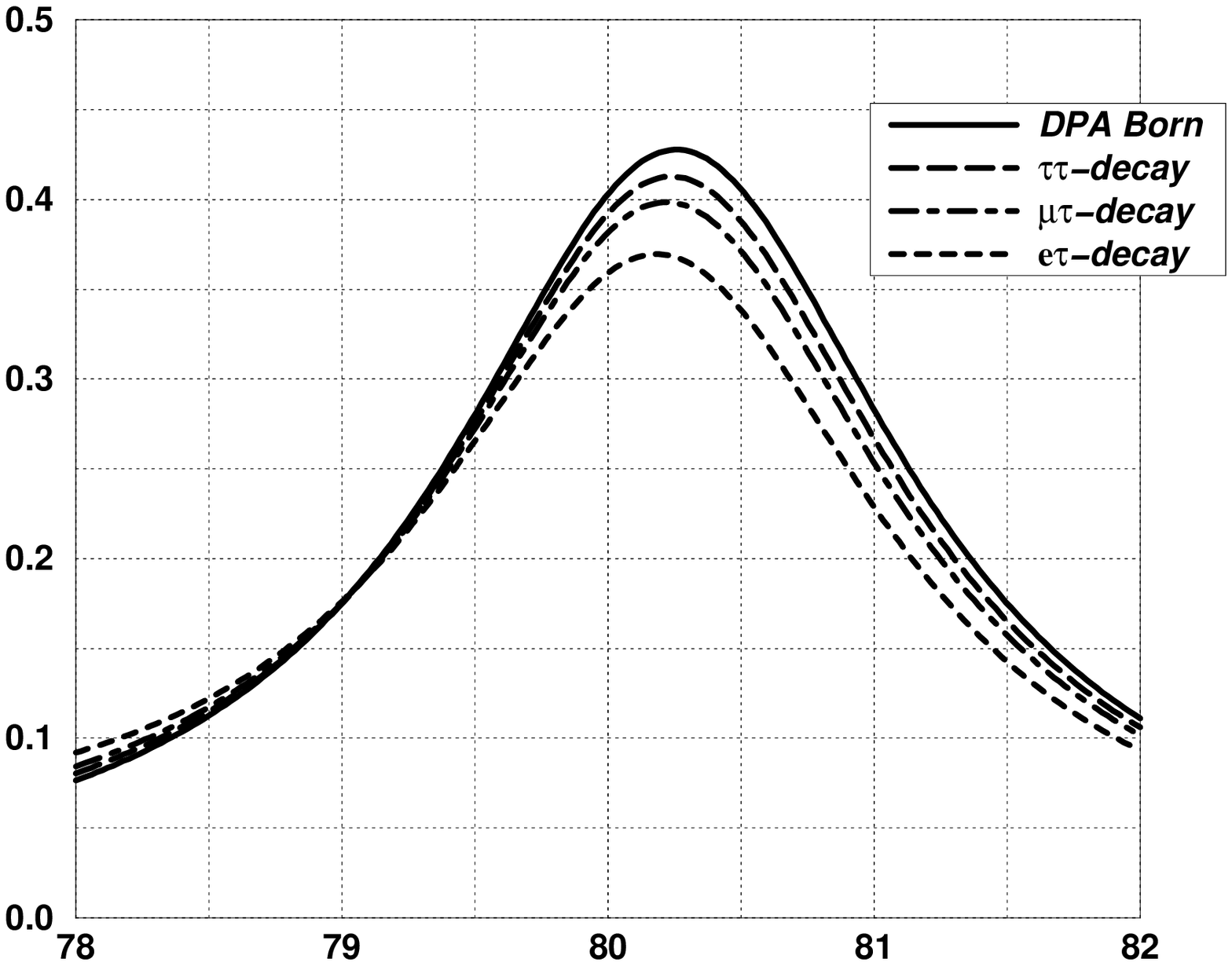}}
  \end{picture}
  \end{center}
  \caption[]{Distortion of the invariant-mass distribution $d\sigma/dM_1^2$
             at $2E=184\GeV$ due to the decay corrections. Three leptonic
             final states are considered:
             $\tau^+\nu_{\tau} \tau^-\bar{\nu}_{\tau}$ ($\tau\tau$-decay),
             $\mu^+\nu_{\mu} \tau^-\bar{\nu}_{\tau}$ ($\mu\tau$-decay), and
             $e^+\nu_{e} \tau^-\bar{\nu}_{\tau}$ ($e\tau$-decay).}
\label{fig:mass/xxx}
\end{figure}

\begin{figure}
  \unitlength 1cm
  \begin{center}
  \begin{picture}(13.4,8)
  \put(-0.2,6.9){\makebox[0pt][c]{\boldmath $\delta$}}
  \put(-0.2,5.9){\makebox[0pt][c]{\bf [\%]}}
  \put(13.7,-0.2){\makebox[0pt][c]{\bf\boldmath $M_1$ [GeV]}}
  \put(-1,-6){\includegraphics{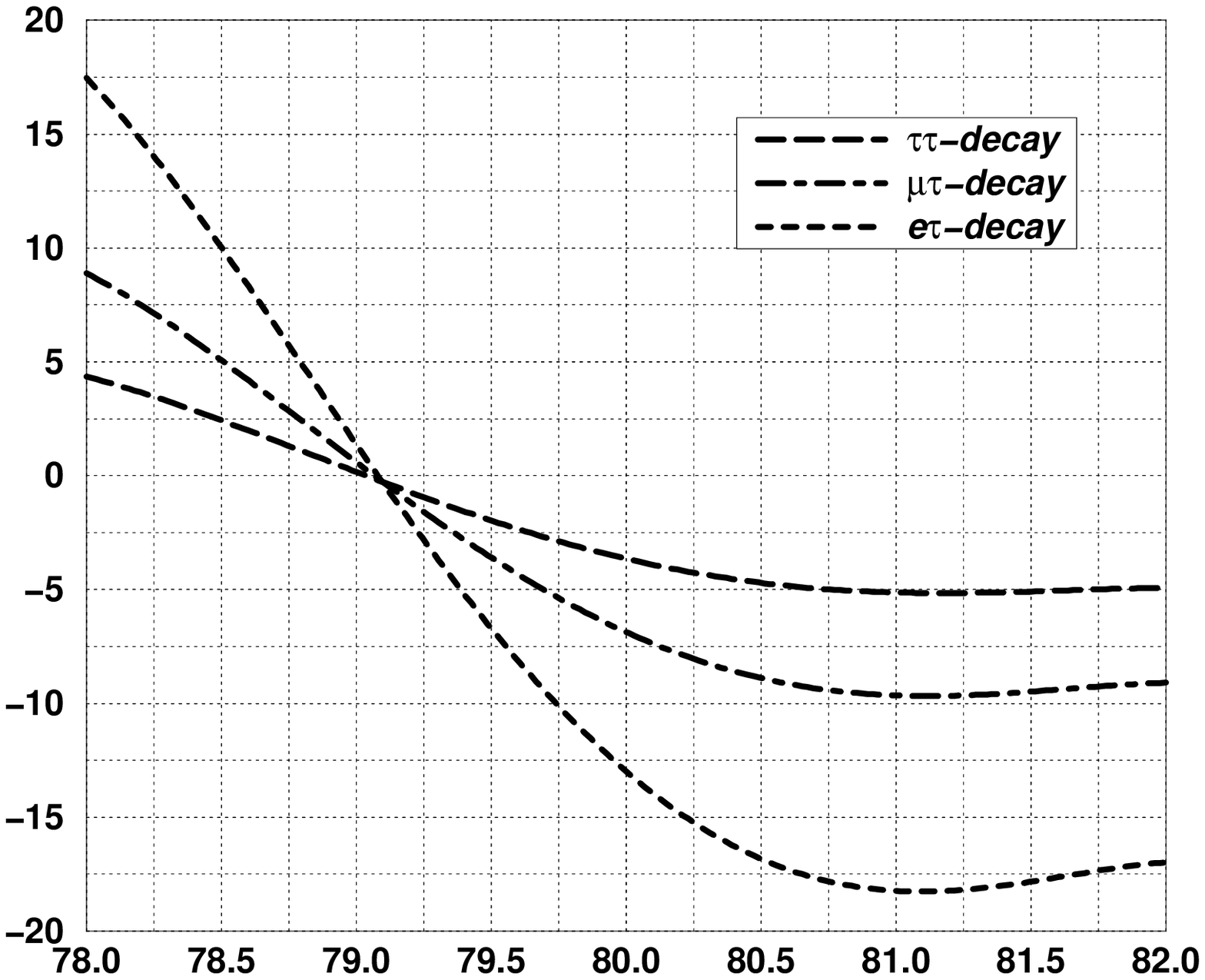}}
  \end{picture}
  \end{center}
  \caption[]{The relative correction factors $\delta$ corresponding to
             Fig.~\protect\ref{fig:mass/xxx}}
\label{fig:mass/xxx/rel}
\end{figure}

The correction to the production stage leads to a constant reduction of the 
Breit--Wigner line shape. The corrections
to the decay stages comprise the factorizable decay corrections  
(columns 3 and 4 in Table ~\ref{tab:2}) and the non-factorizable
corrections (column 5 in Table ~\ref{tab:2}). The latter are very 
small~\cite{nf-corr/bbc,nf-corr/ddr}. We see that also the decay corrections
reduce the Breit--Wigner line shape. The amount of reduction
depends on the particular final state, as can be seen from 
Figs.~\ref{fig:mass/xxx} and \ref{fig:mass/xxx/rel}, where we consider
different leptonic final states.
The differences are caused by the explicit 
fermion-mass dependence for collinear photon radiation 
(see Apps.~\ref{app:collrad} and \ref{app:semi-soft_functions}). For other
final states, involving quarks, the reduction is determined by the masses
and charges of the decay products of the $W^+$ boson 
(see App.~\ref{app:semi-soft_functions}). Besides the reduction, the decay 
corrections also distort the resonance shape, as is clear from 
Fig.~\ref{fig:mass/rel}. This final-state-dependent FSR distortion effect has 
recently been discussed in the literature \cite{fsr}. 
It can be quantified in terms of the shift in the peak position of the
Breit--Wigner line shape, which we find to be $-20$, $-39$, and $-77\MeV$
for $W^+$-boson decays into $\tau^{+}\nu_{\tau}$, $\mu^{+}\nu_{\mu}$,
and $e^{+}\nu_e$, respectively. This is in agreement with the (leading) shifts 
predicted by the $W$-boson version of Eq.~(19) in Ref.~\cite{fsr}, taking into 
account the fact that the correction to the production stage reduces the line 
shape by $12.0\%$. The observed shifts differ by $-5\MeV$ to $-10\MeV$ from 
these predictions as a result of the non-leading terms that are present in our
full DPA calculation.  

It should be stressed that theoretically the large distortion is a genuine 
effect. It would also be as relevant in practice if the invariant mass of 
the fermions could be measured. For various reasons this is problematic. 
One reason is the almost unavoidable inclusion of a collinear photon in the 
measured invariant mass. Such an inclusion would effectively decrease the 
leading logarithm [$\log(M_W^2/m_f^2)$] that dominates the effect. 
What remains of the distortion in practice should be studied with an event 
generator, which simulates the actual measurement.

\subsubsection{Decay-angle distribution}

\begin{figure}
  \unitlength 1cm
  \begin{center}
  \begin{picture}(13.4,8)
  \put(-0.3,6.9){\makebox[0pt][c]{\boldmath $\frac{\displaystyle d\sigma}
                                 {\displaystyle d\cos\theta_3}$}}
  \put(-0.3,5.7){\makebox[0pt][c]{\bf [pb]}}
  \put(13.5,-0.2){\makebox[0pt][c]{\boldmath $\cos\theta_3$}}
  \put(-1,-6){\includegraphics{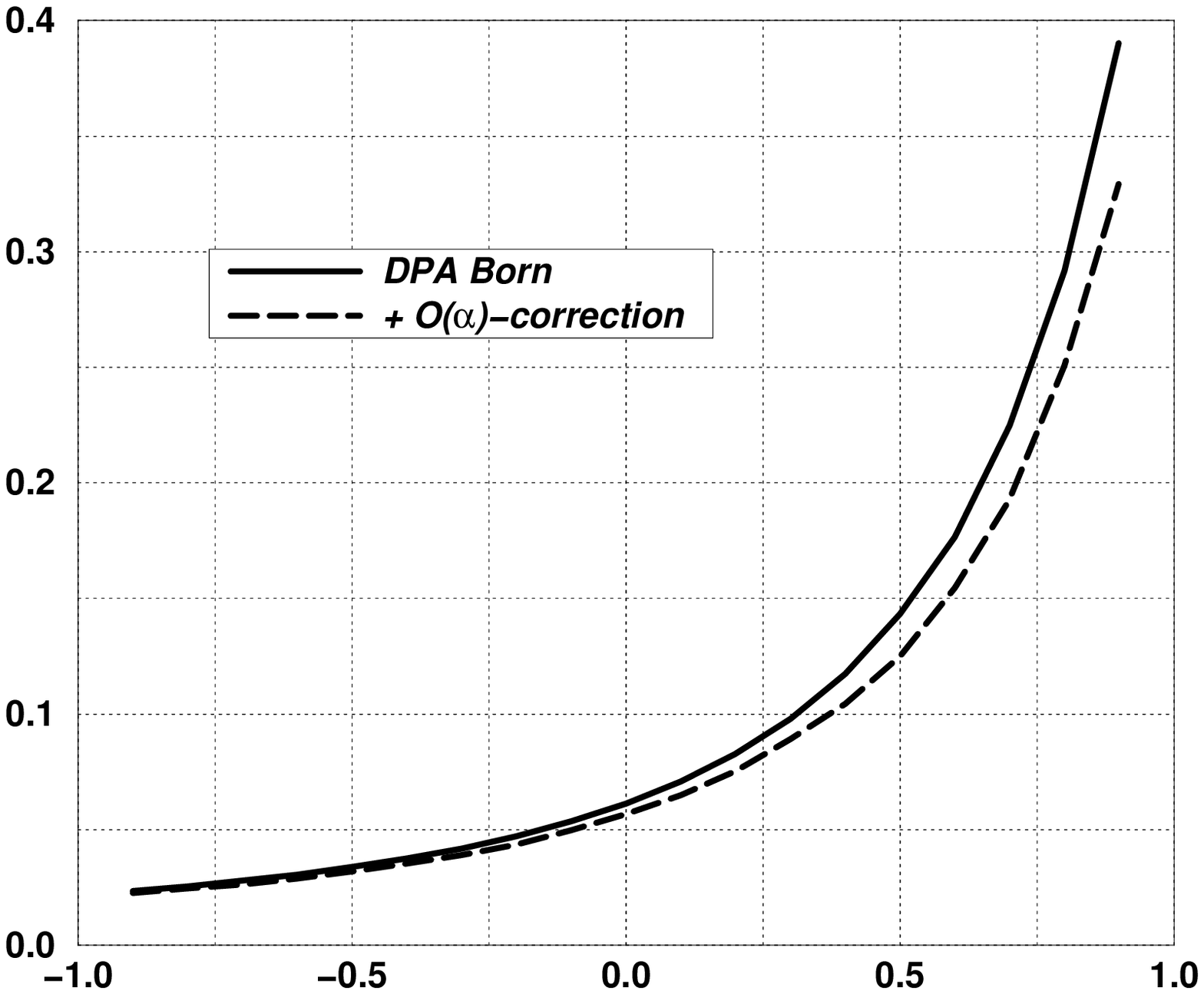}}
  \end{picture}
  \end{center}
  \caption[]{The decay-angle distribution $\,d\sigma/d\cos\theta_3$ for the 
             $\mu^+\nu_{\mu} \tau^-\bar{\nu}_{\tau}$ final state at 
             $2E=184\GeV$. Here $\theta_3 = \angle(\mu^+,W^+)$ in the LAB
             frame.}
\label{fig:dec}
\end{figure}%
\begin{figure}
  \unitlength 1cm
  \begin{center}
  \begin{picture}(13.4,8)
  \put(-0.2,6.2){\makebox[0pt][c]{\boldmath $\delta$}}
  \put(-0.2,5.2){\makebox[0pt][c]{\bf [\%]}}
  \put(13.5,-0.2){\makebox[0pt][c]{\boldmath $\cos\theta_3$}}
  \put(-1,-6){\includegraphics{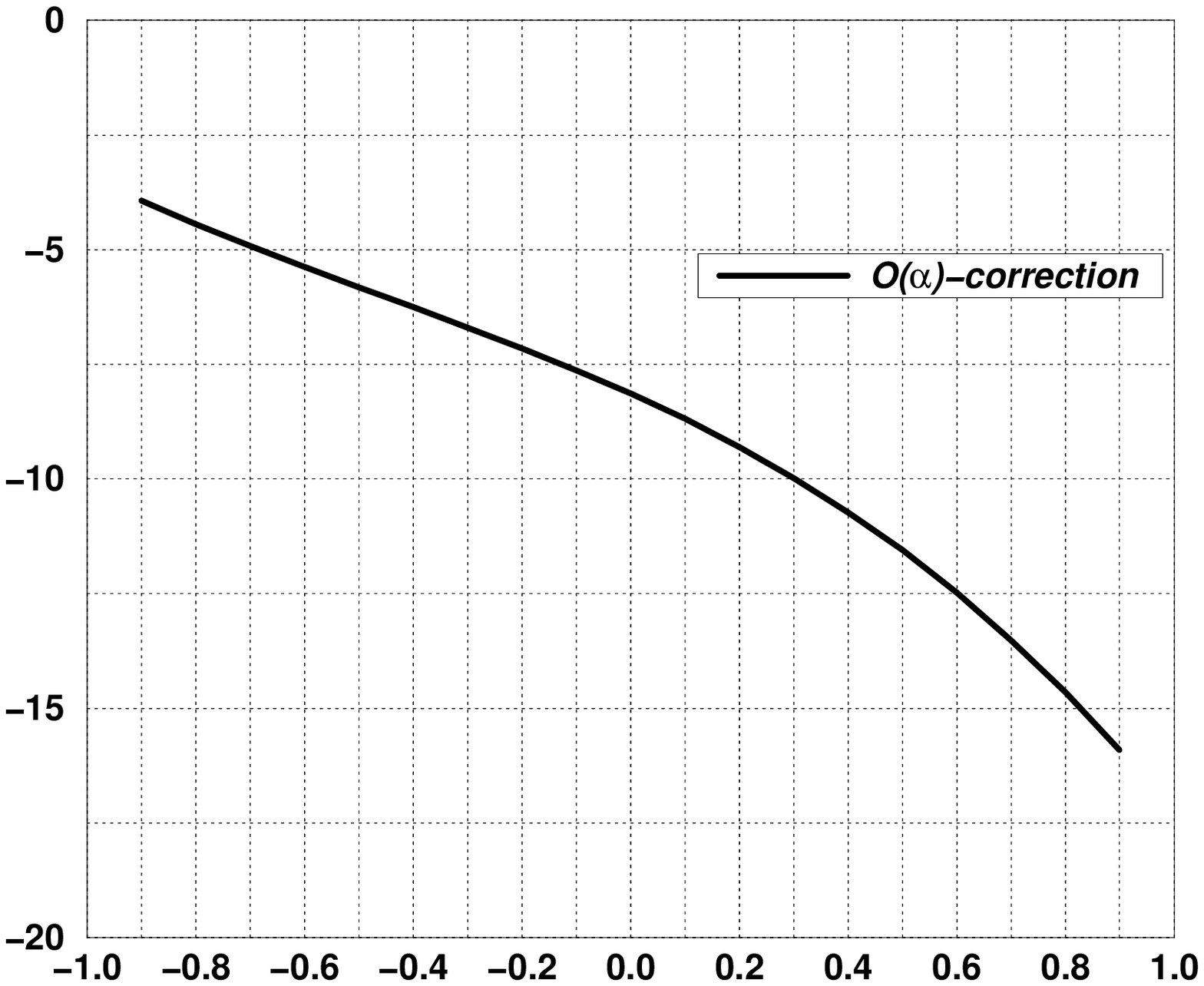}}
  \end{picture}
  \end{center}
  \caption[]{The relative correction factor $\delta$ corresponding to
             Fig.~\protect\ref{fig:dec}.}
\label{fig:dec/rel}
\end{figure}%
In Figs.~\ref{fig:dec} and \ref{fig:dec/rel} the results are shown for the
decay-angle distribution $\,d\sigma/d\cos\theta_3$ for the 
$\mu^+\nu_{\mu} \tau^-\bar{\nu}_{\tau}$ final state at $2E=184\GeV$. 
Here $\theta_3$ is the angle between the $\mu^+$ and the $W^+$ boson in the 
LAB frame [see Eq.~(\ref{pole/kinematics})]. The correction is negative and 
becomes more negative for forward angles. The shape of the relative correction 
factor is again the result of hard-photon boost effects.

\subsection{Double invariant-mass distribution}

\begin{figure}
  \unitlength 1cm
  \begin{center}
  \begin{picture}(13.4,8)
  \put(0,3){\makebox[0pt][c]{\bf\boldmath $\delta$ [\%]}}
  \put(12.5,0.5){\makebox[0pt][c]{\boldmath $M_{2}$ $[\GeV]$}}
  \put(5,0){\makebox[0pt][c]{\boldmath $M_{1}$ $[\GeV]$}}
  \put(1,-2){\includegraphics{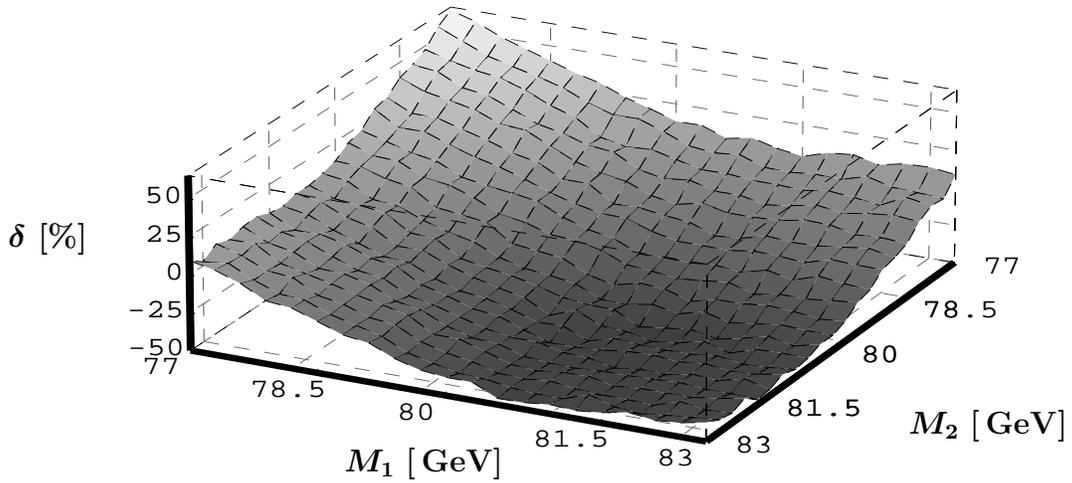}}
  \end{picture}
  \end{center}
  \caption[]{Correction to the double invariant-mass distribution
             $\,d\sigma/dM_1^2 dM_2^2\,$ for the $e^+\nu_{e} e^-\bar{\nu}_{e}$
             final state at $2E=184\GeV$.}
\label{fig:mm/rel/ee}
\end{figure}%
\begin{figure}
  \unitlength 1cm
  \begin{center}
  \begin{picture}(13.4,8)
  \put(0,3){\makebox[0pt][c]{\bf\boldmath $\delta$ [\%]}}
  \put(12.5,0.5){\makebox[0pt][c]{\boldmath $M_{2}$ $[\GeV]$}}
  \put(5,0){\makebox[0pt][c]{\boldmath $M_{1}$ $[\GeV]$}}
  \put(1,-2){\includegraphics{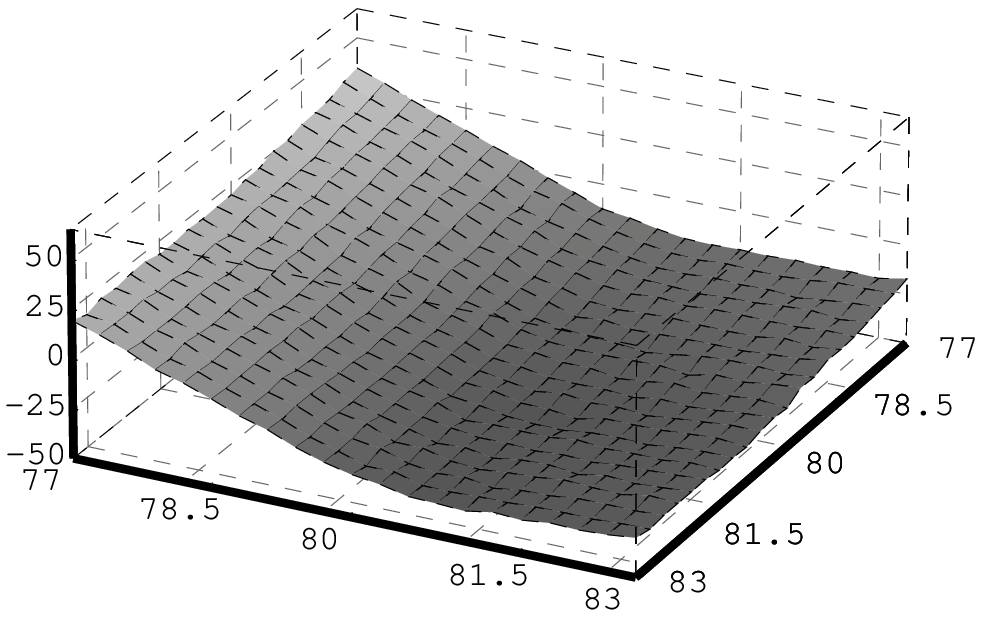}}
  \end{picture}
  \end{center}
  \caption[]{Correction to the double invariant-mass distribution
             $\,d\sigma/dM_1^2 dM_2^2\,$ for the 
             $e^+\nu_{e} \tau^-\bar{\nu}_{\tau}$ final state at $2E=184\GeV$.}
\label{fig:mm/rel/et}
\end{figure}%
Finally we consider the two-dimensional distribution 
$\,d\sigma/dM_1^2 dM_2^2\,$ at $2E=184\GeV$, evaluated
using the contributions specified in Table~\ref{tab:1}. Instead of plotting
the absolute distributions, only the relative correction factors 
$\delta$ are presented. We do this for two specific leptonic final states. 
In Fig.~\ref{fig:mm/rel/ee} the $e^+\nu_{e} e^-\bar{\nu}_{e}$ final state
is considered. This final state has the largest correction. Keeping
one $M_i^2$ fixed, the correction to the other distribution is
qualitatively the same as the correction to the one-dimensional
distribution in Fig.~\ref{fig:mass/xxx/rel}. 
For the $e^+\nu_{e} \tau^-\bar{\nu}_{\tau}$ final state 
(Fig.~\ref{fig:mm/rel/et}) the correction is clearly not symmetric in the 
$M_{i}^{2}$. This was to be expected, since the decay corrections for 
$e^{+}\nu_{e}$ and $\tau^{-}\bar{\nu}_{\tau}$ differ appreciably.

\begin{table}
\[
  \begin{array}{||c||c|c|c||}    \hline \hline
    \raisebox{-3mm}{$\Delta_1$} 
    & \multicolumn{3}{|c||}{\raisebox{-1mm}{decay channel}}\\ 
    \cline{2-4}
         & e^{+}\nu_e & \mu^{+}\nu_{\mu}  & \tau^{+}\nu_{\tau}   \\ 
    \hline  \hline   
    -1/2 & -1.4       & -0.8              & -0.5  \\
     0   & -15.0      & -7.8              & -4.0  \\
     1/2 & -17.3      & -9.0              & -4.6  \\
    \hline \hline 
    \multicolumn{4}{c}{}\\
    \multicolumn{4}{c}{\delta^{+}_{\sss{dec}}(M_1)}
  \end{array} 
  \quad\quad 
   \begin{array}{||c||c|c|c||}    \hline \hline
    \raisebox{-3mm}{$\Delta_1$} 
    & \multicolumn{3}{|c||}{\raisebox{-1mm}{$\Delta_2$}}\\ 
    \cline{2-4}
         & -1/2  &  0    & 1/2   \\ 
    \hline  \hline   
    -1/2 & +0.5  & +0.2  & -0.1  \\
     0   & +0.2  & +0.0  & -0.2  \\
     1/2 & -0.1  & -0.2  & -0.4  \\
    \hline \hline 
    \multicolumn{4}{c}{}\\
    \multicolumn{4}{c}{\delta_{\sss{nf}}(M_1,M_2)}
  \end{array}
\]
\caption[]{Relative correction factors [in \%] for the double invariant-mass 
           distribution $\,d\sigma/dM_1^2 dM_2^2\,$ at $2E=184\GeV$. 
           Left: the corrections from the $W^+$-boson decay stage 
           $\delta_{dec}^{+}(M_1)$ for different leptonic decay channels 
           and three near-resonant invariant masses.
           Right: the non-factorizable corrections 
           $\delta_{\sss{nf}}(M_1,M_2)$.}
\label{tab:3}
\end{table}%
\begin{table}
\[
  \begin{array}{||c||c|c|c||}    \hline \hline
    \raisebox{-3mm}{$\Delta_1$} 
    & \multicolumn{3}{|c||}{\raisebox{-1mm}{$\Delta_2$}}\\ 
    \cline{2-4}
         & -1/2   &  0     & 1/2    \\ 
    \hline  \hline   
    -1/2 & -14.3  & -28.2  & -30.8  \\
     0   & -28.2  & -42.0  & -44.5  \\
     1/2 & -30.8  & -44.5  & -47.0  \\
    \hline \hline 
    \multicolumn{4}{c}{}\\
    \multicolumn{4}{c}{e^+\nu_{e} e^-\bar{\nu}_{e}\ \mbox{final state}}
  \end{array}
  \quad\quad
  \begin{array}{||c||c|c|c||}    \hline \hline
    \raisebox{-3mm}{$\Delta_1$} 
    & \multicolumn{3}{|c||}{\raisebox{-1mm}{$\Delta_2$}}\\ 
    \cline{2-4}
         & -1/2   &  0     & 1/2    \\ 
    \hline  \hline   
    -1/2 & -13.4  & -17.2  & -18.1  \\
     0   & -27.3  & -31.0  & -31.8  \\
     1/2 & -29.9  & -33.5  & -34.3  \\
    \hline \hline 
    \multicolumn{4}{c}{}\\
    \multicolumn{4}{c}{e^+\nu_{e} \tau^-\bar{\nu}_{\tau}\ \mbox{final state}}
  \end{array}
\]
\caption[]{Relative correction factors [in \%] for the double invariant-mass 
           distribution $\,d\sigma/dM_1^2 dM_2^2\,$ at $2E=184\GeV$. 
           Left: the $e^+\nu_{e} e^-\bar{\nu}_{e}$ final state.
           Right: the $e^+\nu_{e} \tau^-\bar{\nu}_{\tau}$ final state.}

\label{tab:4}
\end{table}
In Tables~\ref{tab:3} and \ref{tab:4} we present some explicit values for
the relative correction factor, split up into the separate contributions 
according to 
$$
  \delta_{\sss DPA}(M_1,M_2)
  =
  \delta_{\sss{pr}} + \delta_{\sss{dec}}^{+}(M_1)
  + \delta_{\sss{dec}}^{-}(M_2) + \delta_{\sss{nf}}(M_1,M_2).
$$
The correction from the production stage is constant as a function of the
invariant masses, $\delta_{\sss{pr}}=-12.0\%$. The non-factorizable 
contribution $\delta_{\sss{nf}}(M_1,M_2)$ is given in Table~\ref{tab:3} 
for three near-resonant values for the invariant masses $M_i$. We indicate
these three values by $\Delta_i = (M_i-M_W)/\Gamma_W = -1/2,\,0,\,1/2$.
The non-factorizable corrections do not depend on the particular leptonic
final state. The corrections $\delta_{\sss{dec}}^{\pm}(M_i)$ from the decay 
stages do depend on the particular leptonic final
state as explained before. In Table~\ref{tab:3} we present these 
corrections for the three leptonic decay modes and the three near-resonant 
values for the invariant mass.


\section{Discussion and conclusions}
\label{sec:concl}

In this paper $\OO(\alpha)$ radiative corrections (RC) to 
four-fermion production in $e^{+}e^{-}$ collisions have been studied.
The energy region chosen is that of LEP2, where the four-fermion
final state is predominantly formed through intermediate $W$-pair production. 

Since a complete $\OO(\alpha)$ RC calculation for a two-particle
to four-particle process is beyond present possibilities, a sensible 
approximation scheme has to be used. The smallness of $\Gamma_W/M_W$
offers the possibility to use the double-pole approximation (DPA).
In practice it means that we calculate $\OO(\alpha)$ and 
$\OO(\Gamma_W/M_W)$ corrections but neglect $\OO(\alpha \Gamma_W/M_W)$
corrections. Although this approximation has been advocated in the literature
for quite some time, so far no complete $\OO(\alpha)$ results have been given.
As far as we know, the study that has come closest to achieving a complete
DPA calculation for an actual process involves the factorizable QCD corrections
to the process $e^+e^- \to t\bar{t}$~\cite{Schmidt}. Many of the issues
discussed in the present paper have, however, not been addressed in 
Ref.~\cite{Schmidt}.

We have applied the method to $W$-pair-mediated four-fermion production
for a number of reasons. There is the methodological aspect of dealing
with unstable particles in DPA, involving issues like gauge invariance, 
interactions between different stages of the reaction, RC to density matrices,
and the phase-space mappings. All of these issues play a role in the 
$W$-pair-mediated four-fermion production process. There is also the practical 
aspect of assessing how large $\OO(\alpha)$ corrections can be for certain 
distributions. This is of importance for $W$-pair studies at LEP2.

On the methodological side, we have succeeded in finding a consistent 
prescription for applying the idea of the DPA. The kinematics for calculating 
the poles of the matrix elements is necessarily on-shell kinematics.
Also the phase-space factor in the cross-section is treated in on-shell 
kinematics. The off-shell invariant masses occur only in the Breit--Wigner 
factors. All of this is well defined both for the radiative and non-radiative
parts of the cross-section and therefore our calculation itself is 
unambiguous.

An unavoidable problem is the relation between off-shell four-fermion
events and the on-shell calculated events. This question arises when one 
likes to connect experimental cross-sections to the calculated DPA 
cross-sections. Also here recipes are chosen both for the radiative 
and non-radiative phase-space points. For the latter the mapping 
is natural if one chooses the invariant masses $M_{1,2}$
and angles as variables. All off-shell points can be mapped onto
on-shell points. For the radiative events in an off-shell phase space, 
the photon variables have to be added. A natural choice is the photon 
energy and angles. If the mapping is chosen such that the photon
variables remain the same, one sometimes maps outside the on-shell 
phase space. Different remedies for this problem are possible. One can choose 
a procedure that assigns zero cross-sections to those points. 
On the other hand, the photon energy in the off-shell 
four-fermion\,--\,one-photon phase space may be rescaled in order to obtain a 
physical point in the on-shell phase space. In general, there will be a 
dependence on the chosen mapping between off-shell and on-shell phase spaces. 
However, the induced numerical differences remain within the accuracy of the 
calculation, i.e.~$\OO(\alpha\Gamma_W/M_W)$. At high energies, say above 
$2\TeV$, when peaks in the cross-section become much more pronounced,
a more sizeable implementation dependence may occur. The present 
calculation is primarily meant for LEP2 energies and slightly above,
say up to $500\GeV$.

On the practical side, the results give an insight in the size of RC
for off-shell $W$-pair production. Within the claimed accuracy
it is a complete calculation. It should be stated that imaginary parts
of loop diagrams have been neglected in the expectation that they only 
induce small effects.  Moreover, such terms are characterized by an explicit 
$\sin\phi_{3}$ and/or $\sin\phi_{4}$ dependence in the cross-section, since 
they select the antisymmetric parts of the $\D$-matrices. Integration over 
these azimuthal angles removes the imaginary parts of loop diagrams from 
the cross-section. So they do not contribute to the distributions of 
Sect.~\ref{sec:plots}. It should also be stated that some large corrections
(ISR, FSR), which usually require the inclusion of higher-order terms, are 
considered purely in first order here. The corresponding higher-order terms 
can be determined in a straightforward way within our approach, using the 
usual exponentiation/factorization techniques. 

We have presented the results for leptonic final states. The reason is that 
those states are theoretically well defined. In exactly the same way also 
quark final states can be treated as long as one assumes certain masses for 
the quarks. It is this mass assignment which gives some arbitrariness
in the actual numbers. The RC presented are corrections to ideal
theoretical situations, which cannot be realized experimentally
in the same way. For that purpose the calculation should be implemented 
in an event generator. In principle this is possible. Events can be
generated in the on-shell phase space with a radiatively corrected weight.
The outside Breit--Wigner distributions can then generate the invariant
masses and consequently off-shell events could be constructed from the 
on-shell ones (with certain angles kept fixed).
Event generators offer the possibility to include realistic experimental 
cuts and therefore to study effects like the line-shape deformation in 
practical cases. The actual numbers presented here should give the reader an 
indication of the size of RC in ideal cases, of which remnants survive in 
practical situations.

For some questions the present study could already be useful in its present  
form. An example of this would be the comparison of a DPA Born cross-section
with $CP$-conserving anomalous triple gauge-boson couplings and a normal DPA 
cross-section with RC. The question of how RC mimic anomalous couplings could 
be studied in this way, but goes beyond the size and scope of the present 
paper.

In conclusion, the DPA method for unstable-particle production has been 
shown to give reasonable results. It could also be applied to other 
unstable-particle production processes that undergo electroweak or
QCD radiative corrections.

\newpage


\appendix


\section{Helicity amplitudes for the virtual factorizable corrections}
\label{app:virt}

In this appendix we give the basic ingredients for the calculation of the 
virtual factorizable corrections. The one-loop RC to on-shell $W$-pair 
production have been carried out in the literature in terms of helicity 
amplitudes with a particular choice for the decomposition into basic matrix 
elements and invariant functions~\cite{prod-corr}. This calculation serves 
as our basis for obtaining the RC to the production density matrix 
$\PP_{[\lambda_{1}\lambda_{2}][\lambda_{1}'\lambda_{2}']}(M_W,M_W)$,
defined in Eq.~(\ref{pole/prod}). Therefore we will set up our
conventions in close analogy to what has been used in the numerical
routines of Ref.~\cite{prod-corr}. Once we have fixed the choice of
polarization basis in the production stage, the same choice should be
applied to the decay stages as well, i.e~to $\D_{\lambda_i\lambda_i'}(M_W)$
in Eq.~(\ref{pole/decay}).

\subsection{Virtual corrections to the production stage}
\label{app:virt/prod}

The amplitude $\Pi_{\lambda_{1}\lambda_{2}}(M_W,M_W)$ for on-shell $W$-pair 
production depends on the spinors of the initial-state $e^{\pm}$ and
on the polarization vectors $\e_{i}^{\mu}(p_i,\lambda_i)$ 
of the $W$ bosons. In order to define the latter we first introduce 
\begin{eqnarray}
  \e_{1,2}^{\mu}(p_{1,2},||) 
  &=& \frac{q_{1,2}^{\mu}(M_{W}^{2}+u) - q_{2,1}^{\mu}(M_{W}^{2}+t)
            + p_{1,2}^{\mu}(t-u)}{\sqrt{ut-M_{W}^{4}}\,\sqrt{s-4M_{W}^{2}}},
      \nonumber \\
  \e_{i}^{\mu}(p_{i},\perp) 
  &=& {}- 2\,\frac{\epsilon^{\mu\nu\rho\tau} q_{2\nu} q_{1\rho}
               p_{i\tau}}{\sqrt{s(ut-M_{W}^{4})}},
      \nonumber \\
  \e_{i}^{\mu}(p_{i},L) 
  &=& \frac{s p_{i}^{\mu}-2M_{W}^{2}(q_{1}+q_{2})^{\mu}}
           {M_{W}\sqrt{s(s-4M_{W}^{2})}},
\end{eqnarray}
using the conventions defined in Sect.~\ref{sec:pole-scheme} and 
$\epsilon^{0123}=-1$. The helicity states, defined in the LAB frame,
can be expressed in terms of the energy $E$ and velocity 
$\beta = p/E = \sqrt{1-M_W^2/E^2}$ of the $W$ bosons in the following way:
\begin{eqnarray}
\label{born/helicity}
  \e_1^{\mu}(p_1,\pm 1)
  &=& \frac{1}{\sqrt{2}}\biggl[ \e_1^{\mu}(p_1,||) \pm i\e_1^{\mu}(p_1,\perp)
                        \biggr] 
      \ = \ \frac{1}{\sqrt{2}}\,(0,-1,\pm i,0),
      \nonumber \\
  \e_1^{\mu}(p_1,0) 
  &=& \e_1^{\mu}(p_1,L) \ = \ \frac{E}{M_W}\,(\beta,0,0,1),
      \nonumber \\
  \e_2^{\mu}(p_2,\pm 1)
  &=& \frac{1}{\sqrt{2}}\biggl[ \e_2^{\mu}(p_2,||) \mp i\e_2^{\mu}(p_2,\perp)
                        \biggr] 
      \ = \ \frac{1}{\sqrt{2}}\,(0,1,\pm i,0),
      \nonumber \\
  \e_2^{\mu}(p_2,0) 
  &=& \e_2^{\mu}(p_2,L) \ = \ \frac{E}{M_W}\,(\beta,0,0,-1).
\end{eqnarray}
This polarization basis satisfies the usual identities
\be
  p_{i}^{\mu}\,\e_{i,\,\mu}(p_{i},\lambda_{i}) = 0, \quad\quad
  \e_{i}^{\mu}(p_{i},\lambda_{i})\,\e_{i,\,\mu}^{*}(p_{i},\lambda_{i}') 
  =
  -\,\delta_{\lambda_{i}\lambda_{i}'}.
\ee

In its most general form, the amplitude $\Pi_{\lambda_1\lambda_2}(M_W,M_W)$ 
can be written as a sum of invariant functions $\C_{j}(\sigma)$ multiplied 
by Lorentz-invariant basic matrix elements 
$\M_{j}^{\sigma}(\lambda_{1},\lambda_{2})$. The basic matrix elements
are simple, purely kinematical objects and contain the complete
dependence on the $W$-boson polarizations. The invariant functions
contain the dynamical information, i.e.~details of the model, but are
independent of the $W$-boson polarizations. Both parts depend on the
helicity of the electron, $\lambda_{e^-}\!\equiv \sigma/2$ ($\sigma = \pm 1$).
In view of our massless treatment of the $e^{\pm}$, the helicity of
the positron is fixed to $\lambda_{e^+}\! = -\lambda_{e^-}$ in the
virtual corrections. For a specific helicity of the electron the 
decomposition of the helicity amplitudes reads
\be
  \Pi_{\sigma;\lambda_1\lambda_2}(M_W,M_W)
  =
  \sum_{j=1}^{9} \C_j(\sigma) \M_j^{\sigma}(\lambda_1,\lambda_2\bigr),
\ee
with
\begin{eqnarray}
  \M_{1}^{\sigma}(\lambda_1,\lambda_2)
  &=& \Bigl[ \bar{v}(q_{1}) \ps_{1} \omega_{\sigma} u(q_{2}) \Bigr]
      (\e_{1}\e_{2}),
      \nonumber \\[1mm]
  \M_{2}^{\sigma}(\lambda_1,\lambda_2)
  &=& \Bigl[ \bar{v}(q_{1}) \ps_{1} \omega_{\sigma} u(q_{2}) \Bigr]
      (p_{1} \e_{2})(p_{2} \e_{1}),
      \nonumber \\[1mm]
  \M_{3}^{\sigma}(\lambda_1,\lambda_2)
  &=& \bar{v}(q_{1}) 
      \biggl[ \es_{1} (p_{1} \e_{2}) - \es_{2} (p_{2} \e_{1}) \biggr] 
      \omega_{\sigma} u(q_{2}),
      \nonumber \\[1mm]
  \M_{4}^{\sigma}(\lambda_1,\lambda_2)
  &=& \bar{v}(q_{1})\es_{1} (\ps_{1} - \qs_{1}) \es_{2} 
      \omega_{\sigma} u(q_{2}),
      \nonumber \\[1mm]
  \M_{5}^{\sigma}(\lambda_1,\lambda_2)
  &=& \bar{v}(q_{1}) 
      \biggl[ \es_{1} (q_{1} \e_{2}) - \es_{2} (q_{2} \e_{1}) \biggr]
      \omega_{\sigma} u(q_{2}),
      \nonumber \\[1mm]
  \M_{6}^{\sigma}(\lambda_1,\lambda_2)
  &=& \Bigl[ \bar{v}(q_{1}) \ps_{1} \omega_{\sigma} u(q_{2}) \Bigr]
      (q_{1}\e_{2}) (q_{2} \e_{1}),
      \nonumber \\[1mm]
  \M_{7}^{\sigma}(\lambda_1,\lambda_2)
  &=& \Bigl[ \bar{v}(q_{1}) \ps_{1} \omega_{\sigma} u(q_{2}) \Bigr]
       \Bigl[ (p_{1}\e_{2})(q_{2} \e_{1}) + (p_{2}\e_{1})(q_{1} \e_{2}) \Bigr],
      \nonumber \\[1mm]
  \M_{8}^{\sigma}(\lambda_1,\lambda_2)
  &=& i \Bigl[ \bar{v}(q_{1}) \gamma^{\mu} \omega_{\sigma} u(q_{2}) \Bigr]
        \epsilon_{\mu\nu\rho\tau}\,\e_{2}^{\nu}\,\e_{1}^{\rho}\,
        (p_{1} - p_{2})^{\tau},
      \nonumber \\[1mm]
  \M_{9}^{\sigma}(\lambda_1,\lambda_2)
  &=& i \Bigl[ \bar{v}(q_{1}) \gamma^{\mu} \omega_{\sigma} u(q_{2}) \Bigr]
        \epsilon_{\mu\nu\rho\tau}\,p_{2}^{\nu}\,p_{1}^{\rho}\,
        \bigl[ \e_{1}^{\tau} (p_{1} \e_{2}) - \e_{2}^{\tau} (p_{2} \e_{1})
        \bigr].
\end{eqnarray}
The helicity projectors
\be
  \omega_{\sigma} = \frac{1 + \sigma \gamma_{5}}{2},
\ee
with $\gamma_5 = i \gamma^0 \gamma^1 \gamma^2 \gamma^3$, project on
right- and left-handed massless fermions. Note that our set of 18 basic
matrix elements $\M_j^{\sigma}$ is overcomplete. Because of the massless
treatment of the fermions, CP invariance implies the relation%
\footnote{We have fixed the overall phase of the matrix elements such
          that this relation holds. The density matrix will of course not be
          affected by this choice.}%
\be
  \M_j^{\sigma}(\lambda_1,\lambda_2) = \M_j^{\sigma}(-\lambda_2,-\lambda_1),
\ee
resulting in only 12 independent matrix elements. The last three pairs,
$\M_{7,8,9}^{\sigma}$, have been kept for convenience and can be
expressed in terms of the others according to 
\be 
  \M_7^{\sigma} = \frac{s}{4}\,\sigma\M_8^{\sigma} + \M_2^{\sigma}
                  + \frac{t-u}{4}\,\M_3^{\sigma}, \quad\quad
  \M_9^{\sigma} = -\,\frac{s}{2}\,\M_8^{\sigma}, \quad\quad
  \sigma\M_8^{\sigma} = 2\Bigl(\M_1^{\sigma} + \M_4^{\sigma} 
                               + \M_5^{\sigma} \Bigr) - 3\M_3^{\sigma}.
\ee

For the lowest-order matrix element only a few of these basic matrix 
elements are relevant: 
\be
  \Pi^0_{\sigma;\lambda_1\lambda_2}(M_W,M_W)
  =
  \delta_{(\sigma,-1)}\, G_{1}(t)\, \M_4^{\sigma}(\lambda_1,\lambda_2)
  + 2\, G_{2}(s,\sigma)\Bigl[ \M_3^{\sigma}(\lambda_1,\lambda_2)
                            - \M_1^{\sigma}(\lambda_1,\lambda_2)
                       \Bigr],
\ee
where the coefficients $G_{1,2}$ are defined as 
\be
\label{born/coupling}
  G_{1}(t) = -\, \frac{i e^{2}}{2 s_{W}^{2}}\,\frac{1}{t}, \quad\quad 
  G_{2}(s,\sigma) = i e^{2} \biggl( 
                        \frac{Q_e}{s} 
                      + \frac{c_{W}}{s_{W}}\,\frac{g_{e}(\sigma)}{s-M_{Z}^{2}}
                            \biggr).
\ee
Here $s_W$ and $c_W$ are the sine and cosine of the weak mixing angle 
$\theta_{W}$:
\be
  c_W = \cos\theta_W = \frac{M_W}{M_Z}, \quad\quad 
  s_W = \sin\theta_W = \sqrt{1-c_W^2},
\ee
and $g_e(\sigma)$ denotes the coupling between the $Z$ boson and electrons:
\be
     g_{e}(+1) = -\,\frac{s_{W}}{c_{W}}\,Q_{e}, \quad\quad
     g_{e}(-1) = -\,\frac{1 + 2 Q_{e}s_{W}^{2}}{2 s_{W} c_{W}},
\ee
where $Q_{e}=-1$ is the charge of the electron.

For the virtual corrections to on-shell $W$-pair production we need the 
complete list of basic matrix elements $\M_j^{\sigma}$. For the $W$-boson 
helicity states (\ref{born/helicity}) the non-vanishing matrix elements read:
\begin{eqnarray}
  \M_4^\sigma(\pm 1,\mp 1) &=& 2E^2(\cos\theta \mp \sigma)\sin\theta 
                               \nonumber \\
  \M_5^\sigma(\pm 1,\mp 1) &=& -\,2E^2(\cos\theta \mp \sigma)\sin\theta 
                               \nonumber \\
  \M_6^\sigma(\pm 1,\mp 1) &=& E^4\beta\sin^3\theta 
                               \\[2ex]
  \M_1^\sigma(\pm 1,\pm 1) &=& 2E^2\beta\sin\theta 
                               \nonumber \\
  \M_4^\sigma(\pm 1,\pm 1) &=& 2E^2(\cos\theta - \beta)\sin\theta
                               \nonumber \\ 
  \M_5^\sigma(\pm 1,\pm 1) &=& -\,2E^2\cos\theta\sin\theta 
                               \nonumber \\ 
  \M_6^\sigma(\pm 1,\pm 1) &=& E^4\beta\sin^3\theta 
                               \\[2ex]
  \M_3^\sigma(\pm 1, 0 )   &=&  \M_3^\sigma( 0 ,\mp 1) =
          \frac{\sqrt{2}E}{M_W}\, 2E^2\beta\, (\cos\theta \mp \sigma) 
          \nonumber \\ 
  \M_4^\sigma(\pm 1, 0 )   &=& \M_4^\sigma( 0 ,\mp 1) =
          \frac{\sqrt{2}E}{M_W}\, E^2\,
          \Bigl[ 2\beta - 2 \cos\theta \mp \sigma (1 - \beta^2) \Bigr]
          (\cos\theta \mp \sigma) 
          \nonumber \\ 
  \M_5^\sigma(\pm 1, 0 )   &=&  \M_5^\sigma( 0 ,\mp 1) =
          \frac{\sqrt{2}E}{M_W}\, E^2 
          (\beta + 2\cos\theta \pm \sigma)(\cos\theta \mp \sigma) 
          \nonumber \\ 
  \M_6^\sigma(\pm 1, 0 )   &=&  \M_6^\sigma( 0 ,\mp 1) =
          -\,\frac{\sqrt{2}E}{M_W}\, E^4\beta\,(\beta + \cos\theta)\sin^2\theta
          \nonumber \\ 
  \M_7^\sigma(\pm 1, 0 )   &=&  \M_7^\sigma( 0 ,\mp 1) =
          -\,\frac{\sqrt{2}E}{M_W}\, 2E^4\beta^2 \sin^2\theta 
          \nonumber \\ 
  \M_8^\sigma(\pm 1, 0 )   &=&  \M_8^\sigma( 0 ,\mp 1) =
          \pm\,\frac{\sqrt{2}E}{M_W}\, 2E^2\beta^2 (\cos\theta \mp \sigma)
          \nonumber \\
  \M_9^\sigma(\pm 1, 0 )   &=&  \M_9^\sigma( 0 ,\mp 1) =
          \mp\,\frac{\sqrt{2}E}{M_W}\, 4E^4\beta^2 (\cos\theta \mp \sigma) 
          \\[2ex]
  \M_1^\sigma( 0 , 0 ) &=& \frac{E^2}{M_W^2}\, 2E^2\beta\, (1 + \beta^2)
                           \sin\theta
                           \nonumber \\
  \M_2^\sigma( 0 , 0 ) &=& \frac{E^2}{M_W^2}\, 8E^4\beta^3\sin\theta 
                           \nonumber \\
  \M_3^\sigma( 0 , 0 ) &=& \frac{E^2}{M_W^2}\, 8E^2\beta\sin\theta 
                           \nonumber \\
  \M_4^\sigma( 0 , 0 ) &=& \frac{E^2}{M_W^2}\, 2E^2\, 
                           [3\beta - \beta^3 - 2\cos\theta]\sin\theta 
                           \nonumber \\
  \M_5^\sigma( 0 , 0 ) &=& \frac{E^2}{M_W^2}\, 4E^2(\beta + \cos\theta)
                           \sin\theta 
                           \nonumber \\
  \M_6^\sigma( 0 , 0 ) &=& \frac{E^2}{M_W^2}\, 2E^4\beta\,
                           (\beta + \cos\theta)^2 \sin\theta 
                           \nonumber \\
  \M_7^\sigma( 0 , 0 ) &=& \frac{E^2}{M_W^2}\, 8E^4\beta^2(\beta + \cos\theta)
                           \sin\theta.
\end{eqnarray}
From this list and the invariant functions of Ref.~\cite{prod-corr}
the density matrix 
$\PP_{[\lambda_{1}\lambda_{2}][\lambda_{1}'\lambda_{2}']}(M_W,M_W)$
follows in a straightforward way.

\subsection{Virtual corrections to the decay stages}
\label{app:virt/decay}

Since we have chosen a specific polarization basis for the calculation
of the production stage, both at lowest order and at virtual one-loop order, 
the same basis has to be used for describing the on-shell $W$-boson decays.  
All results presented in this subsection are therefore given in the LAB
frame, rather than the often used rest frame of the decaying $W$ boson.

Like in the on-shell $W$-pair-production case, we again write the decay
matrix element as a sum of invariant functions $\E_j^{(\pm)}$
multiplied by Lorentz-invariant basic matrix elements 
$\M_j^{(\pm)}(\lambda_i)$:
\be
  \Delta_{\lambda_1}^{(+)}(M_W) = \sum\limits_j \E_j^{(+)}
                                  \M_j^{(+)}(\lambda_1), \quad\quad
  \Delta_{\lambda_2}^{(-)}(M_W) = \sum\limits_j \E_j^{(-)}
                                  \M_j^{(-)}(\lambda_2).
\ee 
In the most general case of the decay of a $W$ boson into massive quarks,
there are four basic matrix elements~\cite{decay-corr}. For the decay
of the $W^{-}$ boson, $W^-(p_2) \to f_2(k_2)\bar{f}_2'(k_2')$, they are
given by
\begin{eqnarray}
  \M_{0}^{(-)}(\lambda_{2}) 
  &=& \bar{u}(k_{2})\, \es_2^{*}\, \omega_{-}\, v(k_{2}'), \nonumber \\[1mm]
  \M_{1}^{(-)}(\lambda_{2}) 
  &=& \bar{u}(k_{2})\, \es_2^{*}\, \omega_{+}\, v(k_{2}'), \nonumber \\[1mm]
  \M_{2}^{(-)}(\lambda_{2}) 
  &=& \bigl[ \bar{u}(k_{2})\, \omega_{-}\, v(k_{2}') \bigr]
      (\e_2^{*} k_{2}), \nonumber \\[1mm]
  \M_{3}^{(-)}(\lambda_{2}) 
  &=& \bigl[ \bar{u}(k_{2})\, \omega_{+}\, v(k_{2}') \bigr]
      (\e_2^{*} k_{2}),
\end{eqnarray}
with similar expressions for the decay of the $W^+$ boson.
For massless particles in the final state only $\M_{0}^{(\pm)}$ occurs.
At lowest-order we then obtain
\be
  \Delta_{\lambda_i}^{0}(M_W) 
  = 
  \frac{ieV_{f_i' f_i}}{\sqrt{2} s_W}\,\M_{0}^{(\pm)}(\lambda_i),
\ee
with $V_{f' f}$ the quark mixing matrix.

For the decay density matrix $\D_{\lambda_i\lambda_i'}(M_W)$ 
it is useful to have the expressions for 
\be
  \M_{00}^{(\pm)}(\lambda_i,\lambda_i') 
  = 
  \sum\limits_{\sss{fermion helicities}}
  \M_0^{(\pm)}(\lambda_i)\,\M_{0}^{(\pm)\,*}(\lambda_i').
\ee
For the decay of the $W^-$ boson one finds 
\begin{eqnarray}
  \label{M00}
  \M_{00}^{(-)}(0,0)
  &=& \frac{M_W^4}{E^2}\, \frac{\sin^2\theta_4}{(1 - \beta \cos\theta_4)^2},
      \nonumber \\
  \M_{00}^{(-)}(\pm 1,\pm 1)
  &=& \frac{M_W^2}{2}\, (1 \pm \beta)^2\,
      \frac{(1 \mp \cos\theta_4)^2}{(1 - \beta \cos\theta_4)^2},
      \nonumber \\
  \M_{00}^{(-)}(\pm 1, 0) 
  &=& \pm\, \frac{M_W^3}{\sqrt{2} E}\, (1 \pm \beta)\, 
      {\mbox e}^{\mp i \phi_4}\,
      \frac{(1 \mp \cos\theta_4)\sin\theta_4}{(1 - \beta \cos\theta_4)^2},
      \nonumber \\
  \M_{00}^{(-)}(0, \pm 1) 
  &=& \pm\, \frac{M_W^3}{\sqrt{2} E}\, (1 \pm \beta)\,
      {\mbox e}^{\pm i \phi_4}\,
      \frac{(1 \mp \cos\theta_4)\sin\theta_4}{(1 - \beta \cos\theta_4)^2},
      \nonumber \\
  \M_{00}^{(-)}(\pm 1,\mp 1) 
  &=&  -\, \frac{M_W^4}{2E^2}\, {\mbox e}^{\mp 2 i \phi_4}\,
       \frac{\sin^2\theta_4}{(1 - \beta \cos\theta_4)^2}.
\end{eqnarray}
The expressions for the charge-conjugate process, describing the decay of 
the $W^+$ boson, can be obtained through the simple relation
\be
  \M_{00}^{(+)}(\lambda_1,\lambda_1')
  =
  \M_{00}^{(-)}(-\lambda_1,-\lambda_1'),
     \quad\quad \mbox{with} \quad 
     \phi_4 \to \phi_3,\ 
     \cos\theta_4 \to \cos\theta_3,\ 
     \sin\theta_4 \to -\sin\theta_3.
\ee 
These expressions can be combined with the invariant functions from 
Ref.~\cite{decay-corr} to obtain the decay density matrices including 
virtual RC. It can be seen from Eq.~(\ref{M00}) that the matrices 
$\M_{00}^{(\pm)}(\lambda_i,\lambda_i')$ contain asymmetric imaginary parts 
proportional to $\sin\phi_{3,4}$. These terms will be responsible for picking
up imaginary loop effects present in the invariant functions, which do
not depend on $\phi_{3,4}$. The symmetric parts of 
$\M_{00}^{(\pm)}(\lambda_i,\lambda_i')$ are real and depend on 
$\cos\phi_{3,4}$. Upon integration over the azimuthal angles $\phi_{3,4}$ the 
matrices $\M_{00}^{(\pm)}(\lambda_i,\lambda_i')$ become real and diagonal, and
the same holds for the corresponding decay density matrices.


\section{The Weyl\,--\,van der Waerden formalism for the real-photon
factorizable corrections}
\label{app:hard_rad}

Like in the case of the virtual factorizable corrections, also for the
real-photon factorizable corrections our choice for the
polarization basis and the calculational scheme is guided by existing
calculations in the literature. As mentioned in Sect.~\ref{sec:pole-scheme},
there is no objection against having different choices for the polarization
basis in different contributions to the RC, provided that the contribution
to the density matrix is calculated consistently within the chosen approach.   
We adopt the conventions of Ref.~\cite{wwgamma} and calculate the real-photon
RC in the Weyl\,--\,van der Waerden formalism.

\subsection{The Weyl\,--\,van der Waerden formalism for massive gauge bosons}
\label{app:wvdw}

Before giving the results for the various matrix elements, we first give
a few essential details of the Weyl\,--\,van der Waerden formalism for massive 
gauge bosons. We follow the conventions of Ref.~\cite{wwgamma} and define
the two-dimensional Weyl spinor for a massless particle with light-like
momentum $q$ as
\be
  q_A = \left( \begin{array}{c}
                 \sqrt{q_0-q_3} \\
                 -\,(q_1+iq_2)/\sqrt{q_0-q_3}
               \end{array}
        \right), \quad\quad
  q_{\dot{A}} = (q_A)^*.
\ee
The indices can be raised and lowered by the spinor metric
\be
  \epsilon^{AB} = \left( \begin{array}{cc}
                           0  & 1 \\
                           -1 & 0
                         \end{array}
                  \right)
                = \epsilon_{AB} 
                = \epsilon^{\dot{A}\dot{B}}
                = \epsilon_{\dot{A}\dot{B}}
\ee
according to 
\be
  q^A = q_B\, \epsilon^{BA}, \quad\quad
  q_A = \epsilon_{AB}\,q^B.
\ee
The spinor products 
\be 
  \la qk \ra   = q_A k^A = -\,q^A k_A, \quad\quad
  \la qk \ra^* = q_{\dot{A}}k^{\dot{A}} = -\,q^{\dot{A}}k_{\dot{A}}
\ee
are hence antisymmetric. 
In the Weyl representation for the $\gamma$-matrices we obtain the following 
set of translation rules into two-dimensional spinor language:
\begin{eqnarray}
  u(q,+) 
  &=& v(q,-)
      \ = \ \left( \begin{array}{c} q_A \\ 0 \end{array} \right),
      \nonumber \\
  u(q,-) 
  &=& v(q,+)
      \ = \ \left( \begin{array}{c} 0 \\ q^{\dot{A}} \end{array} \right),
      \nonumber \\
  \gamma^{\mu} 
  &=& \left( \begin{array}{cc}
               0 & \sigma^{\mu}_{\dot{B}A} \\
               \sigma^{\mu\,\dot{A}B} & 0 
             \end{array}
      \right),
      \nonumber \\[1mm]
  Q_{\mu}\sigma^{\mu}_{\dot{A}B} 
  &=& Q_{\dot{A}B} \ \ \ \ \Rightarrow \ \ \ \ 
      2QK \ = \ Q^{\dot{A}B}K_{\dot{A}B},
      \nonumber \\[2ex]
  q_{\mu}\sigma^{\mu}_{\dot{A}B} 
  &=& q_{\dot{A}B} \ = \ q_{\dot{A}}q_B \ \ \ \ \Rightarrow \ \ \ \ 
      2qk \ = \ |\la qk \ra |^2,
\end{eqnarray}
with $Q,K$ arbitrary Lorentz vectors and $q,k$ light-like ones. The Dirac
spinors $u(q,\pm)$ denote right-handed ($+$) and left-handed ($-$) states.
The matrices $\sigma^{\mu\,\dot{A}B} = (\sigma^0,\vec{\sigma})$ consist of the
$2\times 2$ unit matrix $\sigma^0$ and the standard Pauli matrices 
$\sigma^i$ ($i=1,2,3$).

For a photon with momentum $k$ we use the polarization vectors describing 
the two helicity eigenstates
\be
\label{wvdw/photpol}
  \e_{\gamma}^{\dot{A}B}(+1) 
  = 
  \sqrt{2}\,\frac{k^{\dot{A}}b^{B}}{\la kb\ra},
  \quad\quad  
  \e_{\gamma}^{\dot{A}B}(-1)
  =  
  \sqrt{2}\,\frac{b^{\dot{A}}k^{B}}{\la kb\ra^{*}}
  = \e_{\gamma}^{\dagger\,\dot{A}B}(+1).
\ee
To handle the massive $W^{\pm}$ bosons we first decompose their massive momenta
$p_{1,2}$ into a sum of two light-like momenta: 
\be
\label{wvdw/decomp}
  p_{1,2}^{\mu} = p_{3,4}^{\mu} + c_{1,2}\,m_{3,4}^{\mu}, \quad\quad
  c_1\,(2p_3m_3) = c_2\,(2p_4m_4) = M_W^2.
\ee  
Note that the so-defined light-like vectors $m_{3,4}$ can be chosen freely.
An orthogonal basis for the three physical polarizations of the
massive $W^{\pm}$ bosons is now given by
\begin{eqnarray}
\label{wvdw/wpol}
  \e_1^{\dot{A}B}(+1) 
  &=& \sqrt{2}\,\frac{p_3^{\dot{A}}m_3^{B}}{\la p_3m_3\ra}, \quad\quad  
      \e_{1}^{\dot{A}B}(-1) 
      =  \sqrt{2}\,\frac{m_3^{\dot{A}}p_3^{B}}{\la p_3m_3\ra^{*}}, \quad\quad  
      \e_{1}^{\dot{A}B}(0) 
      =  \frac{1}{M_{W}}\,\bigl( p_3-c_1 m_3 \bigr)^{\dot{A}B},
      \nonumber \\
  \e_2^{\dot{A}B}(+1) 
  &=& \sqrt{2}\,\frac{p_4^{\dot{A}}m_4^{B}}{\la p_4m_4\ra}, \quad\quad  
      \e_{2}^{\dot{A}B}(-1) 
      =  \sqrt{2}\,\frac{m_4^{\dot{A}}p_4^{B}}{\la p_4m_4\ra^{*}}, \quad\quad  
      \e_{2}^{\dot{A}B}(0) 
      =  \frac{1}{M_{W}}\,\bigl( p_4-c_2 m_4 \bigr)^{\dot{A}B},\quad\quad
\end{eqnarray}
with $\e_i^{\dot{A}B}(-1) = \e_i^{\dagger\,\dot{A}B}(+1)$. It should be
stressed that the so-obtained polarization basis does not correspond to the 
helicity eigenstates. However, the corresponding states transform like
helicity eigenstates under a parity transformation, which is very useful 
for practical calculations.

\subsubsection{An example: lowest-order on-shell W-pair production}
\label{app:wvdw/prod-born}

As an example we apply the above method to the lowest-order on-shell 
production stage. To this end we choose $m_{3,4} = p_{4,3}$ in
Eqs.~(\ref{wvdw/decomp}) and (\ref{wvdw/wpol}), and write 
$c_1 = c_2 = M_W^2/(2p_3p_4) \equiv c$. 

The complete Born amplitude of the process is of the form
\be
  \Pi^0_{\sigma;\lambda_1\lambda_2}(M_W,M_W)
  =
    G_1(t)\,\M_t(\sigma,\lambda_1,\lambda_2)
  + G_2(s,\sigma)\M_s(\sigma,\lambda_1,\lambda_2),
\ee
where the coefficients $G_{1,2}$ are defined in Eq.~(\ref{born/coupling}),
and 
\begin{eqnarray}
  \M_t(\sigma,\lambda_1,\lambda_2) 
  &=& \bar{v}(q_1)\, \es_1 (\qs_2 - \ps_2)\es_2\, \omega_{-}u(q_2),
      \nonumber \\ [1mm]   
  \M_s(\sigma,\lambda_1,\lambda_2) 
  &=& \bar{v}(q_1)\Vs(p_1,\e_1,p_2,\e_2)\,\omega_{\sigma}u(q_2),
\end{eqnarray}
where
\be
\label{TGC}
  V^{\mu}(p_1,\e_1,p_2,\e_2) 
  = 
  2 \e^{\mu}_1 (\e_2p_1) - 2 \e^{\mu}_2 (\e_1p_2) 
  + (p^{\mu}_2 - p^{\mu}_1) (\e_1\e_2).
\ee
These matrix elements can be translated into two-dimensional representation,
e.g.~for $\sigma=-1$ one obtains
\begin{eqnarray}
  \M_t(\sigma=-1,\lambda_1,\lambda_2) 
  &=& q_1^A\, \e_{1\,\dot{B}A} (q_2-p_2)^{\dot{B}C}
      \e_{2\,\dot{D}C}\, q_2^{\dot{D}},
      \nonumber \\[1mm]
  \M_s(\sigma=-1,\lambda_1,\lambda_2)
  &=& q_1^A\, V_{\dot{B}A} (p_1,\e_1,p_2,\e_2)\, q_2^{\dot{B}}.
\end{eqnarray}
For $\sigma=+1$ we can make use of the relations 
\be
  \M_t(\sigma=+1,\lambda_1,\lambda_2) = 0, \quad\quad 
  \M_s(\sigma=+1,\lambda_1,\lambda_2) 
  = 
  \M_s^*(\sigma=-1,-\lambda_1,-\lambda_2),
\ee
where the last identity is the result of parity conservation of the
$s$-channel matrix element $\M_s$, since all parity violation is
contained in the coefficient $G_2(s,\sigma)$. From CP invariance one 
obtains two more relations:
\be
  \M_{s,t}(q_1,-\sigma,q_2,\sigma,p_1,\lambda_1,p_2,\lambda_2)
  =
  -\,\M_{s,t}^{*}(q_2,-\sigma,q_1,\sigma,p_2,-\lambda_2,p_1,-\lambda_1),
\ee
so only 6 independent polarization states remain. The independent matrix 
elements read:
\begin{eqnarray}
\M_t(-1,+1,+1) 
   &=& 
        \frac{2}{\la p_3p_4 \ra^2}\, \la q_1p_4 \ra\la q_2p_4 \ra^*
        \Bigl[ 2q_1p_3 - M_W^2 \Bigr],     
        \nonumber \\
\M_t(-1,+1,-1) 
   &=& 
        -\,\frac{\la q_1p_4 \ra\la q_2p_4 \ra\la q_2p_3\ra^{*\,2}}{p_3p_4}, 
        \nonumber \\
\M_t(-1,-1,+1) 
   &=& 
        \frac{\la q_1p_3 \ra^{2}\la q_1p_4 \ra^*\la q_2p_4 \ra^*}{p_3p_4},
        \nonumber \\
\M_t(-1,+1,0) 
   &=& 
        \frac{\sqrt{2}\la q_1p_4 \ra\la q_2p_3 \ra^*}{M_W \la p_3p_4 \ra}\,
        \Bigl[ 2q_2p_4 - 2 c\, q_2p_3 + M_W^2 \Bigr],        
        \nonumber \\
\M_t(-1,-1,0) 
    &=& 
        \frac{\sqrt{2}\la q_1p_3 \ra\la q_2p_4 \ra^*}{M_W \la p_3p_4 \ra^*}\,
        \Bigl[ 2q_2p_4 - 2 c\, q_2p_3 - M_W^2 \Bigr],      
        \nonumber \\
\M_t(-1,0,0) 
    &=& 
        \la q_1p_3 \ra\la q_2p_3 \ra^*
        \biggl\{
        1-c + \frac{2}{M_W^2}(1+c)\Bigl[ q_2p_4 - c\, q_2p_3 \Bigr]
        \biggr\}
\end{eqnarray}
for the $t$-channel matrix elements, and 
\begin{eqnarray}
\M_s(-1,+1,+1) 
    &=& 
        -\,2 \la q_1p_3 \ra\la q_2p_3 \ra^* 
        \frac{\la p_3p_4 \ra^*}{\la p_3p_4 \ra}\,(1-c),  
        \nonumber \\[1mm]
\M_s(-1,+1,-1) 
    &=& 
        \M_s(-1,-1,+1) \ = \ 0,                
        \nonumber \\[1mm]
\M_s(-1,+1,0) 
    &=&  
        \frac{\sqrt{2}}{M_W}\,\la q_1p_4 \ra\la q_2p_3 \ra^*\la p_3p_4 \ra^*
        (1-c^2),   
        \nonumber \\
\M_s(-1,-1,0) 
    &=&  
        \frac{\sqrt{2}}{M_W}\,\la q_1p_3 \ra\la p_3p_4 \ra\la q_2p_4 \ra^*
        (1-c^2),       
        \nonumber \\
\M_s(-1,0,0) 
    &=&  
        \la q_1p_3 \ra\la q_2p_3 \ra^*\biggl( \frac{1}{c}+3-3c-c^2 \biggr)
\end{eqnarray}
for the $s$-channel matrix elements.

\subsection{Non-collinear photon radiation from the production stage}
\label{app:mtrx_rad/prod}

Using the above example as guideline, we
now address the process of non-collinear real-photon radiation from the
production stage:
\be
  e^{+}(q_{1},\sigma_{1})e^{-}(q_{2},\sigma_{2}) 
  \to
  W^{+}(p_{1},\lambda_{1}) W^{-}(p_{2},\lambda_{2}) \gamma(k,\lambda).  
\ee
Since we are dealing with non-collinear radiation and massless initial-state
electrons and posi\-trons, we can ignore the possibility of helicity flip in 
the initial state. Therefore again the condition 
$\sigma_{1}=-\sigma_{2}=-\sigma$ applies. As a first step we extend the
list of kinematical invariants of Sect.~\ref{sec:pole-scheme}:
\begin{eqnarray}
  s &=& (q_{1} + q_{2})^{2}, \quad\quad
        \,\,t \ = \ (q_{1} - p_{1})^{2}, \quad\quad 
        \;u \ = \ (q_{1} - p_{2})^{2},
        \nonumber \\
  s^{\prime} &=& (p_{1} + p_{2})^{2}, \quad\quad  
       t^{\prime} \ = \ (q_{2} - p_{2})^{2}, \quad\quad
       u^{\prime} \ = \ (q_{2} - p_{1})^{2}.
\end{eqnarray}
The complete matrix element can now be written in the form
\be
  \Pi_{\gamma}(M_W,M_W)  
  = 
  -\, e \Bigl[  G_{1}(t^{\prime})\M_1^{\gamma} 
              + G_{1}(t)\M_2^{\gamma}
              + G_{2}(s^{\prime},\sigma)\M_3^{\gamma} 
              + G_{2}(s,\sigma)\M_4^{\gamma}
        \Bigr],
\ee
where the functions $G_{1,2}$ are the same as the ones defined in 
Eq.~(\ref{born/coupling}). The basic matrix elements $\M_j^{\gamma}$ 
are invariant under gauge transformations of the radiated photon. 
They are given by
\begin{eqnarray}
  \M_1^{\gamma} \!\!\!&=&\!\!\! \bar{v}(q_1) \Biggl\{
                       \es_{\gamma}\frac{\qs_{1}\! -\! \ks}{2q_{1}k}\,
                       \es_{1}(\qs_{2}\! -\! \ps_{2})\es_{2}
                       +
                       \Bigl[ - 2 \es_{1} (\e_{\gamma} p_{1})  
                               + 2 \es_{\gamma} (\e_{1}k)
                               - 2 \ks (\e_{\gamma}\e_{1})
                       \Bigr]\, \frac{\qs_{2}\! -\! \ps_{2}}{2p_{1}k}\,\es_{2}
                                         \Biggr\}\, \omega_{-} u(q_2),
                       \nonumber \\
  \M_2^{\gamma} \!\!\!&=&\!\!\! \bar{v}(q_1) \Biggl\{
                       \es_{1}(\qs_{1}\! -\! \ps_{1})\es_{2}\,
                       \frac{\qs_{2}\! -\! \ks}{2q_{2}k}\,\es_{\gamma}
                       +
                       \es_{1}\frac{\ps_{1}\! -\! \qs_{1}}{2p_{2}k}\, 
                       \Bigl[   2 \es_{2}(\e_{\gamma} p_{2}) 
                               - 2 \es_{\gamma} (\e_{2}k)
                               + 2 \ks (\e_{\gamma}\e_{2})
                       \Bigr]
                                         \Biggr\}\, \omega_{-} u(q_2),
                       \nonumber \\
  \M_3^{\gamma} \!\!\!&=&\!\!\! \bar{v}(q_1) \Biggl\{
                        \es_{\gamma}\frac{\qs_{1}\! -\! \ks}{2q_{1}k}\,
                        \Vs (p_{1},\e_{1},p_{2},\e_{2})
                        - \Vs (p_{1},\e_{1},p_{2},\e_{2})\,
                        \frac{\qs_{2}\! -\! \ks}{2q_{2}k}\,\es_{\gamma}
                                         \Biggr\}\, \omega_{\sigma}u(q_2),
                        \nonumber \\[1mm]
  \M_4^{\gamma} \!\!\!&=&\!\!\! \bar{v}(q_1) \biggl\{
                       - 2\es_{\gamma} (\e_{1}\e_{2})
                       + \Vs_{b}(p_{1},\e_{1},p_{2},\e_{2})
                       + \Vs_{b}(p_{2},\e_{2},p_{1},\e_{1})
                                         \biggr\}\, \omega_{\sigma}u(q_2),
\end{eqnarray}  
where we introduced the shorthand notation
\begin{eqnarray}
  \lefteqn{ \frac{p_{1}k}{2}\, V^{\mu}_{b}(p_{1},\e_{1},p_{2},\e_{2}) 
              \ =\  -\, \frac{\e_{\gamma}p_{1}}{2}\, 
                      V^{\mu}(p_{1}+k,\e_{1},p_{2},\e_{2}) 
                    + \Bigl[   p^{\mu}_{1}(\e_{1}\e_{\gamma})
                              + \e^{\mu}_{\gamma}(\e_{1}k) 
                      \Bigr] (\e_{2}p_{1}+\e_{2}k)
          }\quad\quad\quad\quad
    \nonumber \\[1mm]
  & & +\, \e^{\mu}_{2} \Bigl[   (\e_{1}\e_{\gamma})(kp_{1} + kp_{2})
                              - (\e_{1}k)(\e_{\gamma}p_{2}) 
                       \Bigr]
      + p^{\mu}_{2} \Bigl[   (\e_{2}\e_{\gamma})(\e_{1}k)
                           + (\e_{1}\e_{\gamma})(\e_{2}p_{1})
                    \Bigr].\quad\quad
\end{eqnarray}
The vertex function $V$ can be taken from Eq.~(\ref{TGC}).
Note that the term $-2\ks (\e_{\gamma}\e_{1})$ between the square brackets of
$\M_1^{\gamma}$ originally had the form $(\ps_{1}-\ks)(\e_{\gamma}\e_{1})$. 
The difference $(\ps_{1}+\ks)(\e_{\gamma}\e_{1})$ cancels against similar
terms in $\M_4^{\gamma}$. This cancellation is a consequence of the 
lowest-order Ward identity of the $W^+$ boson. In the same way also the Ward
identity of the $W^-$ boson has been used to simplify $\M_2^{\gamma}$.

For the calculation in the Weyl\,--\,van der Waerden formalism we choose
$m_{3,4} = k$ in Eqs.~(\ref{wvdw/decomp}) and (\ref{wvdw/wpol}). Furthermore
we choose the free gauge parameter $b$ in Eq.~(\ref{wvdw/photpol}) to be 
equal to $q_1$.
Like in the case without photon radiation, we can exploit some symmetry
relations. First of all CP invariance implies the relations
\begin{eqnarray}
  \M_1^{\gamma}(q_{1},-\sigma,q_{2},\sigma,
                p_{1},\lambda_{1},p_{2},\lambda_{2},k,\lambda)
  &=& \M_2^{\gamma\,*}(q_{2},-\sigma,q_{1},\sigma,
                       p_{2},-\lambda_{2},p_{1},-\lambda_{1},k,-\lambda),
      \nonumber \\[1mm]
  \M_{3,4}^{\gamma}(q_{1},-\sigma,q_{2},\sigma,
                    p_{1},\lambda_{1},p_{2},\lambda_{2},k,\lambda)
  &=& \M_{3,4}^{\gamma\,*}(q_{2},-\sigma,q_{1},\sigma,
                           p_{2},-\lambda_{2},p_{1},-\lambda_{1},k,-\lambda).
\end{eqnarray}
The matrix elements for right-handed electrons are again completely determined:
\begin{eqnarray}
  \M_1^{\gamma}(\sigma=+1) 
  &=& \M_2^{\gamma}(\sigma=+1) \ = \ 0, 
      \nonumber \\[1mm] 
  \M_{3,4}^{\gamma}(\sigma=+1,\lambda_1,\lambda_2,\lambda) 
  &=& \M_{3,4}^{\gamma\,*}(\sigma=-1,-\lambda_1,-\lambda_2,-\lambda),
\end{eqnarray}
where the last identity is the result of parity conservation of the
$s$-channel matrix elements $\M_{3,4}^{\gamma}$. Due to the symmetry 
(antisymmetry) property of the quartic (triple) gauge boson vertex under 
the exchange of the $W^+$ and $W^-$ bosons, one can derive two more relations:
\begin{eqnarray}
  \M_3^{\gamma}(p_{1},\lambda_{1},p_{2},\lambda_{2},k,\lambda)
  &=& -\, \M_{3}^{\gamma}(p_{2},\lambda_{2},p_{1},\lambda_{1},k,\lambda),
      \nonumber \\[1mm]
  \M_4^{\gamma}(p_{1},\lambda_{1},p_{2},\lambda_{2},k,\lambda)
  &=& +\, \M_{4}^{\gamma}(p_{2},\lambda_{2},p_{1},\lambda_{1},k,\lambda).
\end{eqnarray}

After all these preparations we now list the independent matrix
elements for $\sigma=-1$. In order to keep the results as compact as 
possible we use the shorthand notations 
$\la r_{i}r_{j} \ra = \la ij \ra$ and $(r_{i}r_{j}) = (ij)$, with
$r_i=(q_1,q_2,p_3,p_4,k)$ for $i=(1,2,3,4,5)$.
For the amplitude $\M_1^{\gamma}(\lambda_1,\lambda_2,\lambda)$ we find:
\begin{eqnarray}
\M_1^{\gamma}(+1,+1,+1) 
     &=& 
         -\, 4 \sqrt{2}\,
         \frac{\la15\ra\la24\ra^{*}}{\la35\ra^{2}\la45\ra}\,
         \Bigl[ (35)-(13)-(15) \Bigr],  \nonumber \\
\M_1^{\gamma}(+1,-1,+1) 
     &=& 
         -\, 2 \sqrt{2}\,
         \frac{\la25\ra^{*}}{\la35\ra^{2}\la45\ra^{*}}
         \biggl[
         \la24\ra\Bigl(\la13\ra\la23\ra^{*} + \la15\ra\la25\ra^{*}\Bigr)
         - c_{2} \la13\ra\la45\ra\la35\ra^{*}\biggr],      \nonumber \\
\M_1^{\gamma}(-1,+1,+1) 
     &=& 
         -\, 2 \sqrt{2}\,
         \frac{\la13\ra^{2}\la24\ra^{*}}{\la15\ra\la45\ra}
         \Biggl[1 - \frac{(15)}{(35)}\Biggr],        \nonumber \\
\M_1^{\gamma}(-1,-1,+1) 
     &=& 
         -\, \sqrt{2}\,
         \frac{\la13\ra^{2}\la24\ra\la25\ra^{*2}}{\la15\ra\la45\ra^{*} (35)},
         \nonumber \\
\M_1^{\gamma}(+1,0,+1) 
     &=& 
         -\, \frac{2}{\la35\ra^{2} M_{W}}
         \biggl\{
         \Bigl[ \la13\ra\la23\ra^{*}+\la15\ra\la25\ra^{*} \Bigr]
         \Bigl[t^{\prime} + 4(24) \Bigr]
         \nonumber \\
     & &
         -\, 2c_{2}\la13\ra\la45\ra\la24\ra^{*}\la35\ra^{*}
         \biggr\},                  \nonumber \\
\M_1^{\gamma}(-1,0,+1) 
     &=&  
         -\,\frac{\la13\ra^{2}\la25\ra^{*}}{\la15\ra(35)M_{W}}\,
         \Bigl[ t^{\prime} + 4(24) \Bigr],       \nonumber \\
\M_1^{\gamma}(0,+1,+1) 
     &=& 
         -\,4\,\frac{\la13\ra\la24\ra^{*}}{\la35\ra\la45\ra M_{W}}
         \biggl\{
         -(13)+\bigl(1 - c_{1}\bigr)\Bigl[ (35)-(15) \Bigr]
         \biggr\},      \nonumber \\
\M_1^{\gamma}(0,-1,+1) 
     &=& 
         -\,2\,\frac{\la13\ra\la25\ra^{*}}{\la35\ra\la45\ra^{*}M_{W}}
         \biggl[
         \frac{\la13\ra}{\la15\ra}
         \Bigl( \la24\ra\la23\ra^{*}-c_{2}\la45\ra\la35\ra^{*} \Bigr)
         + \bigl(1-c_{1}\bigr)\la24\ra\la25\ra^{*}
         \biggr],        \nonumber \\
\M_1^{\gamma}(0,0,+1) 
     &=& 
         -\,\frac{\sqrt{2}\la13\ra}{\la35\ra M_{W}^{2}}
1         \Biggl[
         -2 c_{2}\frac{\la13\ra}{\la15\ra}\la45\ra\la24\ra^{*}\la35\ra^{*}
         \nonumber \\
     & &   
         +\, \Bigl[ t^{\prime}+4(24) \Bigr]
         \biggl(\frac{\la13\ra}{\la15\ra}\la23\ra^{*} 
                + (1 - c_{1})\la25\ra^{*} \biggr)
         \Biggr],             \nonumber \\
\M_1^{\gamma}(+1,+1,-1) 
     &=& 
         \sqrt{2}\,\frac
         {\la15\ra^{2}\la13\ra^{* 2}\la24\ra^{*}}
         {\la45\ra\la15\ra^{*} (35)},            \nonumber \\
\M_1^{\gamma}(+1,-1,-1) 
     &=& 
         -\, \sqrt{2}\,\frac
         {\la15\ra\la13\ra^{*}\la25\ra^{*}}
         {\la15\ra^{*}\la45\ra^{*} (35)}\,
         \Bigl[ \la24\ra\la23\ra^{*} - c_{2}\la45\ra\la35\ra^{*} \Bigr],
         \nonumber \\
\M_1^{\gamma}(-1,+1,-1) 
     &=& 
         -\,\frac
         {8\sqrt{2}\la24\ra^{*}}
         {\la45\ra\la15\ra^{*}\la35\ra^{* 2}}\,
         \Bigl[ (15)-(35) \Bigr]\Bigl[ (35)-(13)-(15) \Bigr],    \nonumber \\
\M_1^{\gamma}(-1,-1,-1) 
     &=&
         \frac
         {4\sqrt{2}\la24\ra\la25\ra^{* 2}}
         {\la35\ra^{* 2}\la45\ra^{*}\la15\ra^{*}}\,
         \Bigl[ (35)-(13)-(15) \Bigr],        \nonumber \\
\M_1^{\gamma}(+1,0,-1) 
     &=& 
         -\,\frac
         {\la15\ra\la13\ra^{*}}
         {\la15\ra^{*}(35)M_{W}}
         \biggl\{
         -2c_{2}\la45\ra\la24\ra^{*}\la35\ra^{*}
         +\la23\ra^{*}\Bigl[ t^{\prime} + 4(24) \Bigr]
         \biggr\},         \nonumber \\
\M_1^{\gamma}(-1,0,-1) 
     &=& 
         \frac
         {4 \la25\ra^{*}}
         {\la15\ra^{*}\la35\ra^{* 2}M_{W}}\,
         \Bigl[ t^{\prime}+4(24) \Bigr]
         \Bigl[ (35)-(13)-(15) \Bigr],        \nonumber \\
\M_1^{\gamma}(0,+1,-1) 
     &=& 
         -\,\frac
         {4 \la15\ra\la13\ra^{*}\la24\ra^{*}}
         {\la45\ra\la15\ra^{*}\la35\ra^{*}M_{W}}
         \biggl\{
         c_{1}\Bigl[ (15) - (35) \Bigr]
         -(13)-(15)+(35)
         \biggr\},        \nonumber \\
\M_1^{\gamma}(0,-1,-1) 
     &=&
         \frac
         {2 \la25\ra^{*}}
         {\la15\ra^{*}\la35\ra^{*}\la45\ra^{*}M_{W}}
         \biggl\{
         c_{1}\la15\ra\la24\ra\la13\ra^{*}\la25\ra^{*}
         \nonumber \\
     & &
         +\, 2\Bigl( \la24\ra\la23\ra^{*} - c_{2}\la45\ra\la35\ra^{*} \Bigr)
         \Bigl[ (35) - (13) - (15) \Bigr]
         \biggr\},             \nonumber \\
\M_1^{\gamma}(0,0,-1) 
     &=&
         \frac
         {\sqrt{2}}
         {\la15\ra^{*}\la35\ra^{*}M_{W}^{2}}
         \Biggl\{
         -\, 4c_2 \la24\ra^{*}\la35\ra^{*}\la45\ra
         \Bigl[ (35) - (13) - (15) \Bigr]
         \nonumber \\
     & & 
         +\, \Bigl[ t^{\prime}+4(24) \Bigr]
         \biggl(
         c_{1}\la15\ra\la13\ra^{*}\la25\ra^{*}
         + 2\la23\ra^{*}\Bigl[ (35) - (13) - (15) \Bigr] 
         \biggr) 
         \Biggr\}. 
\end{eqnarray}
The independent matrix elements $\M_3^{\gamma}(\lambda_1,\lambda_2,\lambda)$ 
read:
\begin{eqnarray}
\M_3^{\gamma}(+1,+1,+1) 
     &=& 
         -\,\frac{4\sqrt{2}\la15\ra\la34\ra^{*}}
         {\la25\ra\la35\ra\la45\ra}\,
         \Bigl[ -(12)+(35)+(45) \Bigr],   \nonumber \\
\M_3^{\gamma}(+1,-1,+1) 
     &=& 
         \frac{2\sqrt{2}}
         {\la25\ra\la35\ra\la45\ra^{*}}
         \Biggl\{
         2\frac{\la12\ra}{\la15\ra}(45)\la14\ra\la23\ra^{*}
         +c_{2}\la12\ra\la45\ra\la25\ra^{*}\la35\ra^{*}
         \nonumber \\
     & &
         -\, \la35\ra^{*}
         \biggl[ 
             \la34\ra\Bigl( \la12\ra\la23\ra^{*}+\la15\ra\la35\ra^{*}\Bigr)
             -2(45)\la14\ra
         \biggr]
         \Biggr\},                        \nonumber \\
\M_3^{\gamma}(-1,-1,+1) 
     &=& 
         -\,\frac
         {2\sqrt{2}\la12\ra^{2}\la34\ra\la25\ra^{* 2}}
         {\la15\ra\la25\ra\la35\ra^{*}\la45\ra^{*}},    \nonumber \\
\M_3^{\gamma}(+1,0,+1)
     &=&
         \frac
         {- 4}
         {\la25\ra\la35\ra M_{W}}
         \biggl\{
          -c_{2}\la12\ra\la45\ra\la25\ra^{*}\la34\ra^{*} 
         \nonumber \\
     & &
         +\, \Bigl[ \la12\ra\la23\ra^{*} + \la15\ra\la35\ra^{*} \Bigr]
         \Bigl[ (34) + c_{1}(45) - c_{2}(35) \Bigr]
         \biggr\},         \nonumber \\
\M_3^{\gamma}(-1,0,+1) 
    &=& 
       -\, \frac
         {4 \la12\ra\la25\ra^{*}}
         {\la15\ra\la25\ra\la35\ra^{*}M_{W}}
         \biggl\{
         -c_{2}\la15\ra\la34\ra\la45\ra^{*}
         + \la13\ra\Bigl[ (34) + c_{1}(45) - c_{2}(35) \Bigr]
         \biggr\},           \nonumber \\
\M_3^{\gamma}(0,0,+1) 
    &=& 
        -\,\frac
         {\sqrt{2}}
         {\la25\ra M_{W}^{2}}
         \biggl\{ 
         -\,2\,\frac{\la13\ra}{\la15\ra}\,\la14\ra\la34\ra^*
         \Bigl[ (34) + c_2(35) + c_1(45) \Bigr]
         \nonumber \\[1mm]
    & &
         +\,\la13\ra\la35\ra^*
         \Bigl[ 2c_2^2(35) - 4c_1^2(45) + (2c_2-4c_1)(34) - (3c_1-2c_2)M_W^2
         \Bigr]
         \nonumber \\[1mm]
    & &
         -\,\la14\ra\la45\ra^*
         \Bigl[ 2c_1^2(45) - 4c_2^2(35) + (2c_1-4c_2)(34) - (3c_2-2c_1)M_W^2
         \Bigr]
         \biggr\}. 
         \nonumber \\
\end{eqnarray}
For $\M_4^{\gamma}(\lambda_1,\lambda_2,\lambda)$ we obtain:
\begin{eqnarray}
\M_4^{\gamma}(+1,+1,+1) 
    &=&  
         -\,4\sqrt{2}\,
         \frac{\la15\ra^{2}\la12\ra^{*}}{\la35\ra^{2}\la45\ra^{2}}
         \Bigl[ (34)+(35)+(45) \Bigr],     \nonumber \\
\M_4^{\gamma}(+1,-1,+1) 
    &=&
         \frac{2\sqrt{2}}{\la35\ra^{2}\la45\ra^{*}}
         \biggl\{
           c_{1}\la15\ra\la34\ra\la25\ra^{*}\la35\ra^{*}
           + 2 \frac{\la14\ra}{\la45\ra}\la34\ra\la23\ra^{*} (35)
         \nonumber \\
    & &  
         +\, 2 \la14\ra\la25\ra^{*}\Bigl[ (34) + (45) \Bigr]
         \biggr\},          \nonumber \\
\M_4^{\gamma}(-1,-1,+1) 
    &=& 
         \frac{\sqrt{2}}{2}\,
         \frac{\la12\ra\la34\ra^{2}\la25\ra^{* 2}}{(35)(45)},   \nonumber \\
\M_4^{\gamma}(+1,0,+1) 
    &=& 
         \frac{4}{M_{W}\la35\ra^{2}}
         \Biggl\{
           \frac{\la15\ra}{\la45\ra}\la34\ra\la23\ra^{*}
           \Bigl[ (34)+(35)(1-c_{2})+(45)(1+c_{1}) \Bigr]
         \nonumber \\
    & &   
         -\, \la15\ra\la25\ra^{*}
         \Bigl[ (34)(1+2c_{2})+(35)(1+c_{2})+(45)(1+c_{1})+M_{W}^{2} \Bigr]
         \Biggr\},         \nonumber \\
\M_4^{\gamma}(-1,0,+1) 
    &=& 
         \frac{2\la25\ra^{*}}{M_{W}\la45\ra(35)}
         \biggl\{
         - c_{2}\la14\ra\la35\ra\la34\ra\la45\ra^{*}           
         \nonumber \\
    & &  
         +\, \la13\ra\la34\ra
         \Bigl[ (34)+(35)(1-c_{2})+(45)(1+c_{1}+2c_{2}) \Bigr]
         \biggr\},            \nonumber \\
\M_4^{\gamma}(0,0,+1) 
    &=&
         \frac{2\sqrt{2}\la34\ra\la15\ra\la25\ra^{*}}{M_W^2\la35\ra\la45\ra}
         \Biggl[
              -  2 c_{1}(1+c_{1})(45) + c_{2}(1+c_{2})(35)
         \nonumber \\
    & & 
         +\, (c_{2}-2c_{1})(34)
         + \biggl(c_{2}-\frac{3}{2}c_{1}-\frac{1}{2}\biggr)M_{W}^{2}
         \Biggr]
         \nonumber   \\ 
    & &
         -\, \frac{2\sqrt{2}\la14\ra}{M_{W}^{2}\la45\ra}
         \Biggl[
           \la25\ra^{*}
         + \la24\ra^{*}\frac{\la34\ra}{\la35\ra}
         \Biggr]
         \Bigl[ (34)+(35)(1+c_{2})+(45)(1+c_{1}) \Bigr]. \nonumber \\
\end{eqnarray}

\subsubsection{Lowest-order decay of the W bosons}

Having fixed the polarization choice for the real-photon factorizable 
corrections to the production stage, we now calculate the lowest-order
decay parts accordingly, since they are needed for obtaining the DPA limit
of the full matrix element $\M_0$ in Eq.~(\ref{prodrad}).
The matrix elements for the $W$-boson decays are given by  
\be
  \Delta^0_{\lambda_i}(M_W) 
  = 
  \frac{ieV_{f_i' f_i}}{\sqrt{2}s_{W}}\,\M^{(\pm)}_{0}(\lambda_i). 
\ee 
Using again $m_{3,4} = k$ in Eqs.~(\ref{wvdw/decomp}) and (\ref{wvdw/wpol}),
one ends up with
$$
  \M_{0}^{(+)}(+1) 
  = 
  \sqrt{2}\,\frac{\la k_1'p_3 \ra\la k_1k \ra^{*}}{\la p_3k \ra^{*}},
  \quad\quad  
  \M_{0}^{(+)}(-1) 
  = 
  \sqrt{2}\,\frac{\la k_1'k \ra\la k_1p_3 \ra^{*}}{\la p_3k \ra},
$$
\be
  \M_{0}^{(+)}(0) 
  = 
  \frac{1}{M_W}\Bigl( \la k_1'p_3 \ra\la k_1p_3 \ra^{*} 
                      - c_1\la k_1'k \ra\la k_1k \ra^{*} \Bigr)
\ee
for the $W^+$ boson, and 
$$
  \M_{0}^{(-)}(+1) 
  = 
  \sqrt{2}\,\frac{\la k_2p_4 \ra\la k_2'k \ra^{*}}{\la p_4k \ra^{*}},
  \quad\quad    
  \M_{0}^{(-)}(-1) 
  = 
  \sqrt{2}\,\frac{\la k_2k \ra\la k_2'p_4 \ra^{*}}{\la p_4k \ra},
$$
\be
  \M_{0}^{(-)}(0) 
  = 
  \frac{1}{M_W}\Bigl( \la k_2p_4 \ra\la k_2'p_4 \ra^{*} 
                      - c_2\la k_2k \ra\la k_2'k \ra^{*} \Bigr)
\ee
for the $W^{-}$ boson.

\subsection{Non-collinear photon radiation from the decay stages}
\label{app:mtrx_rad/decay}

Next we address the process of non-collinear real-photon radiation from the
decay stages. We start off with the decay of the $W^+$ boson:
\be 
\label{wvdw/decay/rad}
  W^{+}(p_{1},\lambda_1) \to \bar{f}_1(k_1) f_1'(k_1') \gamma(k,\lambda).
\ee 
We do not explicitly write the helicities of the final-state fermions.
The final-state fermions are treated as being massless, hence for 
non-collinear radiation their helicities are fixed by the left-handed 
interaction with the $W$ bosons: $\lambda_{f_1} = - \lambda_{f_1'} = +1$.
The matrix element for process (\ref{wvdw/decay/rad}) can be written as
\be
  \Delta_{\gamma}^{(+)}(M_W) = \frac{ie^{2}V_{f_1' f_1}}{\sqrt{2}s_W}
                               \biggl[ -\, Q_{f_1}\M^{\gamma\,(+)}_1
                                       + Q_{f_1'}\M^{\gamma\,(+)}_2
                                       + \M^{\gamma\,(+)}_3 \biggr],
\ee
with 
\begin{eqnarray}
  \M^{\gamma\,(+)}_1
   &=& \bar{u}(k_1')\,\es_1^*\,\omega_{-}\frac{\ks_1+\ks}{2k_1k}\,
       \es_{\gamma}\,v(k_1)
       \ = \ k_1^{\prime A}\,\e_{1\,\dot{B}A}^{\dagger}
       \frac{(k_1+k)^{\dot{B}C}}{2k_1k}\,\e_{\gamma\,\dot{D}C}\,k_1^{\dot{D}},
       \nonumber \\
  \M^{\gamma\,(+)}_2
  &=& \bar{u}(k_1')\,\es_{\gamma}\,\frac{\ks_1'+\ks}{2k_1'k}\,
      \es_1^*\,\omega_{-}v(k_1)
      \ = \ k_1^{\prime A}\,\e_{\gamma\,\dot{B}A}
      \frac{(k_1'+k)^{\dot{B}C}}{2k_1'k}\,\e_{1\,\dot{D}C}^{\dagger}
      \,k_1^{\dot{D}},
       \nonumber \\
  \M^{\gamma\,(+)}_3 
   &=& \bar{u}(k_1')\,\frac{\Vs(-p_1,\e_1^*,k,\e_{\gamma})}{2p_1k}\,
       \omega_{-}v(k_1)
       \ = \ k_1^{\prime A}\,
       \frac{V(-p_1,\e_1^*,k,\e_{\gamma})_{\dot{B}A}}{2p_1k}\,k_1^{\dot{B}}.
\end{eqnarray}
Here the vertex function $V$ can be taken from Eq.~(\ref{TGC}), but 
$(\ks+\ps_1)$ can be replaced by $2\ks$ as a result of the lowest-order
Ward identity of the $W^+$ boson.

For the calculation in the Weyl\,--\,van der Waerden formalism we choose
the same polarization basis as adopted for the on-shell W-pair example in
Sect.~\ref{app:wvdw/prod-born}, i.e.~$m_{3,4}=p_{4,3}$. For the definition
of the photon polarizations we choose the free gauge parameter $b$ 
in Eq.~(\ref{wvdw/photpol}) to be equal to $k_1$. 
A straightforward calculation gives the following results for the amplitudes
$\M^{\gamma\,(+)}_1(\lambda_1,\lambda)$:
\begin{eqnarray}
\M^{\gamma\,(+)}_1(+1,+1) &=& 
          -\,2\,\frac{\la k_1'p_3 \ra\la p_4k \ra^*}{\la k_1k \ra
          \la p_3p_4\ra^*}, \nonumber \\[1mm]
\M^{\gamma\,(+)}_1(+1,-1) &=& \M^{\gamma\,(+)}_1(-1,-1) 
                              \ = \ \M^{\gamma\,(+)}_1(0,-1) \ = \ 0, 
          \nonumber \\[1mm]
\M^{\gamma\,(+)}_1(-1,+1) &=& 
          -\,2\,\frac{\la k_1'p_4 \ra\la p_3k \ra^*}{\la k_1k \ra
          \la p_3p_4 \ra}, 
          \nonumber \\
\M^{\gamma\,(+)}_1(0,+1) &=& 
            \frac{\sqrt{2}}{\la k_1k \ra M_W} 
            \Bigl[   c\,\la k_1'p_4 \ra\la p_4k \ra^*
                   - \la k_1'p_3 \ra\la p_3k \ra^* \Bigr].
\end{eqnarray}
The corresponding expressions for $\M^{\gamma\,(+)}_{2,3}(\lambda_1,\lambda)$ 
read:
\begin{eqnarray}
\M^{\gamma\,(+)}_2(+1,+1) &=& 
            2\,\frac{\la k_1k_1' \ra\la k_1'p_3 \ra\la k_1p_4 \ra^*}
            {\la k_1k \ra\la k_1'k \ra\la p_3p_4 \ra^*}, \nonumber \\
\M^{\gamma\,(+)}_2(+1,-1) &=& 
           -\,2\, \frac{c\la p_3p_4 \ra\la k_1p_4 \ra^{*\,2}}
           {\la k_1k \ra^*\la k_1'k \ra^*\la p_3p_4 \ra^*}, \nonumber \\
\M^{\gamma\,(+)}_2(-1,+1) &=& 
            2\, \frac{\la k_1k_1' \ra\la k_1'p_4 \ra\la k_1p_3 \ra^*}
            {\la k_1k \ra\la k_1'k \ra\la p_3p_4 \ra}, \nonumber \\
\M^{\gamma\,(+)}_2(-1,-1) &=& 
            2\, \frac{\la k_1p_3 \ra^{*\,2}}{\la k_1k \ra^*\la k_1'k \ra^*}, 
            \nonumber \\
\M^{\gamma\,(+)}_2(0,+1) &=& 
            \frac{\sqrt{2}\la k_1k_1' \ra}{M_W\la k_1k \ra\la k_1'k \ra}
            \Bigl[  \la k_1'p_3 \ra\la k_1p_3 \ra^*
                   - c\,\la k_1'p_4 \ra\la k_1p_4 \ra^* \Bigr], \nonumber \\
\M^{\gamma\,(+)}_2(0,-1) &=& 
            -\,2\sqrt{2}c\,\frac{\la p_3p_4 \ra\la k_1p_3 \ra^*
             \la k_1p_4 \ra^*}{M_W\la k_1k \ra^*\la k_1'k \ra^*}, 
\end{eqnarray}
and
\begin{eqnarray}
2p_1k\,\M^{\gamma\,(+)}_3(+1,+1) &=&
             -\,\frac{2\la k_1'p_3\ra}{\la k_1k \ra\la p_3p_4 \ra^*}
             \Bigl[ \la k_1k_1' \ra\la k_1p_4 \ra^*\la k_1'k \ra^* 
                    - 2k_1k\la p_4k \ra^* 
             \Bigr],      
             \nonumber \\
2p_1k\,\M^{\gamma\,(+)}_3(+1,-1) &=&
             2c\,\frac{\la k_1'k \ra\la p_3p_4 \ra\la k_1p_4 \ra^{*\,2}}
                         {\la k_1k \ra^*\la p_3p_4 \ra^*},   
             \nonumber \\
2p_1k\,\M^{\gamma\,(+)}_3(-1,+1) &=&
             -\,\frac{2\la k_1'p_4 \ra}{\la k_1k \ra\la p_3p_4 \ra}
             \Bigl[   \la k_1k_1' \ra\la k_1p_3 \ra^*\la k_1'k \ra^* 
                    - 2k_1k\la p_3k \ra^* 
             \Bigr], 
             \nonumber \\
2p_1k\,\M^{\gamma\,(+)}_3(-1,-1) &=&
             -\,2\,\frac{\la k_1'k \ra\la k_1p_3 \ra^{*\,2}}{\la k_1k \ra^*}, 
             \nonumber \\
2p_1k\,\M^{\gamma\,(+)}_3(0,+1) &=&
             -\,\frac{\sqrt{2}}{\la k_1k \ra M_W}
             \biggl[   \la k_1k_1' \ra\la k_1'k \ra^*
             \Bigl(   \la k_1'p_3 \ra\la k_1p_3 \ra^*
                    - c\,\la k_1'p_4 \ra\la k_1p_4 \ra^* \Bigr)
             \nonumber \\[1mm]
                                & & 
             -\,2k_1k\Bigl( \la k_1'p_3 \ra\la p_3k \ra^*
             -c\,\la k_1'p_4 \ra\la p_4k \ra^* \Bigr)
             \biggr], 
             \nonumber \\
2p_1k\,\M^{\gamma\,(+)}_3(0,-1) &=&
             2\sqrt{2}c\,\frac{\la k_1'k \ra\la p_3p_4\ra\la k_1p_3 \ra^*
             \la k_1p_4 \ra^*}{M_W \la k_1k \ra^*}. 
\end{eqnarray}
The expressions for the charge-conjugate process, describing the decay
of the $W^-$ boson, can be obtained as follows:
\be
  \Delta_{\gamma}^{(-)}(M_W) = -\,\frac{ie^{2}V_{f_2' f_2}}{\sqrt{2}s_W}
                               \biggl[ -\, Q_{f_2}\M^{\gamma\,(-)}_1
                                       + Q_{f_2'}\M^{\gamma\,(-)}_2
                                       + \M^{\gamma\,(-)}_3 \biggr],
\ee
where
\be
  \M_j^{\gamma\,(-)}(\lambda_2,\lambda) 
  = 
  \Bigl[ \M_j^{\gamma\,(+)}(-\lambda_2,-\lambda) \Bigr]^*,  
  \quad\quad \mbox{with} \quad 
  (k_1,k_1',p_3,p_4) \to (k_2,k_2',p_4,p_3).
\ee

When the above matrix elements for real-photon radiation from the
decay stages are combined with the lowest-order matrix element for the
production stage, presented in App.~\ref{app:wvdw/prod-born}, one obtains 
the DPA limit of the full matrix elements $\M_{\pm}$ in Eqs.~(\ref{dec+rad}) 
and (\ref{dec-rad}).

\subsection{Radiation of collinear photons}
\label{app:collrad}

Up to now we have only discussed the case of non-collinear photon radiation,
which allowed us to neglect the fermion masses and the possibility of 
spin-flip in the initial state. The picture changes, however, if the 
radiated photons are sufficiently collinear with one of the external fermions.
In such cases factorization takes place, i.e.~the matrix element squared 
including collinear radiation can be approximately written in terms of the 
lowest-order matrix element squared and collinear factors.

Let us first consider collinear photon radiation in the direction of
one of the light fermions in the production stage of the process,
e.g.~the positron. In that case the matrix element squared can be written in 
the following form~\cite{ww-review}:
\be
  \sum_{\lambda} 
  |\M_{\sss{coll},\,e^+}(q_1,\sigma_1,q_2,\sigma_2,k,\lambda)|^2 
  \approx 
  e^2\,f_{\sss{coll}}^{(\sigma_2)}(q_1,\sigma_1,k)\,
  |\M^0_{\sss{DPA}}(x_1q_1,-\sigma_2,q_2,\sigma_2)|^2,
\ee
where $x_1 = (E-k_0)/E$ is the ratio of the positron energy after and before 
photon radiation, $\sigma_{1,2}$ are the helicities of the $e^{\pm}$,
and
\be
  f_{\sss{coll}}^{(\sigma_2)}(q_1,\sigma_1,k)
  = \delta_{(\sigma_1,-\sigma_2)}\,
    \Biggl[   \frac{1+x_1^2}{x_1(1-x_1)}\,\frac{1}{q_1k}
            - \frac{1+x_1^2}{2x_1}\,\frac{m_e^2}{(q_1k)^2}
    \Biggr]
    + \delta_{(\sigma_1,\sigma_2)}\,\frac{(1-x_1)^2}{2x_1}\,
    \frac{m_e^2}{(q_1k)^2}.
\ee
The last term in this collinear factor gives rise to the so-called spin-flip,
which allows the positron to have the same helicity as the electron.
Note that we have only indicated the momenta and helicities of the relevant 
particles ($e^{\pm},\gamma$) and that the photon helicities are summed over,
as the photon cannot be detected anyway. Collinear radiation in the
direction of the initial-state electron can be obtained in the same way,
with the role of the $e^+$ and $e^-$ interchanged. If the initial-state 
particles are not polarized, as is the case at LEP2, the collinear factor 
takes on the well known form
\be 
  \sum_{\sigma_1} f_{\sss{coll}}^{(\sigma_2)}(q_1,\sigma_1,k) 
  =
  \frac{1+x_1^2}{x_1(1-x_1)}\,\frac{1}{q_1k} - \frac{m_e^2}{(q_1k)^2}.
\ee
When the photon angles are integrated out, the terms $\propto 1/q_1k$
yield contributions of the large-logarithmic type [$\propto \log(s/m_e^2)$],
whereas the term $\propto m_e^2/(q_1k)^2$ gives rise to additional $\OO(1)$
contributions, which would have been neglected in a massless treatment of 
the initial state.

In the case of collinear photon radiation in the direction of one of the 
final-state fermions, say the fermion $f_2$ from the $W^-$ decay, the
factorization reads
\be
  \sum |\M_{\sss{coll},\,f_2}(k_1,k_1',k_2,k_2',k)|^2 
  \approx 
  e^2 Q_{f_2}^2\, \Biggl[   \frac{1+y_2^2}{1-y_2}\,\frac{1}{k_2k}
                          - \frac{m_{f_2}^2}{(k_2k)^2}
                  \Biggr]       
  |\M^0_{\sss{DPA}}(k_1,k_1',k_2/y_2,k_2')|^2,
\ee
where the summation is performed over all final-state helicities and 
$y_2 = E_4/(E_4+k_0)$ is the ratio of the $f_2$ energy after and before 
photon radiation. The other final-state collinear factors can be obtained 
in the same way.


\section{Special integrals for semi-soft photon radiation}
\label{app:semi-soft_functions}

In this appendix we have a closer look at the inclusive treatment of
the photon in shifted Breit--Wigner resonances $1/[D_i+2kp_i]$ in the
vicinity of the $M_i^2$ resonance (see Sect.~\ref{sec:corr}).
We start off with factorizable real-photon radiation, involving the ratios
\be 
  \frac{|D_i|^2}{|D_i+2kp_i|^2} 
  = 
  \biggl[ \frac{1}{D_i^{*}+2kp_i} - \frac{1}{D_i+2kp_i} \biggr]
  \frac{|D_i|^2}{2iM_W\Gamma_W}
  =
  \frac{|D_i|^2}{M_W\Gamma_W}\,\real\frac{i}{D_i+2kp_i}.
\ee
In order to study the phenomenon of hard-photon suppression we consider 
the generic integrals
\be 
  I_n 
  = 
  \frac{|D_i|^2}{M_W\Gamma_W}\,\real
  \left\{\,
  \int\limits_{\lambda_s<k_0<\Lambda} \frac{d\vec{k}}{(2\pi)^3 2k_0}\,
  \frac{M_W^{n-2}}{k_0^n}\,\frac{i}{D_i+2kp_i}
  \right\}
\ee
for $n=1$ or 2. The integration is performed over the photon angles and 
the photon-energy range $\lambda_s < k_0 < \Lambda$, where $\lambda_s$
is a soft-photon cut-off ($\lambda_s \ll \Gamma_W$).       
For $n=2$ this integral quantifies the influence of the shifted resonance
on the $M_i^2$ distribution in the vicinity of the pole $M_i^2=M_W^2$.
For $n=1$ it quantifies the effect of $\OO(k)$ shifts in the definition 
of the DPA residues. In the latter case we find
\be
  I_1
  =
  -\, \frac{|D_i|^2}{16 \pi^2 E \beta M_W^2 \Gamma_W} \,
  \imag \Biggl\{   \Li\biggl( \frac{-1+\beta}{z} \biggr)
                 - \Li\biggl( \frac{-1-\beta}{z} \biggr) 
        \Biggr\},
\ee
with $z = D_i/(2E\Lambda)$ and $\Li$ the usual dilogarithm. One can 
immediately read off that $I_1$ is suppressed by $\OO(\Gamma_W/E)$, 
irrespective of the precise value for $\Lambda$. For $n=2$ the integral reads
\be
  I_2
  =
  -\, \frac{1}{4 \pi^2} \,\imag
  \Biggl\{ \frac{D_i^*}{M_W \Gamma_W}
           \Biggl[ 1 + \log\biggl( \frac{D_i}{2E\lambda_s} \biggr) 
                   + \frac{z+1-\beta}{2\beta}\,\log(z+1-\beta)
                   - \frac{z+1+\beta}{2\beta}\,\log(z+1+\beta)
           \Biggr]
  \Biggr\}.
\ee
This type of integral will lead to an $\OO(1)$ contribution.
The dependence on the cut-off $\Lambda$, however, is suppressed by
$\OO(\Gamma_W^2/\Lambda^2)$. So, the more energetic the photon is the more 
suppressed its contribution will be. Hence, as soon as $\Lambda$ is
taken to be much larger than $\Gamma_W$ it can safely be replaced by 
infinity. 

Based on the latter observation, we can now list the relevant integrals 
needed for the inclusive treatment of final-state radiation effects involving
shifted Breit--Wigner resonances (see columns 3,4 of Table~\ref{tab:1} and
column 3 of Table~\ref{tab:2}). For the radiation from the $W^+$-boson
decay stage the following four integrals are required:
\begin{eqnarray}
  && \int\limits_{k_0>\lambda_s}\frac{d\vec{k}}{(2\pi)^3 2k_0}
     \frac{1}{(2kk_1)^2 [D_1+2kp_1]}
     = \frac{1}{16\pi^2 m_{f_1}^2 D_1}
       \log\biggl( \frac{D_1 E_3}{\lambda_s M_W^2} \biggr),
       \nonumber \\
  && \int\limits_{k_0>\lambda_s}\frac{d\vec{k}}{(2\pi)^3 2k_0}
     \frac{1}{(2kk_1)(2kp_1)[D_1+2kp_1]}
     = -\,\frac{1}{16\pi^2 M_W^2 D_1}
       \Biggl[ \log\biggl( \frac{\lambda_s M_W^2}{D_1 E_3} \biggr) 
               \log\biggl( \frac{M_W^2}{m_{f_1}^2} \biggr)
       \nonumber \\
  && \hphantom{\int\limits_{k_0>\lambda_s}\frac{d\vec{k}}{(2\pi)^3 2k_0}
     \frac{1}{(2kk_1)(2kp_1)[D_1+2kp_1]} =}
       +\,\Li\biggl( 1-\frac{1-\beta}{2}\,\frac{E}{E_3} \biggr)
       +  \Li\biggl( 1-\frac{1+\beta}{2}\,\frac{E}{E_3} \biggr)
       \Biggr],
       \nonumber \\
  && \int\limits_{k_0>\lambda_s}\frac{d\vec{k}}{(2\pi)^3 2k_0}
     \frac{1}{(2kp_1)^2 [D_1+2kp_1]}
     = -\,\frac{1}{16\pi^2 M_W^2 D_1}
       \Biggl[ \log\biggl( \frac{2\lambda_s M_W}{D_1} \biggr) + 1
               +\,\frac{1}{2\beta}\log\biggl( \frac{1-\beta}{1+\beta} \biggr)
       \Biggr],
       \nonumber \\
  && \int\limits_{k_0>\lambda_s}\frac{d\vec{k}}{(2\pi)^3 2k_0}
     \frac{1}{(2kk_1)(2kk_1')[D_1+2kp_1]}
     = \int\limits_{k_0>\lambda_s}\frac{d\vec{k}}{(2\pi)^3 2k_0}
       \frac{1}{(2kp_1)[D_1+2kp_1]}\Biggl[ \frac{1}{2kk_1} + \frac{1}{2kk_1'}
                                   \Biggr].
       \nonumber \\
\end{eqnarray}
From these integrals one can determine the correction factor corresponding 
to the $|\I_{+}^2|$ term in Eq.~(\ref{corr/real/mtrx}):
\begin{eqnarray}
  \lefteqn{
           -\,\int\limits_{k_0>\lambda_s}
           \frac{d\vec{k}}{(2\pi)^3 2k_0}\,|\I_{+}^2|
           = -\,\frac{\alpha}{\pi}\,\real
           \Biggl\{ \frac{iD_1^*}{M_W\Gamma_W}
           \Biggl[ -\,\log\biggl( \frac{2\lambda_s M_W}{D_1} \biggr) -1
              - \frac{1}{2\beta}\log\biggl( \frac{1-\beta}{1+\beta} \biggr)
          }\quad\quad  
      \nonumber \\
  & &    +\, Q_{f_1}^2 \Biggl\{ 
              \log\biggl( \frac{\lambda_s M_W^2}{D_1 E_3} \biggr) 
              \biggl[ \log\biggl( \frac{M_W^2}{m_{f_1}^2} \biggr) - 1 \biggr]
              + \Li\biggl( 1-\frac{1-\beta}{2}\,\frac{E}{E_3} \biggr)
              + \Li\biggl( 1-\frac{1+\beta}{2}\,\frac{E}{E_3} \biggr)
                       \Biggr\} 
      \nonumber \\
  & &    +\, Q_{f_1'}^2 \Biggl\{ 
              \log\biggl( \frac{\lambda_s M_W^2}{D_1 E_3'} \biggr) 
              \biggl[ \log\biggl( \frac{M_W^2}{m_{f_1'}^2} \biggr) - 1 \biggr]
              + \Li\biggl( 1-\frac{1-\beta}{2}\,\frac{E}{E_3'} \biggr)
              + \Li\biggl( 1-\frac{1+\beta}{2}\,\frac{E}{E_3'} \biggr)
                        \Biggr\}    
      \Biggr]
      \Biggr\}.
      \nonumber \\ 
\end{eqnarray}
Here $E_3'$ denotes the energy of $f_1'$, i.e.~$E_3' = E-E_3$.
The correction factor corresponding to the $|\I_{-}^2|$ term in 
Eq.~(\ref{corr/real/mtrx}) is obtained by replacing $(D_1,f_1,f_1',E_3,E_3')\,$
by $\,(D_2,f_2,f_2',E_4,E_4')$.

Finally we study the hard-photon suppression for the non-factorizable 
corrections. To this end we consider the integrals
\be 
  J_n 
  = 
  \real\left\{\,
  \int\limits_{\lambda_s<k_0<\Lambda} \frac{d\vec{k}}{(2\pi)^3 2k_0}\,
  \frac{M_W^{n-2}}{k_0^n}\,\frac{D_i}{D_i+2kp_i}
  \right\}
\ee 
for $n=1$ and $2$.
From the results for $I_{1,2}$ one straightforwardly obtains
\be
  J_1
  =
  \frac{1}{16 \pi^2 E \beta M_W} \,
  \real\Biggl\{ D_i \Biggl[   \Li\biggl( \frac{-1+\beta}{z} \biggr)
                            - \Li\biggl( \frac{-1-\beta}{z} \biggr) 
                    \Biggr]
       \Biggr\}
\ee
and 
\be
  J_2
  =
  \frac{1}{4 \pi^2} \,\real
  \Biggl\{ 1 + \log\biggl( \frac{D_i}{2E\lambda_s} \biggr)
           + \frac{z+1-\beta}{2\beta}\,\log(z+1-\beta) 
           - \frac{z+1+\beta}{2\beta}\,\log(z+1+\beta)
  \Biggr\}.
\ee
Again a suppression of $\OO(\Gamma_W/E)$ is observed for $n=1$,
whereas for $n=2$ the dependence on the cut-off $\Lambda$ is suppressed by 
$\OO(M_W\Gamma_W/[E\Lambda])$. So, again $\Lambda$ can be replaced by
infinity if it is sufficiently large. For explicit expressions for the
non-factorizable corrections we refer to the 
literature~\cite{nf-corr/bbc,nf-corr/ddr}.


\end{document}